\documentclass[aps,reprint,onecolumn,pre,tighten,nofootinbib,eqsecnum]{revtex4-2}
\usepackage{comment} 
\usepackage{graphicx}
\usepackage{hyperref}
\usepackage{dcolumn}
\usepackage[utf8]{inputenc}
\usepackage{bm}
\usepackage{amsmath, amsfonts, amssymb}
\usepackage{mathrsfs}
\usepackage{color}
\usepackage{multirow}
\usepackage{tikz}
\usetikzlibrary{shapes.geometric}
\usetikzlibrary{shapes.misc}
\allowdisplaybreaks
\newcommand{\vt}{v_\mathrm{T}}

\newcommand{\erfc}{\mathrm{erfc}}

\begin{document}
\title{Kinetic theory of dry granular particles under simple shear: steady flows and relaxation dynamics}

%%%%%%%%%%%%%%%%%%%%%%%%%%%%%%
\author{Shunsuke Iizuka}
\email[e-mail: ]{shunsuke@st.go.tuat.ac.jp}
\altaffiliation{Present address: Tokio Marine \& Nichido Systems Co.,Ltd., 2-1-1 Tsurumaki, Tama, Tokyo 206-8510, Japan}
\affiliation{Department of Mechanical Systems Engineering,
Tokyo University of Agriculture and Technology, 
2-24-16 Naka-cho, Koganei, Tokyo 184-8588, Japan}

%%%%%%%%%%%%%%%%%%%%%%%%%%%%%%
\author{Satoshi Takada}
\email[Corresponding author, e-mail: ]{takada@go.tuat.ac.jp}
\affiliation{Department of Mechanical Systems Engineering,
Tokyo University of Agriculture and Technology, 
2-24-16 Naka-cho, Koganei, Tokyo 184-8588, Japan}
%%%%%%%%%%%%%%%%%%%%%%%%%%%%%%
\author{Hisao Hayakawa}
\email[e-mail: ]{hisao@yukawa.kyoto-u.ac.jp}
\affiliation{Center for Gravitational Physics and Quantum Information, Yukawa Institute for Theoretical Physics, 
Kyoto University,
Kitashirakawa Oiwakecho, Sakyo-ku, Kyoto 606-8502, Japan}
%%%%%%%%%%%%%%%%%%%%%%%%%%%%%%
\date{\today}
%%%%%%%%%%%%%%%%%%%%%%%%%%%%%%
\begin{abstract}
We develop a kinetic theory for dry granular particles under simple shear flow using the Enskog approximation combined with Grad's moment method for moderately dense granular gases with the volume fraction $\varphi\le 0.5$. 
The collisional moments and stresses are expressed in terms of generalized hypergeometric functions, retaining all orders in the dimensionless shear-rate parameter within the adopted Grad closure.
Within this framework, we derive a complete set of equations governing the kinetic stress and the collisional contribution to the stress tensor. 
The steady-state solutions reproduce the shear viscosity over a broad parameter range and improve the predictions of the kinetic temperature relative to the anisotropic Gaussian theory at the parameter values examined.
Temperature still shows appreciable deviations under strongly inelastic conditions.
Additionally, the theory captures the overall trends of the normal stress differences, while their quantitative accuracy depends on density and restitution coefficient.
We further examine relaxation following an instantaneous reduction in $\varphi$ from $0.35$ to $0.20$ at a restitution coefficient of $0.90$. 
The theory reproduces the temperature and viscosity relaxation for this protocol.
\end{abstract}
\maketitle

%%%%%%%%%%%%%%%%%%%%%%%%%%%%%%
\section{Introduction}
%%%%%%%%%%%%%%%%%%%%%%%%%%%%%%
Granular materials, such as sand, snow, and soil, are ubiquitous and have a wide range of applications across science and engineering. 
Despite their prevalence, understanding the behavior of granular systems remains a significant challenge due to the absence of a well-defined equilibrium state. 
Unlike conventional fluids or solids, granular materials can exhibit diverse phases depending on their density and the nature of external forces. 
However, these phases cannot be fully described by standard thermodynamic principles~\cite{Andreotti13}. 
Even the simplest phase, a dilute granular gas, is inherently unstable and cannot be sustained without continuous external energy input~\cite{Goldhirsch93,Haff83,Brilliantov10,Garzo,Hayakawa2025}.

One of the most successful theoretical frameworks for describing granular flows in the gaseous or liquid-like regimes is granular kinetic theory. 
This approach introduces the concept of granular temperature, which quantifies the velocity fluctuations of individual grains~\cite{Ogawa78}. 
Building on this idea, many researchers have developed the kinetic theory and reached several points of consensus~\cite{Hayakawa2025,Savage81,Jenkins83,Lun84,Jenkins85a,Jenkins85b,Lun91,Goldstein95,Brey98,Garzo99,Brilliantov10,Garzo}:
1) Granular flows can be modeled by compressible hydrodynamic equations that include an energy dissipation term in the temperature equation, provided the grain density is not too high.
2) The freely cooling state~\cite{Goldhirsch93,Haff83}, characterized by a decrease in granular temperature due to inelastic collisions, is unstable and leads to the formation of cluster-like structures.
3) To reach a steady state, external forcing or energy input is required to compensate for the continuous energy loss.
4) While dilute granular gases can be accurately described by the inelastic Boltzmann equation~\cite{Brey98,Brilliantov10}, moderately dense systems require the Enskog correction to account for excluded volume effects~\cite{Jenkins83,Lun84,Jenkins85a,Jenkins85b,Lun91,Goldstein95,Garzo99,Garzo}.

Several researchers have applied granular kinetic theory not only to simple shear flows but also to granular flows on inclined planes~\cite{Lun94,Chialvo13,Mitarai05,Mitarai07,Jenkins10,Berzi18,Berzi20}. 
These studies have demonstrated that the kinetic theory developed by Garz\'{o} and Dufty~\cite{Garzo99} provides reliable predictions for volume fractions $\varphi < 0.49$, i.e., below the Alder transition point~\cite{Wood57,Alder57}, particularly in the context of simple shear flows~\cite{Chialvo13}.
Approximate simple shear conditions are often realized in the bulk regions of granular flows down inclined slopes~\cite{Mitarai05,Mitarai07,Silbert01}, although such situations are often realized only for $\varphi>0.49$. 
Consequently, a major focus within the granular kinetic theory community has been to extend theoretical frameworks to incorporate correlation effects at higher volume fractions ($\varphi \ge 0.49$)~\cite{Jenkins10,Berzi18,Berzi20,Suzuki15}.

However, this emphasis is somewhat paradoxical, as the kinetic theory developed by Garz\'{o} and Dufty~\cite{Garzo99} was originally formulated for freely cooling states rather than shear-driven flows.\footnote{Concerning this point, when one of the authors showed some figures from Ref.~\cite{Chialvo13} on the occasion of his seminar in 2015 at Extremadura, Garz\'{o} remarked, ``This is outside the range of applicability, so you should not trust it too much.''
He also suggested that we should rely on Ref.~\cite{Santos04} instead of Ref.~\cite{Garzo99} for describing sheared granular flows.}
While simple shear flow is spatially uniform, the viscosity in their framework~\cite{Garzo99} only arises in the context of inhomogeneous flows. 
In steady uniform shear, viscous heating must balance collisional cooling. 
At finite inelasticity, this balance fixes the reduced shear rate and prevents an independently controlled Navier-Stokes limit.
We therefore employ a finite-shear moment description rather than directly applying the Navier-Stokes transport coefficients.
In reality, sheared granular systems exhibit normal stress differences that are intrinsically coupled to the shear stress~\cite{Santos04,Garzo,Campbell89,Goldhirsch96,Alam03,Lutsko04,Lutsko05,Saha14,Saha16}, a feature not accounted for in the original Garz\'{o}-Dufty theory~\cite{Garzo99}. 
Interestingly, similar normal stress differences have also been reported in studies of inertial suspensions~\cite{Tsao95,Sangani96,Koch99,Koch01,Garzo12,Saha17,Alam19,Saha20,Hayakawa19,Hayakawa17,Takada20,Sugimoto20,Takada22a,Takada22b}. 
This suggests that a kinetic theory tailored specifically for shear flows, rather than one based on freely cooling states, should be adopted to accurately describe these systems.

There are, at least, two appropriate theoretical tools to describe sheared granular flows.
One is using Grad's approximation~\cite{Grad49,Garzo,Garzo13,Santos98,Montanero99,Santos04,Hayakawa19,Hayakawa17,Takada20,Sugimoto20,Takada22a,Takada22b}.
Another is the anisotropic Gaussian theory, which describes sheared granular flows~\cite{Lutsko04,Saha14, Saha16,Saha17,Alam19,Saha20}.
Although both theoretical frameworks can describe an anisotropic stress tensor,
Saha and Alam~\cite{Saha14} suggested that their theory for two-dimensional systems is slightly superior to the theory based on Grad's approximation, at least for one of the normal stress differences, the pressure, and the stress ratio~\cite{Grad49,Garzo2012}.
The comparison in Ref.~\cite{Saha14} concerns dilute two-dimensional systems. 
We therefore assess the two closures for three-dimensional flows using separate comparisons of temperature, viscosity, and normal stress differences.\footnote{We note that Alam et al.~\cite{Alam19} also demonstrated the superiority of anisotropic Gaussian theory over an old Grad-like theory~\cite{Sangani96}, but the modern Grad-like theory~\cite{Garzo2012,Hayakawa19, Hayakawa17, Takada20, Sugimoto20, Takada22a, Takada22b} includes the improved treatments from the old one~\cite{Sangani96}.
Therefore, the demonstration in Ref.~\cite{Alam19} is not directly connected with the superiority of the anisotropic Gaussian theory.
This is partially because the present authors know that Grad's approximation gives precise results, at least for inertial suspensions~\cite{Hayakawa19, Hayakawa17, Takada20, Sugimoto20, Takada22a, Takada22b}.}

So far, most previous theoretical studies are only interested in steady granular flows.
This is strange because we have recognized the existence of anomalous relaxation processes and memory effects~\cite{Keim_2019}, at least in jammed granular systems, even for frictionless grains~\cite{Kumar16,Babu21,Nagasawa19,Kawasaki24}. 
Thus, we analyze relaxation processes to reach steady states in simple shear.

In this paper, we formulate the kinetic theory for dense granular flows under a simple shear based on the Enskog kinetic theory with Grad's approximation~\cite{Grad49,Jenkins85a,Jenkins85b,Santos98,Montanero99,Garzo12,Hayakawa19,Hayakawa17,Takada20,Sugimoto20,Takada22a}.
Within Grad's approximation, we first sum up all the expansions in terms of the dimensionless shear rate introduced in previous literature~\cite {Santos98,Montanero99,Takada20}.
We compare the predictions with event-driven simulations for $\varphi\le 0.5$ without introducing fitting parameters. The theory reproduces the viscosity reasonably well over much of the range examined. 
Compared with the anisotropic Gaussian theory, it gives smaller temperature errors at all points listed in Table \ref{table:comparison} and smaller viscosity errors at most of these points, although substantial temperature deviations remain under strongly inelastic conditions. 
The relative accuracy depends on the observable: the anisotropic Gaussian theory provides a quantitative description of the normal stress differences as good as those predicted by Grad's approximation. 
We also examine the relaxation of the temperature and viscosity following a prescribed reduction in the volume fraction.

The organization of this paper is as follows:
In Sec.~\ref{Sec:setup}, we explain the setup of our analyzed system.
In Sec.~\ref{Sec:kinetics}, we outline the kinetic theory for sheared dry granular flows, including the framework and an approximate kinetic theory.
In Sec.~\ref{sec:steady_rheology}, we present our theoretical results for steady rheology by comparing our results with the simulation and the other theoretical predictions. 
In Sec.~\ref{Sec:relaxation}, we analyze relaxation processes to reach a steady state.
In Sec.~\ref{Sec:Discussion}, we discuss our results and future directions.
In Sec.~\ref{Sec:Summary}, we summarize our obtained results.
Technical derivations and additional numerical checks are provided in the appendices.

%%%%%%%%%%%%%%%%%%%%%%%%%%%%%%
\section{Setup}\label{Sec:setup}
%%%%%%%%%%%%%%%%%%%%%%%%%%%%%%

%%%%%%%%%%%%%%%%%%%%%%%%%%%%%%
\begin{figure}
    \centering
    \includegraphics[width=0.5\linewidth]{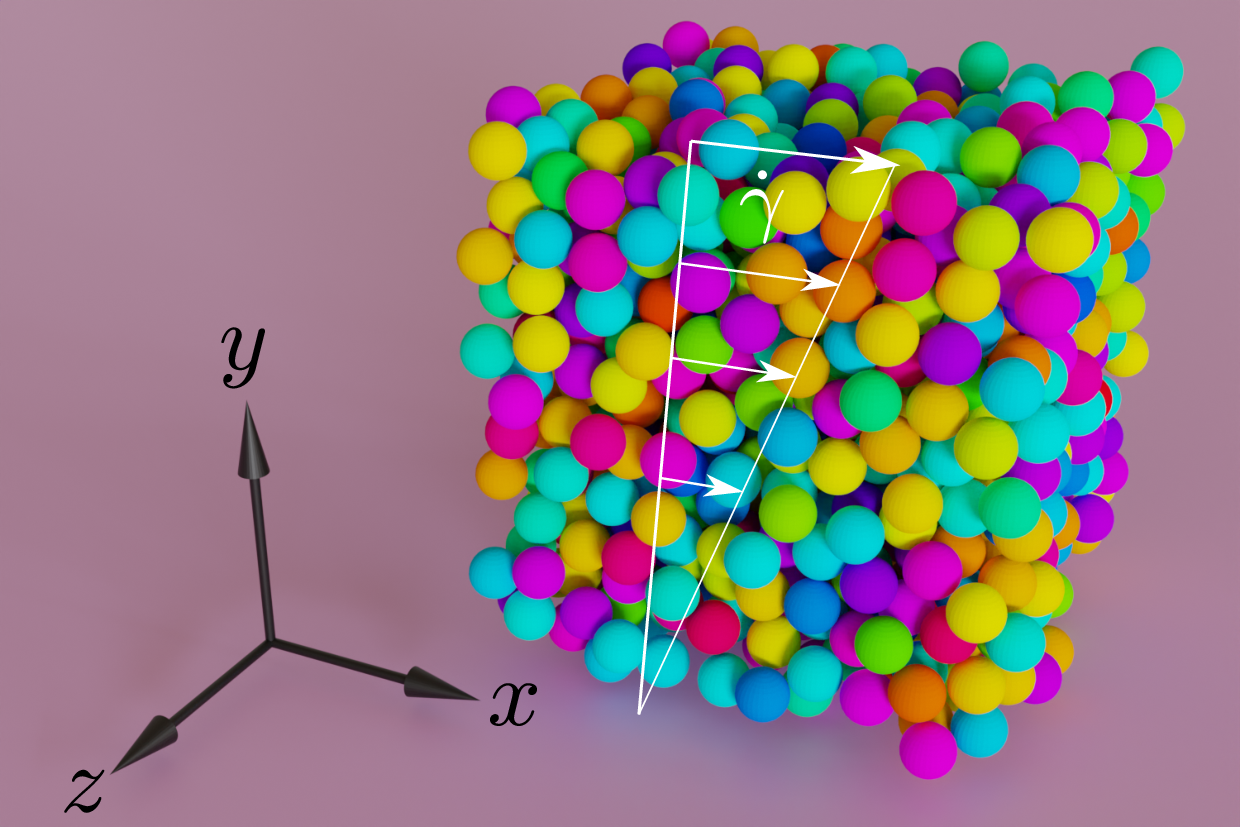}
    \caption{Setup of the system for the volume fraction $\varphi=0.45$.
    Monodisperse particles are distributed randomly at $t=0$.
    Then, a shear is applied to the system with the fixed shear rate $\dot\gamma$.}
    \label{fig:setup}
\end{figure}
%%%%%%%%%%%%%%%%%%%%%%%%%%%%%%

\begin{figure}[hbtp]
    \centering
    \includegraphics[width=0.25\linewidth]{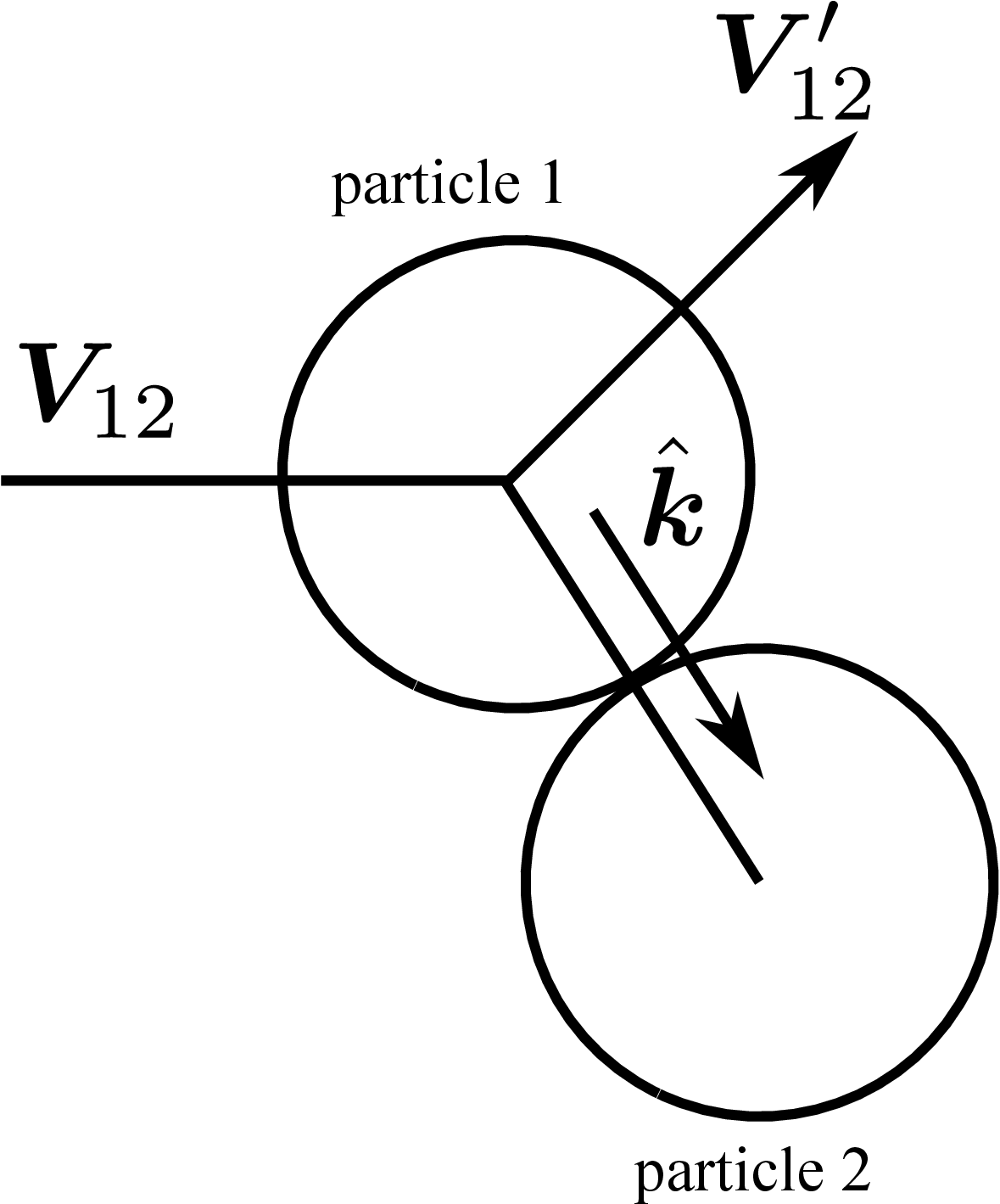}
    \caption{A schematic illustration of the interparticle collision between two particles, viewed in the reference frame fixed to particle 2.
    }
    \label{fig:collision}
\end{figure}

We consider a collection of monodisperse particles whose mass and diameter are, respectively, given by $m$ and $d$, distributed in a three-dimensional system under the influence of a simple shear with the shear rate $\dot\gamma$.
We assume that the interaction among particles is hard-core, and there is energy loss during a collision.
We introduce the restitution coefficient $e$ satisfying $0< e < 1$ to characterize inelastic collisions. 
Although the real restitution coefficient $e$ should depend on the impact speed and the tangential loss of energy~\cite{Bridges84,Kuwabara87, Brilliantov96,Labous97,Louge02,Kuninaka04,Kuninaka09,Saitoh10,Murakami14}, we ignore such complicated features of inelastic collisions for simplicity.
The post-collisional velocities $(\bm v_1^\prime,\bm v_2^\prime)$ are related to the pre-collisional velocities $(\bm v_1,\bm v_2)$ by
\begin{equation}\label{eq:collision_rule}
    \bm{v}_1^\prime 
    = \bm{v}_1 
    -\frac{1+e}{2} \left(\bm{v}_{12}\cdot \hat{\bm{k}}\right)\hat{\bm{k}},\quad
    %%%
    \bm{v}_2^\prime 
    = \bm{v}_2 
    +\frac{1+e}{2} \left(\bm{v}_{12}\cdot \hat{\bm{k}}\right)\hat{\bm{k}},
\end{equation}
where $\bm{v}_{12}:= \bm{v}_1-\bm{v}_2$ is the relative velocity and $\hat{\bm{k}}:= \bm{r}_{12}/|\bm{r}_{12}|$ is the unit vector parallel to the relative position $\bm{r}_{12}:= \bm{r}_1-\bm{r}_2$.
See Fig.~\ref{fig:collision} for the schematic picture of the collision.

%%%%%%%%%%%%%%%%%%%%%%%%%%%%%%
%%%%%%%%%%%%%%%%%%%%%%%%%%%%%%

To suppress the boundary effect and simplify the theoretical treatment, we assume that the shear flow exists in a bulk region under a periodic boundary condition.
To simulate such a system, we impose the Lees-Edwards boundary condition~\cite{Lees72} with the aid of the SLLOD dynamics \cite{Evans, Evans84} to suppress the shear band and inhomogeneity near the boundaries.\footnote{The main reason for adopting the SLLOD is the simplification of the theoretical treatment.
We also note that the preparation of the initial state may not use the SLLOD dynamics under the Lees-Edwards boundary condition, as shown in Secs.  \ref{sec:steady_rheology} and \ref{Sec:relaxation}.
}
In this paper, we perform event-driven molecular dynamics simulations~\cite{Poschel,Hayakawa19,Takada20}.
Each simulation contains $N=10^3$ particles in a cubic box of side length $L=[\pi N/(6\varphi)]^{1/3}d$, where $\varphi$ is the particle volume fraction.\footnote{See Appendix~\ref{sec:finite_size} for an assessment of finite-size effects.}
A typical simulation snapshot at $\varphi=0.45$ is shown in Fig.~\ref{fig:setup}.

For each parameter set, we perform $100$ independent realizations starting from distinct nonoverlapping particle configurations.
The initial peculiar velocities are sampled from an isotropic Maxwellian distribution at temperature $T_0$, with $\Delta T=\delta T=0$ and vanishing initial kinetic shear stress.
The initial temperature $T_0$ is chosen as the temperature predicted by the dilute-limit Bagnold solution~\cite{Santos04}.

Steady-state observables are time-averaged over $80/\dot\gamma \leq t \leq 100/\dot\gamma$, corresponding to a measurement interval of duration $\Delta t_{\mathrm{av}}=20/\dot\gamma$.
The kinetic temperature and kinetic stress are sampled at uniform time intervals of $\Delta t_{\mathrm{samp}}=0.01/\dot\gamma$.
For each realization, the time-averaged kinetic stress is calculated as
\begin{equation}
    \overline{\sigma_{\alpha\beta}^{k}}
    =-\frac{m}{L^3M}
    \sum_{s=1}^{M}\sum_{i=1}^{N}
    V_{i,\alpha}(t_s)V_{i,\beta}(t_s),
\end{equation}
where $L^3$ is the system volume, $M$ is the number of samples in the measurement interval, and $\bm{V}_i=\bm{v}_i-\dot\gamma y_i\bm{e}_x$ is the peculiar velocity relative to the imposed shear flow.
The overbar denotes a time average within one realization.

The collisional stress is obtained by accumulating the collision impulses over the same measurement interval:
\begin{equation*}
    \overline{\sigma_{\alpha\beta}^{c}}
    =-\frac{1}{L^3\Delta t_\mathrm{av}}
    \sum_{c}J_{c,\alpha}r_{c,\beta},
\end{equation*}
where the sum runs over all binary collisions during the measurement interval.
Here, $\bm{r}_c$ is the contact vector directed from particle $j$ to particle $i$, and $\bm{J}_c=m(\bm{v}_i^{+}-\bm{v}_i^{-})$ is the collision impulse acting on particle $i$, with the superscripts $-$ and $+$ denoting velocities immediately before and after the collision, respectively.
The contact vector and relative velocity are evaluated with the appropriate Lees--Edwards image corrections.
Each collision is counted once.
The negative signs in the kinetic and collisional contributions follow the Cauchy-stress convention adopted in this paper.

For each realization, the time-averaged total stress is $\overline{\sigma_{\alpha\beta}}=\overline{\sigma_{\alpha\beta}^{k}}+\overline{\sigma_{\alpha\beta}^{c}}$.
The viscosity is calculated by dividing the time-averaged shear stress by the imposed shear rate.
The normal stress differences are calculated from differences between the time-averaged diagonal stresses, normalized by the time-averaged pressure $\overline{P}=-\sum_{\alpha}\overline{\sigma_{\alpha\alpha}}/3$.
Thus, the normalized normal stress differences are ratios of time-averaged stresses.
The reported values are averages over $100$ independent realizations, and the error bars represent the standard error of the mean across these realizations.
The sampling unit is one realization rather than an individual time sample.

Because particle trajectories are advanced from one collision event to the next, the simulations do not use a fixed integration time step.
The sampling interval specifies the output frequency rather than a time step for integrating the equations of motion.

We control the volume fraction $\varphi$ and restitution coefficient.
The shear rate $\dot\gamma$ sets the time scale rather than serving as an independent control parameter because the shear stress and pressure obey Bagnold scaling, i.e., they are proportional to $\dot\gamma^2$~\cite{Bagnold54} for the uniform simple shear flows.

We further examine relaxation following an instantaneous reduction in the volume fraction from $0.35$ to $0.20$ at a restitution coefficient of $0.90$.
The relaxation data are collected using time windows of width $\Delta t=0.1/\dot\gamma$.
The theory reproduces the temperature and viscosity relaxation for this protocol, although the instantaneous Enskog closure does not describe the initial recovery of near-contact pairs after the expansion.

%%%%%%%%%%%%%%%%%%%%%%%%%%%%%%
\section{Kinetic theory}\label{Sec:kinetics}
%%%%%%%%%%%%%%%%%%%%%%%%%%%%%%
We formulate the kinetic theory for uniform shear flow and close the resulting moment equations using the Enskog and Grad approximations.

%%%%%%%%%%%%%%%%%%%%%%%%%%%%%%
\subsection{Framework}\label{sec:framework}
%%%%%%%%%%%%%%%%%%%%%%%%%%%%%%
We consider the time evolution of the one-body distribution function $F(\bm{r}_1,\bm{v}_1,t)$ of the particle position $\bm{r}_1$ and its velocity $\bm{v}_1$ at time $t$.
In the absence of external body forces, its evolution is governed by
\begin{equation}
    \left(
    \frac{\partial}{\partial t}
    +\bm v_1\cdot\nabla_{{\bm r}_1}
    \right)F(\bm{r}_1,\bm{v}_1,t)
    =J_{\mathrm E}\left(F^{(2)}\right), 
    \label{eq:Enskog}
\end{equation}
where $F^{(2)}:=F^{(2)}(\bm{r}_1,\bm{r}_2,\bm{v}_1,\bm{v}_2;t)$ denotes the two-particle distribution function.
We define the local peculiar velocity by $\bm V:=\bm v-\bm u(\bm r)$ with a macroscopic velocity field $\bm{u}(\bm{r})$ and write
$F(\bm r,\bm v,t)=f(\bm V,t)$ for a uniform flow.
The collision operator in Eq.~\eqref{eq:Enskog} is
\begin{equation}
    J_{\mathrm E}\left(F^{(2)}\right)
    :=
    d^2\int d\bm v_2\int d\hat{\bm k}\,
    \Theta(\hat{\bm k}\cdot\bm v_{12})
    (\hat{\bm k}\cdot\bm v_{12})
    \left[
    \frac{1}{e^2}
    F^{(2)}(
    \bm r_1,\bm r_1-d\hat{\bm k},
    \bm v_1^{\prime\prime},\bm v_2^{\prime\prime};t)
    -
    F^{(2)}(
    \bm r_1,\bm r_1+d\hat{\bm k},
    \bm v_1,\bm v_2;t)
    \right],
    \label{eq:collision_integral}
\end{equation}
where $\bm v_{12}:=\bm v_1-\bm v_2=\bm{V_1}-\bm{V_2}+\dot{\gamma}(y_1-y_2)\bm{e}_x$.
The double primes denote the restituting velocities associated with the inverse collision.
The collision operator $J_{\mathrm E}$ in Eq.~\eqref{eq:Enskog_USF} is obtained by transforming this physical-velocity operator to the peculiar-velocity representation.
Here, we have introduced the two-body distribution function $F^{(2)}(\bm{r}_1,\bm{r}_2, \bm{v}_1, \bm{v}_2; t)$ and Heaviside's step function $\Theta(x)=1$ for $x\ge0$ and $0$ for $x<0$.
We have also introduced $\hat{\bm{k}}$ for the common normal unit vector at contact.

Based on a careful derivation in Ref.~\cite{Takada22a}, the kinetic equation for the uniform flow satisfying $\bm{u}(\bm{r}_1)=\dot\gamma y_1 \bm{e}_x $ is reduced to
\begin{equation}
    \left(
    \frac{\partial}{\partial t}
    -\dot\gamma V_{1,y}\frac{\partial}{\partial V_{1,x}}
    \right)f(\bm V_1,t)
    =J_{\mathrm E}\left(F^{(2)}\right).
    \label{eq:Enskog_USF}
\end{equation}

Let us consider only uniform sheared cases.
In this case, by multiplying $m V_{1,\alpha} V_{1,\beta}$ with Eq.~\eqref{eq:Enskog_USF} and integrating it over $\bm{V}_1$, one gets the time evolution of the kinetic stress as
\begin{equation}
    \frac{\partial}{\partial t}\sigma_{\alpha\beta}^k
    +\dot{\gamma}\left(\delta_{\alpha x}\sigma_{y\beta}^k+\delta_{\beta x}\sigma_{y\alpha}^k\right)
    = \Lambda_{\alpha\beta},
    \label{eq:stress_evol}
\end{equation}
with the collision moment
\begin{equation}
    \Lambda_{\alpha\beta}
    := -m\int d\bm{V}_1
    V_{1,\alpha} V_{1,\beta} J_\mathrm{E}(F^{(2)}) ,
    \label{eq:def_Lambda}
\end{equation}
where $\overleftrightarrow{\sigma}^k:= \{\sigma_{\alpha\beta}^k\}$ is the kinetic stress tensor:
\begin{equation}
    \sigma^k_{\alpha\beta}(\bm{r}_1)
    := -m\int d\bm{v}_1 V_{1,\alpha} V_{1,\beta}
    F(\bm{r}_1,\bm{v}_1,t).
    \label{eq:Pk}
\end{equation}
We note that the collisional contribution to the stress $\overleftrightarrow{\sigma}^c:= \{\sigma^c_{\alpha\beta}\}$ is expressed as~\cite{Garzo,Hayakawa17,Hayakawa2025} 
\begin{align}\label{contact_stress}
    \sigma^c_{\alpha\beta}(\bm{r}_1)
    &:= -\frac{1+e}{4}m d^3
    \int d\bm{v}_1 \int d\bm{v}_2\int d\hat{\bm{k}}
    \Theta\left(\hat{\bm{k}}\cdot\bm{v}_{12}\right)
    \left(\hat{\bm{k}}\cdot\bm{v}_{12}\right)^2
    \hat{k}_\alpha\hat{k}_\beta
    \int_0^1 dx 
    F^{(2)}[\bm{r}_1-xd\hat{\bm{k}}, \bm{r}_1+(1-x)d\hat{\bm{k}},
    \bm{v}_1, \bm{v}_2; t]
\end{align}
corresponds to the contact Cauchy stress in soft particles.
Then, the total stress $\overleftrightarrow{\sigma}$ can be decomposed into two parts as $\overleftrightarrow{\sigma}=\overleftrightarrow{\sigma}^k+\overleftrightarrow{\sigma}^c$.

As will be shown, our approximate framework of the one-body distribution function can be determined once we know the kinetic stress tensor as the solution of Eq.~\eqref{eq:stress_evol}.
Substituting this result into the decoupled approximate expression of Eq.~\eqref{contact_stress} we obtain the collisional contribution to the stress $\overleftrightarrow{\sigma}^c$.

The kinetic stress can be determined by solving Eq.~\eqref{eq:stress_evol}.
An explicit set of dynamic equations for $T$, $\Delta T$, $\delta T$ defined as
\begin{align}
    T := -\frac{\sigma_{xx}^k+\sigma_{yy}^k+\sigma_{zz}^k}{3n}, \quad
    \Delta T := -\frac{\sigma_{xx}^k-\sigma_{yy}^k}{n},\quad
    \delta T := -\frac{\sigma_{xx}^k-\sigma_{zz}^k}{n},
\end{align}
and $\sigma_{xy}^k$ is written as
\begin{subequations}
\begin{align}
    \frac{\partial}{\partial t}T
    &= \frac{2}{3n}\dot{\gamma}\sigma_{xy}^{k}
    -\frac{1}{3n}\Lambda_{\alpha\alpha},\label{eq:t_T}\\
    %%%%
    \frac{\partial}{\partial t}\Delta T
    &= \frac{2}{n}\dot{\gamma}\sigma_{xy}^{k}
    -\frac{1}{n}\left(\Lambda_{xx}-\Lambda_{yy}\right),\label{eq:t_dT}\\
    %%%%
    \frac{\partial}{\partial t}\delta T
    &= \frac{2}{n}\dot{\gamma}\sigma_{xy}^{k}
    -\frac{1}{n}\left(2\Lambda_{xx}+\Lambda_{yy}-\Lambda_{\alpha\alpha}\right),\label{eq:t_deltaT}\\
    %%%%
    \frac{\partial}{\partial t}\sigma_{xy}^{k}
    &=-\dot{\gamma}\sigma_{yy}^k
    + \Lambda_{xy}.\label{eq:t_Pxyk}
\end{align}  
\end{subequations}

Now, let us rewrite these equations in dimensionless form.
As indicated by Bagnold~\cite{Bagnold54}, this system has only an intrinsic time scale $\dot\gamma^{-1}$ if the interaction between contacted particles is hard-core. 
Because we are interested in a monodisperse system, the mass and the length scale should be $m$ and $d$, respectively.
Therefore, it is sufficient to analyze the dimensionless temperatures defined as
\begin{equation}
    \theta 
    := \frac{T}{m d^2\dot{\gamma}^2},\quad
    %%%
    \Delta\theta 
    := \frac{\Delta T}{m d^2\dot{\gamma}^2},\quad
    %%%
    \delta\theta 
    := \frac{\delta T}{m d^2\dot{\gamma}^2},
    \label{eq:def_theta_deltheta}
\end{equation}
and the dimensionless kinetic stress defined as 
\begin{equation}
    \Pi_{\alpha\beta}^*
    := -\frac{\sigma_{\alpha\beta}^k}{nT}
    - \delta_{\alpha\beta}.
    \label{eq:def_Pi}
\end{equation}
In addition, the dimensionless time $\tau$ is also introduced as
\begin{equation}
    \tau := \dot\gamma t.
\end{equation}
After we adopt the dimensionless variables, we should recognize that the shear rate $\dot\gamma$ does not play any roles at all.
Thus, the dimensionless temperatures and stress tensor depend only on $\varphi$ and $e$ in a steady uniform state.

A set of equations for the dimensionless kinetic stress is given by
\begin{subequations}\label{eq:dynamic_eqs}
\begin{align}
    \frac{\partial}{\partial \tau}\theta
    &= -\frac{2}{3} \theta\Pi_{xy}^*
    -\frac{1}{3}\Lambda_{\alpha\alpha}^{*},
    \label{eq:eq_theta}\\
    %%%%
    \frac{\partial}{\partial \tau}\Delta \theta
    &= -2\theta\Pi_{xy}^*
    -\left(\Lambda_{xx}^{*}-\Lambda_{yy}^{*}\right),
    \label{eq:eq_Del_theta}\\
    %%%%
    \frac{\partial}{\partial \tau}\delta \theta
    &= -2\theta\Pi_{xy}^*
    -\left(2\Lambda_{xx}^{*}+\Lambda_{yy}^{*}
    -\Lambda_{\alpha\alpha}^{*}\right),
    \label{eq:eq_del_theta}\\
    %%%%
    \frac{\partial}{\partial \tau}(\theta\Pi_{xy}^*)
    &=-\left(\theta-\frac{2}{3}\Delta \theta
    + \frac{1}{3}\delta \theta\right)
    - \Lambda_{xy}^{*},
    \label{eq:eq_Pi_xy}
\end{align}
\end{subequations}
with
\begin{equation}
    \Lambda_{\alpha\beta}^{*} 
    := \frac{\Lambda_{\alpha\beta}}{nmd^2 \dot{\gamma}^3}.
\end{equation}
Once a set of solutions of Eqs.~\eqref{eq:dynamic_eqs} is obtained, we can evaluate the dimensionless viscosity, which is equivalent to $xy$-component of the dimensionless shear stress as
\begin{equation}\label{eq:def_eta}
    \eta^* 
    = -\theta(\Pi_{xy}^* + \Pi_{xy}^{c*}), 
\end{equation}
with
\begin{equation}\label{def_Pi^c}
    \Pi_{\alpha\beta}^{c*}
    := -\frac{\sigma_{\alpha\beta}^c}{nT}.
\end{equation}
We can also discuss other important quantities: two normal stress differences, which are defined as
\begin{equation}\label{eq:N_1,N_2}
    N_1 := -\frac{\sigma_{xx}-\sigma_{yy}}{P},\quad
    N_2 := -\frac{\sigma_{yy}-\sigma_{zz}}{P},
\end{equation}
with the static pressure
\begin{equation}
    P := -\frac{1}{3}\operatorname{tr} \overleftrightarrow{\sigma}.
\end{equation}
We can evaluate these quantities, which were ignored in the previous studies on the freely cooling state.

%%%%%%%%%%%%%%%%%%%%%%%%%%%%%%
\subsection{Approximate kinetic theory}\label{sec:approximate_kin}
%%%%%%%%%%%%%%%%%%%%%%%%%%%%%%
In general, the dynamic equation for the one-body distribution function depends on the two-body distribution function.
We may write the two-body distribution function as~\cite{Hayakawa17}
\begin{align}
    F^{(2)}(\bm{r}_1, \bm{r}_1\pm d\hat{\bm{k}}, \bm{v}_1, \bm{v}_2;t) 
    &= g_0(|\bm{r}_1-\bm{r}_2|=d, \varphi) 
    \left[F(\bm{r}_1, \bm{v}_1, t)F(\bm{r}_1\pm d\hat{\bm{k}}, \bm{v}_2, t)
    + h(\bm{r}_1, \bm{r}_1\pm d\hat{\bm{k}}, \bm{v}_1, \bm{v}_2; t)\right] \nonumber\\
    %%%
    &\approx 
    g_0(\varphi) f(\bm{v}_1;t)f(\bm{v}_2\mp \dot\gamma  d \hat{k}_y \bm{e}_x;t) ,
    \label{eq:decoupling}
\end{align}
where we have introduced the radial distribution function at contact $g_0(\varphi):=g_0(|\bm{r}|=d, \varphi)$ to express the correlation effect in part.
Its explicit form will be presented in Eq.~\eqref{eq:g0} later.
From the first expression to the final expression, we have ignored the correlation term $h(\bm{r}_1,\bm{r}_1\pm d\hat{\bm{k}},\bm{v}_1,\bm{v}_2;t)$ to get closure.
We have also used the uniform velocity profile under simple shear at $\bm{r}_1\pm d\hat{\bm{k}}$, which can be expressed as $\bm{V}\mp \dot\gamma d\hat{k}_y \bm{e}_x$, where $\bm{V}$ is the peculiar velocity at $\bm{r}_1$.
Then, we obtain the Enskog equation \eqref{eq:Enskog} with
\begin{align}
    J_\mathrm{E}\left[\bm{V}_1|f^{(2)}\right]
    &\approx d^2g_0(\varphi)
    \int d\bm{V}_2\int d\hat{\bm{k}}
    \Theta\left(\bm{v}_{12}\cdot\hat{\bm{k}}\right)
    \left(\bm{v}_{12}\cdot\hat{\bm{k}}\right)
    \left[\frac{1}{e^2}
    f(\bm{V}_1^{\prime\prime};t)f(\bm{V}_2^{\prime\prime}+\dot\gamma d \hat{k}_y \bm{e}_x; t)
    - f(\bm{V}_1;t)f(\bm{V}_2-\dot\gamma d \hat{k}_y \bm{e}_x; t)\right] .
    \label{eq:Enskog_integral}
\end{align}
Note that $\overleftrightarrow{\sigma^c}$ in Eq.~\eqref{contact_stress} is reduced to~\cite{Garzo, Hayakawa2025, Hayakawa17}
\begin{equation}
    \sigma^c_{\alpha\beta}(\bm{r}_1,t)
    \approx -\frac{1+e}{4}md^3 g_0(\varphi)
    \int d\bm{V}_1 \int d\bm{V}_2\int d\hat{\bm{k}}  \Theta\left(\bm{v}_{12}\cdot\hat{\bm{k}}\right)
    \left(\bm{v}_{12}\cdot\hat{\bm{k}}\right)^2
    \hat{k}_\alpha\hat{k}_\beta
    f\left(\bm{V}_1+\frac{1}{2}\dot\gamma d \hat{k}_y \bm{e}_x\right)
    f\left(\bm{V}_2-\frac{1}{2}\dot\gamma d \hat{k}_y \bm{e}_x\right),
    \label{eq:Pc}
\end{equation}
in the Enskog approximation. 

It is known that $g_0(\varphi)$ is approximately expressed as \cite{Carnahan69}:
\begin{equation}
    g_0(\varphi) 
    =\frac{1-\varphi/2}{(1-\varphi)^3},
    \label{eq:g0}
\end{equation}
for $\varphi\lesssim 0.49$.
Thus, we have obtained the Enskog equation as a closure to express the time evolution of the one-body velocity distribution function $f(\bm{V};t)$. 
We note that the radial distribution function should be anisotropic under the influence of shear, at least, for dense situations~\cite{Brady97,Suzuki19,Banetta19}.
However, we ignore such effects for simplicity.

The difficulty in solving the Enskog equation remains because the Enskog equation is a nonlinear integro-partial differential equation.
Therefore, we need to adopt an approximate method to solve the Enskog equation.
Even if we are only interested in the stress tensor, we must first solve the set of equations \eqref{eq:dynamic_eqs}.
To do so, we need to know $\Lambda_{\alpha\beta}^*$ explicitly.

To close the evolution equations for the second-order velocity moments, we adopt Grad's approximation for the velocity distribution function~~\cite{Grad49,Garzo02,Santos04,Garzo13,Hayakawa17,Chamorro15,Hayakawa19,Hayakawa17,Takada20,Sugimoto20,Takada22a,Takada22b}\footnote{
We adopt a Maxwellian reference distribution for simplicity. 
Using a non-Gaussian homogeneous-cooling-state reference distribution would require a corresponding reformulation of the closure.}:
\begin{align}
    f(\bm{V},t)
    &\approx
    f_\mathrm{M}(\bm{V})
    \left(1 + \frac{m}{2T}\Pi^*_{\alpha\beta}V_\alpha V_\beta\right)
    \nonumber\\
    %%%
    &=
    f_\mathrm{M}(\bm{V})
    \left(1+\Pi^*_{\alpha\beta}c_\alpha c_\beta\right).
    \label{eq:Grad}
\end{align}
Here, $c_\alpha:=V_\alpha/v_{\mathrm T}$, and $v_{\mathrm T}:=\sqrt{2T/m}$.
The Maxwellian distribution is defined by
\begin{align}
    f_\mathrm{M}\left(\bm{V}\right)
    := n\left(\frac{m}{2\pi T}\right)^{3/2}
    \exp\left(-\frac{mV^2}{2T}\right).
    \label{eq:Maxwell}
\end{align}
In the present homogeneous shear flow, the heat flux vanishes by symmetry, and the anisotropic correction to the Maxwellian distribution is determined by the kinetic stress tensor.
This polynomial approximation does not guarantee a nonnegative velocity distribution.
For any nonzero symmetric traceless tensor $\Pi_{\alpha\beta}^{*}$, the correction in Eq.~\eqref{eq:Grad} becomes negative at sufficiently large velocities along a direction associated with a negative eigenvalue.
Thus, the occurrence of negative values in the high-velocity tail does not by itself imply that the predicted low-order moments are inaccurate.
Nevertheless, strong anisotropy and deviations from the assumed distribution can limit the accuracy of the closure, particularly for higher-order moments.
Its accuracy must therefore be assessed by comparison with simulation results for the observables of interest.
Although this treatment retains all orders in the dimensionless shear-rate parameter, the approximations inherent in the Grad distribution and the Enskog factorization remain.

After adopting the assumptions \eqref{eq:decoupling} and \eqref{eq:Grad}, we can express the dimensionless collisional moment as
\begin{equation}
    \Lambda_{\alpha\beta}^*
    = \frac{3\sqrt{2}}{\pi}(1+e)\varphi g_0(\varphi)
    \theta^{3/2}\widetilde{\Lambda}_{\alpha\beta} .
    \label{eq:Lambda_tilde_Lambda}
\end{equation}
Although the previous Grad's expansion adopted a truncation of the series expansion, as in Ref.~\cite{Takada20}, the present study demonstrates that $\widetilde{\Lambda}_{\alpha\beta}$ can be written in terms of generalized hypergeometric functions.
Its trace is expressed in terms of the normalized kinetic stress anisotropies $\Pi_{xx}^*$, $\Pi_{yy}^*$, and $\Pi_{xy}^*$ as
\begin{subequations}\label{eq:Lambda_hypergeometric}
\begin{align}
    \widetilde{\Lambda}_{\alpha\alpha}
    &=
    \widetilde{\Lambda}_{\alpha\alpha}^{(0)}(\theta)
    + \widetilde{\Lambda}_{\alpha\alpha}^{(1)}(\theta)
    \left(\Pi_{xx}^*+\Pi_{yy}^*\right)
    + \widetilde{\Lambda}_{\alpha\alpha}^{(2)}(\theta)
    \Pi_{xy}^*\nonumber\\
    %%%
    &\hspace{1em}
    + \widetilde{\Lambda}_{\alpha\alpha}^{(3)}(\theta)
    \left[\left(\Pi_{xx}^*\right)^2
    +\left(\Pi_{yy}^*\right)^2\right]
    + \widetilde{\Lambda}_{\alpha\alpha}^{(4)}(\theta)
    \Pi_{xx}^*\Pi_{yy}^*
    + \widetilde{\Lambda}_{\alpha\alpha}^{(5)}(\theta)
    \left(\Pi_{xy}^*\right)^2.
    \label{eq:Lambda_aa}
\end{align}
Similarly, the $xy$, $xx$, and $yy$ components are given by
\begin{align}
    \widetilde{\Lambda}_{xy}
    &=
    \widetilde{\Lambda}_{xy}^{(0)}(\theta)
    + \widetilde{\Lambda}_{xy}^{(1)}(\theta)
    \left(\Pi_{xx}^*+\Pi_{yy}^*\right)
    + \widetilde{\Lambda}_{xy}^{(2)}(\theta)\Pi_{xy}^*
    + \widetilde{\Lambda}_{xy}^{(3)}(\theta)
    \left(\Pi_{xx}^*+\Pi_{yy}^*\right)\Pi_{xy}^*,
    \\
    %%%
    \widetilde{\Lambda}_{xx}
    &=
    \widetilde{\Lambda}_{xx}^{(0)}(\theta)
    + \widetilde{\Lambda}_{xx}^{(1)}(\theta)\Pi_{xx}^*
    + \widetilde{\Lambda}_{xx}^{(2)}(\theta)\Pi_{yy}^*
    + \widetilde{\Lambda}_{xx}^{(3)}(\theta)\Pi_{xy}^*
    \nonumber\\
    &\hspace{1em}
    + \widetilde{\Lambda}_{xx}^{(4)}(\theta)
    \left(\Pi_{xx}^*\right)^2
    + \widetilde{\Lambda}_{xx}^{(5)}(\theta)
    \left(\Pi_{yy}^*\right)^2
    + \widetilde{\Lambda}_{xx}^{(6)}(\theta)
    \Pi_{xx}^*\Pi_{yy}^*
    + \widetilde{\Lambda}_{xx}^{(7)}(\theta)
    \left(\Pi_{xy}^*\right)^2,
    \\
    %%%
    \widetilde{\Lambda}_{yy}
    &=
    \widetilde{\Lambda}_{yy}^{(0)}(\theta)
    + \widetilde{\Lambda}_{yy}^{(1)}(\theta)\Pi_{xx}^*
    + \widetilde{\Lambda}_{yy}^{(2)}(\theta)\Pi_{yy}^*
    + \widetilde{\Lambda}_{yy}^{(3)}(\theta)\Pi_{xy}^*
    \nonumber\\
    &\hspace{1em}
    + \widetilde{\Lambda}_{yy}^{(4)}(\theta)
    \left(\Pi_{xx}^*\right)^2
    + \widetilde{\Lambda}_{yy}^{(5)}(\theta)
    \left(\Pi_{yy}^*\right)^2
    + \widetilde{\Lambda}_{yy}^{(6)}(\theta)
    \Pi_{xx}^*\Pi_{yy}^*
    + \widetilde{\Lambda}_{yy}^{(7)}(\theta)
    \left(\Pi_{xy}^*\right)^2.
\end{align}
\end{subequations}
The coefficient functions $\widetilde{\Lambda}_{\alpha\alpha}^{(i)}(\theta)$ and $\widetilde{\Lambda}_{\alpha\beta}^{(i)}(\theta)$ are listed in Table~\ref{table:list_of_functions} of Appendix~\ref{sec:Derivation_Lambda}.
They are expressed using the auxiliary functions $\mathcal F_{s,t,u}(\theta)$, $\mathcal G_{s,t,u}(\theta)$, and $\mathcal H_{s,t,u}^{(i)}(\theta)$, defined in Eqs.~\eqref{eq:def_FG} and \eqref{eq:def_H} in terms of generalized hypergeometric functions.

The collisional contributions to the dimensionless pressure, $\Pi^{c*}:=\operatorname{tr}\overleftrightarrow{\Pi^{c*}}/3$, and stress tensor can be expressed in the same manner:
\begin{equation}
    \Pi^{c*}
    = \frac{3}{\pi}(1+e)\varphi g_0(\varphi)
    \widetilde{\Pi}^{c*},
    \qquad
    \Pi_{\alpha\beta}^{c*}
    = \frac{3}{\pi}(1+e)\varphi g_0(\varphi)
    \widetilde{\Pi}_{\alpha\beta}^{c*}.
    \label{eq:Pi_c_tilde_Pi_c}
\end{equation}
Their dependence on the normalized kinetic stress
anisotropies is given by
\begin{subequations}\label{eq:Pi_c_hypergeometric}
\begin{align}
    \widetilde{\Pi}^{c*}
    &=
    \widetilde{\Pi}^{c(0)*}(\theta)
    + \widetilde{\Pi}^{c(1)*}(\theta)\Pi_{xy}^*
    + \widetilde{\Pi}^{c(2)*}(\theta)
    \left(\Pi_{xx}^*+\Pi_{yy}^*\right)\Pi_{xy}^*,
    \\
    %%%
    \widetilde{\Pi}_{xy}^{c*}
    &=
    \widetilde{\Pi}_{xy}^{c(0)*}(\theta)
    + \widetilde{\Pi}_{xy}^{c(1)*}(\theta)
    \left(\Pi_{xx}^*+\Pi_{yy}^*\right)
    + \widetilde{\Pi}_{xy}^{c(2)*}(\theta)\Pi_{xy}^*
    \nonumber\\
    &\hspace{1em}
    + \widetilde{\Pi}_{xy}^{c(3)*}(\theta)
    \left[\left(\Pi_{xx}^*\right)^2
    +\left(\Pi_{yy}^*\right)^2\right]
    + \widetilde{\Pi}_{xy}^{c(4)*}(\theta)
    \Pi_{xx}^*\Pi_{yy}^*
    + \widetilde{\Pi}_{xy}^{c(5)*}(\theta)
    \left(\Pi_{xy}^*\right)^2,
    \\
    %%%
    \widetilde{\Pi}_{xx}^{c*}
    &=
    \widetilde{\Pi}_{xx}^{c(0)*}(\theta)
    + \widetilde{\Pi}_{xx}^{c(1)*}(\theta)\Pi_{xx}^*
    + \widetilde{\Pi}_{xx}^{c(2)*}(\theta)\Pi_{xy}^*
    + \widetilde{\Pi}_{xx}^{c(3)*}(\theta)
    \Pi_{xx}^*\Pi_{xy}^*
    + \widetilde{\Pi}_{xx}^{c(4)*}(\theta)
    \Pi_{yy}^*\Pi_{xy}^*,
    \\
    %%%
    \widetilde{\Pi}_{yy}^{c*}
    &=
    \widetilde{\Pi}_{yy}^{c(0)*}(\theta)
    + \widetilde{\Pi}_{yy}^{c(1)*}(\theta)\Pi_{yy}^*
    + \widetilde{\Pi}_{yy}^{c(2)*}(\theta)\Pi_{xy}^*
    + \widetilde{\Pi}_{yy}^{c(3)*}(\theta)
    \Pi_{xx}^*\Pi_{xy}^*
    + \widetilde{\Pi}_{yy}^{c(4)*}(\theta)
    \Pi_{yy}^*\Pi_{xy}^*.
\end{align}
\end{subequations}
The coefficient functions $\widetilde{\Pi}^{c(i)*}(\theta)$ and $\widetilde{\Pi}_{\alpha\beta}^{c(i)*}(\theta)$ are listed in Table~\ref{table:list_of_functions2} of Appendix~\ref{sec:Derivation_Lambda}.

In Ref.~\cite{Takada20}, the collisional moments and collisional stress were evaluated using expansions in
\begin{equation}
    \epsilon
    := \frac{1}{\sqrt{\theta}}
    = \frac{\dot\gamma d}{\sqrt{T/m}},
\end{equation}
truncated at sixth order.
Expanding the present generalized hypergeometric expressions in $\epsilon$ reproduces those results through sixth order, with the correction noted below.\footnote{
The expression of $\widetilde{\Pi}_{xy}^{c(1)}(\theta)$ in Table II of Ref.~\cite{Takada20} should read
\begin{equation*}
    \widetilde{\Pi}_{xy}^{c(1)*}
    = -\frac{4}{5\sqrt{\pi}}(1+e)
    \left\{
    1+\frac{1}{7\theta}
    \left(\Pi_{xx}^*+\Pi_{yy}^*\right)
    -\frac{1}{28\theta^2}\left(\Pi_{xy}^*\right)^2
    -\frac{1}{84\theta^2}
    \left[
    \left(\Pi_{xx}^*\right)^2
    +\Pi_{xx}^*\Pi_{yy}^*
    +\left(\Pi_{yy}^*\right)^2
    \right]
    \right\}.
\end{equation*}
}
The present formulation retains the full dependence on
$\epsilon$ within the adopted approximations.
The convergence of the truncated expansions and their
errors in $\theta$, $\eta^*$, $N_1$, and $N_2$ are examined in Appendix~\ref{sec:truncation}.

We determine the steady state by solving Eqs.~\eqref{eq:dynamic_eqs} with the collisional moments and stresses given above.
The viscosity is then evaluated using Eq.~\eqref{eq:def_eta}.

%%%%%%%%%%%%%%%%%%%%%%%%%%%%%%
\section{Steady rheology}\label{sec:steady_rheology}
%%%%%%%%%%%%%%%%%%%%%%%%%%%%%%
In this section, we assess the accuracy of the present theory for steady shear flows by comparison with event-driven molecular dynamics simulations.
The theoretical results are obtained by integrating Eqs.~\eqref{eq:dynamic_eqs} to steady state, with the collisional moments evaluated using Eqs.~\eqref{eq:Lambda_tilde_Lambda} and \eqref{eq:Lambda_hypergeometric}.
The collisional stress is evaluated using Eqs.~\eqref{eq:Pi_c_tilde_Pi_c} and \eqref{eq:Pi_c_hypergeometric}.
The viscosity and normal stress differences are then calculated
from Eqs.~\eqref{eq:def_eta} and \eqref{eq:N_1,N_2}, respectively.
Throughout this section, the generalized hypergeometric expressions are evaluated without finite-order truncation in $\epsilon=1/\sqrt{\theta}$.
The convergence of the corresponding finite-order expansions is examined in Appendix~\ref{sec:truncation}.

%%%%%%%%%%%%%%%%%%%%%%%%%%%%%%
\subsection{Comparison with the simulation results}\label{sec:comp_sim}
%%%%%%%%%%%%%%%%%%%%%%%%%%%%%%
In principle, the steady state can be obtained by solving the equation derived by setting the left-hand side of Eqs.~\eqref{eq:dynamic_eqs} to zero. 
However, this set of equations has not been explicitly solved. 
Therefore, we determine the steady state by time integration rather than solving the nonlinear algebraic equations directly. 
Equation~\eqref{eq:dynamic_eqs} is integrated using the Euler method with a time step of $\Delta t = 0.001/\dot\gamma$. 
For the theoretical evolution, we use the isotropic initial moments specified in Sec.~\ref{Sec:setup}: $T=T_0$, $\Delta T=\delta T=0$, and $\sigma_{xy}^{k}=0$.
The evolution for $t\ge0$ is calculated at the imposed shear rate $\dot\gamma$ and restitution coefficient $e$.

%%%%%%%%%%%%%%%%%%%%%%%%%%%%%%
\begin{figure}[hbtp]
    \centering
    \includegraphics[width=\linewidth]{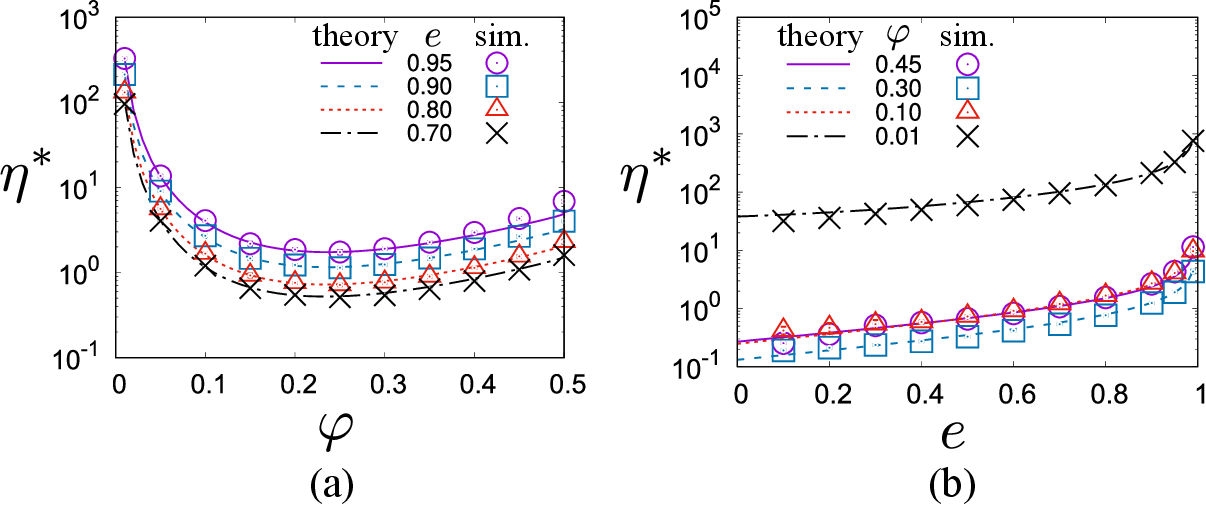}
    \caption{Plots of the viscosity $\eta^*$ against (a) the volume fraction $\varphi$ for $e=0.70$, $0.80$, $0.90$, and $0.95$ and (b) the restitution coefficient $e$ for $\varphi=0.01$, $0.10$, $0.30$, and $0.45$.
    The theoretical results are obtained by integrating Eqs.~\eqref{eq:dynamic_eqs} to steady state, with the collisional moments evaluated using Eqs.~\eqref{eq:Lambda_tilde_Lambda} and \eqref{eq:Lambda_hypergeometric}.
    The viscosity is calculated from Eq.~\eqref{eq:def_eta}, including the collisional stress evaluated using Eqs.~\eqref{eq:Pi_c_tilde_Pi_c} and \eqref{eq:Pi_c_hypergeometric}.
    The lines represent our theory and the marks correspond to the simulation results.
    }
    \label{fig:eta_phi}
\end{figure}
%%%%%%%%%%%%%%%%%%%%%%%%%%%%%%

Figure \ref{fig:eta_phi}(a) presents the volume fraction $\varphi$ dependence of the viscosity $\eta^*$ for various restitution coefficients $e$, while Fig.~\ref{fig:eta_phi}(b) represents $e$ dependence of $\eta^*$ for various $\varphi$.
Our theory reproduces the simulation results well over a wide range of volume fraction and restitution coefficient, the deviations are relatively small.
The nonmonotonic density dependence of $\eta^*$ can be understood qualitatively as a competition between kinetic and collisional momentum transport, accounting for the density dependence of the steady granular temperature. 
In the dilute regime, momentum transport is dominated by particles moving over a mean free path $\ell \sim (nd^2g_0)^{-1}$, giving $\eta^k \sim nm\sqrt{T/m}\,\ell$. 
At fixed restitution coefficient $e<1$, the balance between shear heating and collisional dissipation yields $\theta \propto \varphi^{-2}$ in the dilute limit, so that $\eta^{k*} \propto \varphi^{-2}$. 
Thus, increasing the volume fraction initially lowers the dimensionless viscosity by shortening the mean free path and reducing the steady temperature. 
As the density increases further, direct momentum transfer during collisions becomes increasingly important because both the number of colliding pairs and the contact value $g_0(\varphi)$ increase. 
A simple Enskog scaling estimate gives $\eta^{c*} \sim \varphi g_0(\varphi)\sqrt{\theta}$, which increases when the growth of $\varphi g_0(\varphi)$ outweighs the reduction in temperature. 
The minimum of $\eta^*=\eta^{k*}+\eta^{c*}$ therefore reflects the competition between the decreasing kinetic contribution and the growing importance of collisional momentum transfer.

%%%%%%%%%%%%%%%%%%%%%%%%%%%%%%
\begin{figure}[hbtp]
    \centering
    \includegraphics[width=\linewidth]{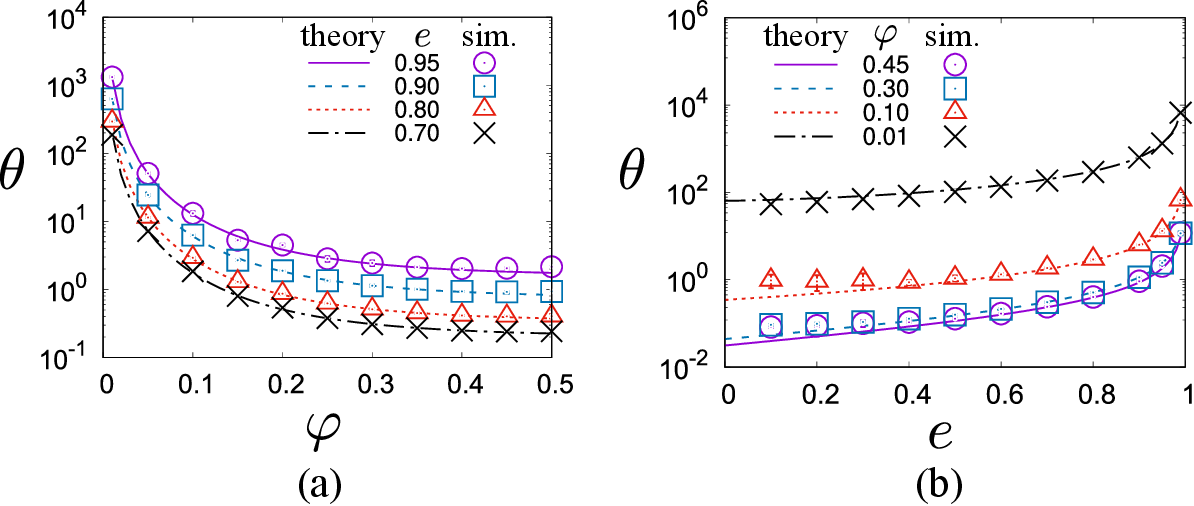}
    \caption{Plots of the kinetic temperature $\theta$ against (a) the volume fraction $\varphi$ for $e=0.70$, $0.80$, $0.90$, and $0.95$ and (b) the restitution coefficient $e$ for $\varphi=0.01$, $0.10$, $0.30$, and $0.45$.
    The theoretical results are evaluated using the same procedure as in Fig.~\ref{fig:eta_phi}.
    The lines represent our theory, and the marks correspond to the simulation results.
    }
    \label{fig:theta_phi}
\end{figure}
%%%%%%%%%%%%%%%%%%%%%%%%%%%%%%

Similarly, Fig.~\ref{fig:theta_phi} plots $\theta$ against (a) the volume fraction $\varphi$ and (b) the restitution coefficient $e$.
Although the theoretical and simulation results appear close on these logarithmic scales, appreciable quantitative deviations remain under strongly inelastic conditions.

\begin{figure}[hbtp]
    \centering
    \includegraphics[width=\linewidth]{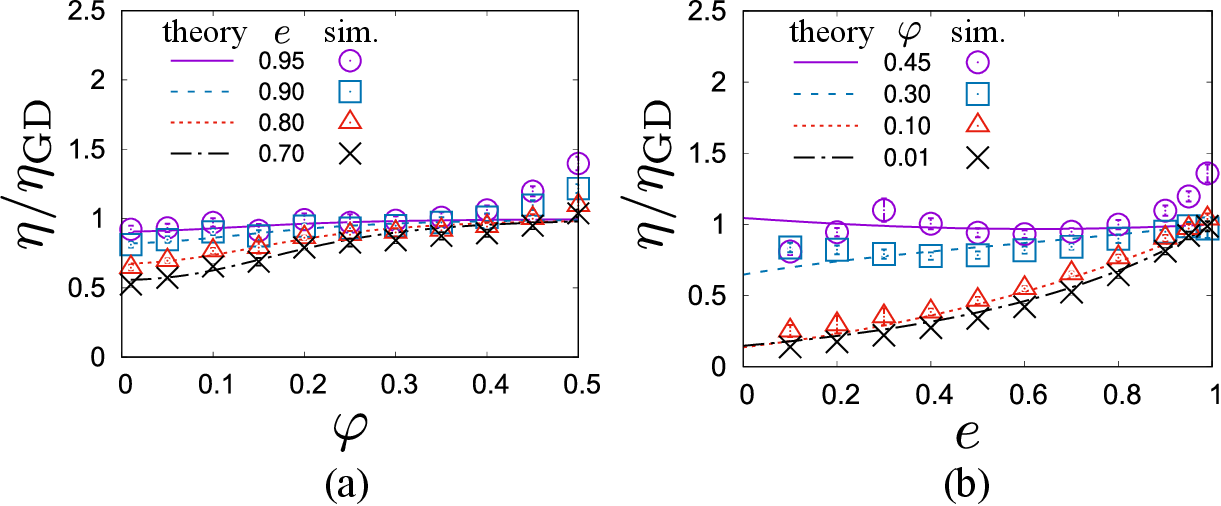}
    \caption{Plots of $\eta/\eta_\mathrm{GD}$ against (a) the volume fraction $\varphi$ for $e=0.70$, $0.80$, $0.90$, and $0.95$ and (b) the restitution coefficient $e$ for $\varphi=0.01$, $0.10$, $0.30$, and $0.45$.
    The theoretical results are evaluated using the same procedure as in Fig.~\ref{fig:eta_phi}.
    The lines represent our theory, and the marks correspond to the simulation results.
    }
    \label{fig:comp_sim_eta}
\end{figure}

\begin{figure}[hbtp]
    \centering
    \includegraphics[width=\linewidth]{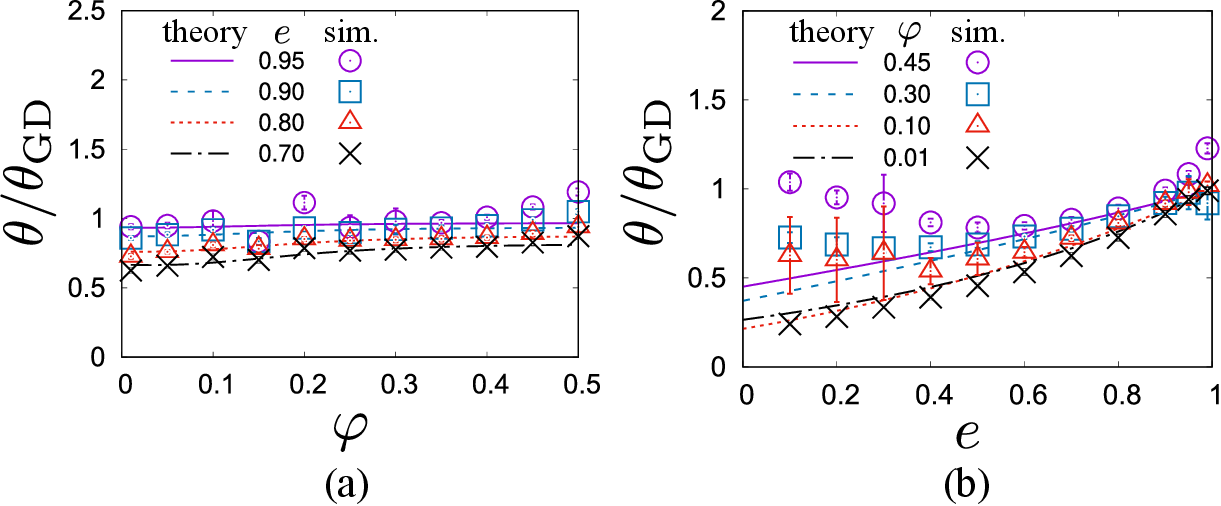}
    \caption{Plots of $\theta/\theta_\mathrm{GD}$ against (a) the volume fraction $\varphi$ for $e=0.70$, $0.80$, $0.90$, and $0.95$ and (b) the restitution coefficient $e$ for $\varphi=0.01$, $0.10$, $0.30$, and $0.45$.
    The theoretical results are evaluated using the same procedure as in Fig.~\ref{fig:eta_phi}.
    The lines represent our theory and the marks correspond to the simulation results.
    }
    \label{fig:comp_sim_theta}
\end{figure}

Although the results in Figs.~\ref{fig:eta_phi} and \ref{fig:theta_phi} successfully reproduce the simulation data, they alone do not reveal any clear discrepancy from the theory of Garz\'{o} and Dufty~\cite{Garzo99}. 
This is because we have used logarithmic plots for $\eta^*$ and $\theta$ in Figs.~\ref{fig:eta_phi} and \ref{fig:theta_phi}. 
In fact, when we plot the ratio of our theoretical and simulation results to the predictions of Garz\'{o} and Dufty theory~\cite{Garzo99}, a clearer difference becomes apparent, as shown in Figs.~\ref{fig:comp_sim_eta} and \ref{fig:comp_sim_theta}.

%%%%%%%%%%%%%%%%%%%%%%%%%%%%%%
%%%%%%%%%%%%%%%%%%%%%%%%%%%%%%

Figure~\ref{fig:comp_sim_eta} shows the normalized viscosity $\eta/\eta_{\mathrm{GD}}$ as a function of (a) the volume fraction $\varphi\le0.5$ and (b) the restitution coefficient $e$, where $\eta_{\mathrm{GD}}$ is the viscosity predicted by Garz\'o and Dufty~\cite{Garzo99}.
Although the prediction of Ref.~\cite{Garzo99} works well even for sheared situations as indicated by Ref.~\cite{Chialvo13}, these figures indicate the significant deviations of the predictions of Ref.~\cite{Garzo99} from the results of the simulation for low $\varphi$ and $e$.
On the other hand, our theory well reproduces the simulation results in wide range of parameters. 
Interestingly, our theory reduces to Garz\'{o} and Dufty~\cite{Garzo99} as $\varphi$ approaches $0.50$, while the simulation results deviate from both theories for low dissipation cases, such as $e=0.95$.

%%%%%%%%%%%%%%%%%%%%%%%%%%%%%%
%%%%%%%%%%%%%%%%%%%%%%%%%%%%%%

Similarly, Fig.~\ref{fig:comp_sim_theta} plots the normalized temperature $\theta/\theta_\mathrm{GD}$ by the prediction $\theta_\mathrm{GD}$ of Ref.~\cite{Garzo99} against (a) $\varphi$ and (b) $e$. 
For the theoretical curves shown, $\theta/\theta_\mathrm{GD}$ is below unity over the range examined, as in the simulation results (see Fig.~\ref{fig:comp_sim_theta} (a)).
We also find that our theory is not quantitatively correct for $\theta/\theta_\mathrm{GD}$ for low $e$ and high $\varphi$, as shown in Fig.~\ref{fig:comp_sim_theta} (b). 
 
Figures~\ref{fig:comp_sim_eta} and \ref{fig:comp_sim_theta}
show that the theory reproduces the viscosity reasonably well over much of the parameter range examined, while appreciable temperature deviations remain under strongly inelastic conditions.
These are intended to highlight the differences between the results for the shear flow system and those of the freely cooling state, which, as previously mentioned, is widely regarded as broadly accurate.
We also indicate an interesting deviation for nearly elastic situations with $e=0.95$.
This may come from the correlation effects~\cite{vanNoije97,vanNoije98,vanNoije00,Otsuki09a,Otsuki09b,Otsuki09c} or the reference distribution $f_\mathrm{M}(\bm{V})$ in Eq.~\eqref{eq:Grad}.
We also indicate that $e=1$ is a singular limit, at which there is no steady state.

%%%%%%%%%%%%%%%%%%%%%%%%%%%%%%
\begin{figure}[hbtp]
    \centering
    \includegraphics[width=\linewidth]{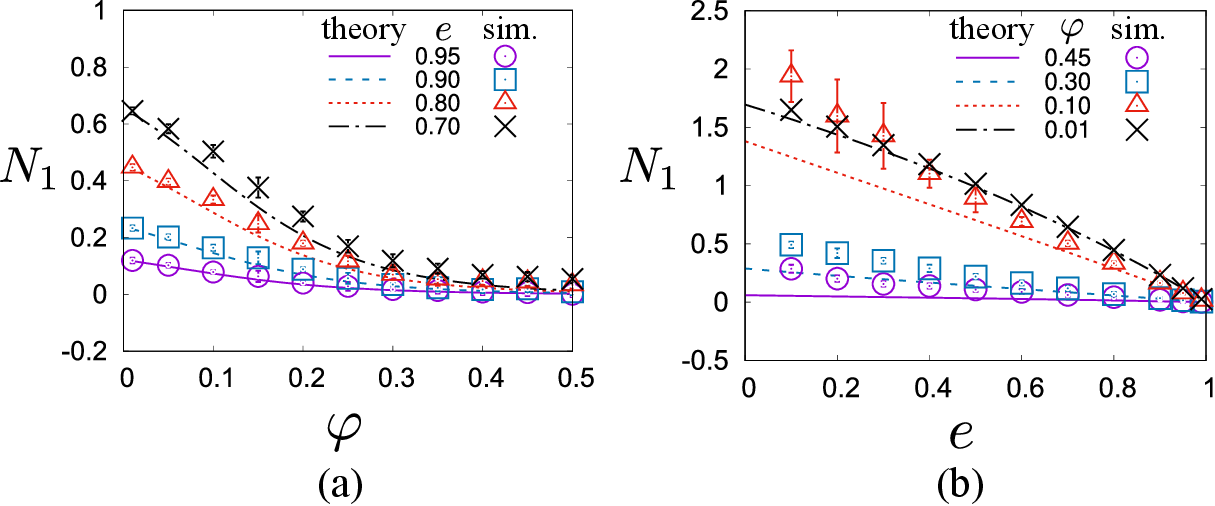}
    \caption{Plots of the first normal stress difference $N_1$ against (a) the volume fraction $\varphi$ for $e=0.70$, $0.80$, $0.90$, and $0.95$ and (b) the restitution coefficient $e$ for $\varphi=0.01$, $0.10$, $0.30$, and $0.45$.
    The theoretical results are evaluated using the same procedure as in Fig.~\ref{fig:eta_phi}.
    The first normal stress difference $N_1$ is evaluated using Eq.~\eqref{eq:N_1,N_2}.
    The lines represent our theory and the marks correspond to the simulation results.}
    \label{fig:comp_sim_N1}
\end{figure}
%%%%%%%%%%%%%%%%%%%%%%%%%%%%%%

Let us now consider the stress differences $N_1$ and $N_2$ defined as Eq.~\eqref{eq:N_1,N_2}. 
Note that the theory by Garz\'{o} and Dufty~\cite{Garzo99} predicts $N_1=N_2=0$, because they are interested in a freely cooling state at Navier-Stokes order. 
Figures \ref{fig:comp_sim_N1}(a) and (b) present $N_1$ versus $\varphi$ and $e$, respectively.
These figures indicate reasonable agreement between the theory and the simulation for $N_1$, at least for nearly elastic cases.

%%%%%%%%%%%%%%%%%%%%%%%%%%%%%%
\begin{figure}[hbtp]
    \centering
    \includegraphics[width=\linewidth]{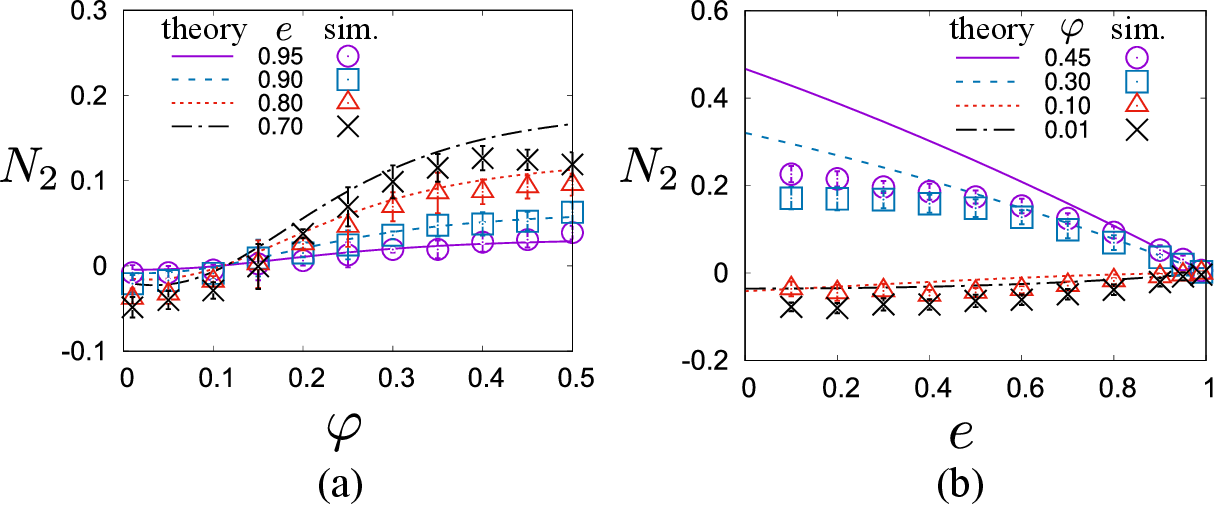}
    \caption{Plots of the second normal stress difference $N_2$ against (a) the volume fraction $\varphi$ for $e=0.70$, $0.80$, $0.90$, and $0.95$ and (b) the restitution coefficient $e$ for $\varphi=0.01$, $0.10$, $0.30$, and $0.45$.
    The theoretical results are evaluated using the same procedure as in Fig.~\ref{fig:eta_phi}.
    The second normal stress difference $N_2$ is evaluated using Eq.~\eqref{eq:N_1,N_2}.
    The lines represent our theory and the marks correspond to the simulation results.}
    \label{fig:comp_sim_N2}
\end{figure}
%%%%%%%%%%%%%%%%%%%%%%%%%%%%%%

Similarly, Fig.~\ref{fig:comp_sim_N2}(a) and (b) present $N_2$ versus $\varphi$ and $e$, respectively. 
The theory reproduces the sign change and the overall parameter dependence of $N_2$, although appreciable quantitative deviations remain in parts of the parameter range.
Because $N_2$ is small in dilute conditions, its relative statistical uncertainty can be large even when the absolute uncertainty is modest.

%%%%%%%%%%%%%%%%%%%%%%%%%%%%%%
\begin{figure}[hbtp]
    \centering
    \includegraphics[width=\linewidth]{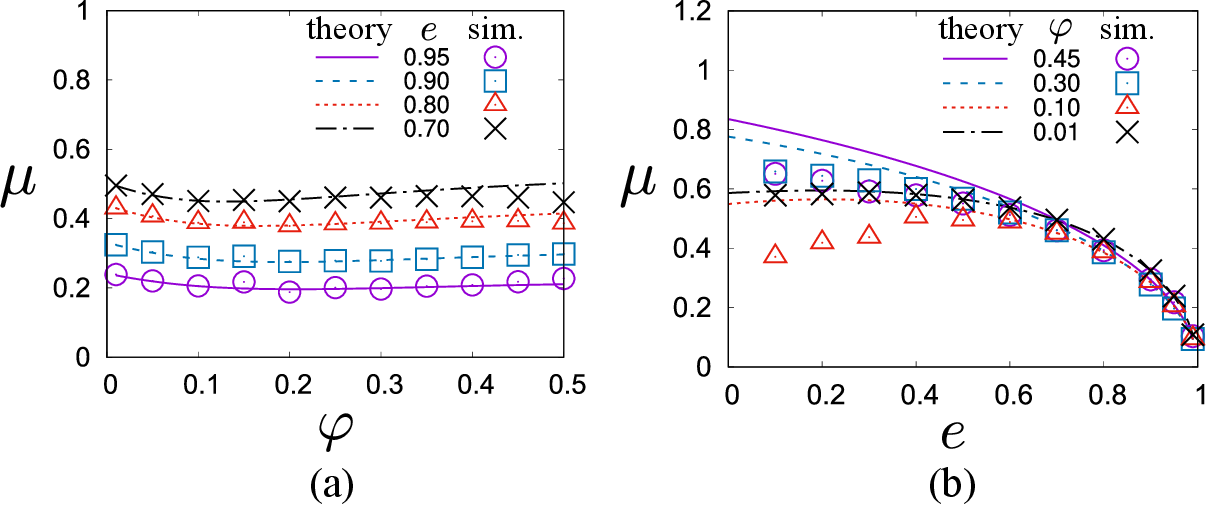}
    \caption{Plots of the macroscopic friction coefficient $\mu=\sigma_{xy}/P$ against (a) the volume fraction $\varphi$ for $e=0.70$, $0.80$, $0.90$, and $0.95$ and (b) the restitution coefficient $e$ for $\varphi=0.01$, $0.10$, $0.30$, and $0.45$.
    The theoretical results are evaluated using the same procedure as in Fig.~\ref{fig:eta_phi}.
    The lines represent our theory and the marks correspond to the simulation results.
    }
    \label{fig:comp_sim_mu}
\end{figure}
%%%%%%%%%%%%%%%%%%%%%%%%%%%%%%

In Fig.~\ref{fig:comp_sim_mu}, we plot the stress ratio $\mu:=\sigma_{xy}/P$ against $\varphi$ (a) and $e$ (b), respectively. 
We stress that the deviations are relatively small for $e=0.90$ and $0.95$ in Fig.~\ref{fig:comp_sim_mu}(a), while we can find deviations for non-dilute $\varphi$ and low $e$ (see Fig.~\ref{fig:comp_sim_mu} (b)).

%%%%%%%%%%%%%%%%%%%%%%%%%%%%%%
\begin{figure}[hbtp]
    \centering
    \includegraphics[width=\linewidth]{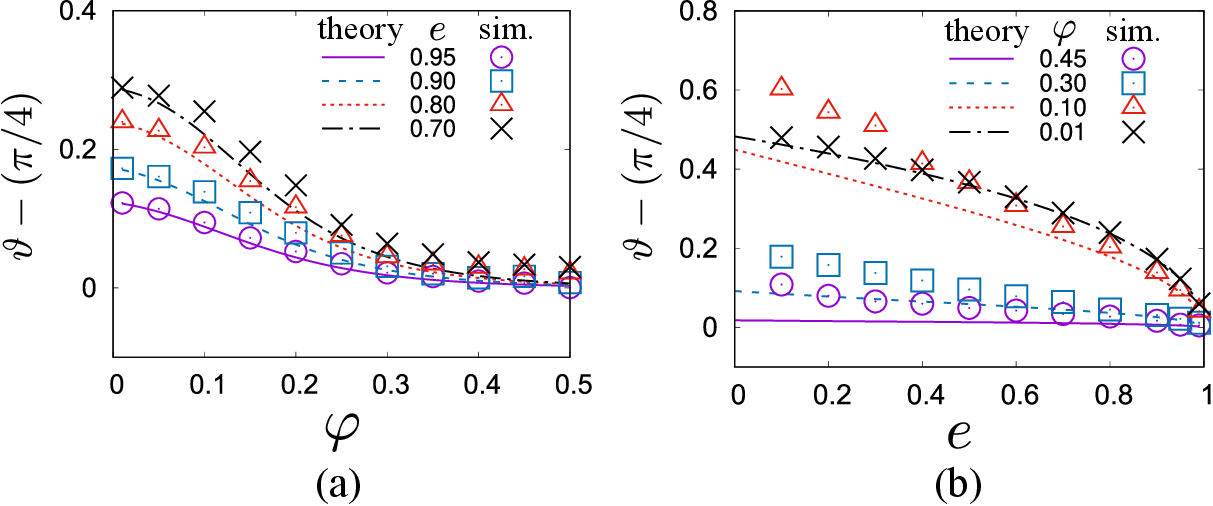}
    \caption{Comparison of the deviation of the principal-axis angle from $\pi/4$, $\vartheta-\pi/4$, as a function of (a) the volume fraction $\varphi$ and (b) the restitution coefficient $e$.
    The theoretical results are evaluated using the same procedure as in Fig.~\ref{fig:eta_phi}.
    The lines represent the theoretical predictions, and the symbols represent the simulation results.
    }
    \label{fig:comp_angle}
\end{figure}
%%%%%%%%%%%%%%%%%%%%%%%%%%%%%%

Let us consider the principal-axis angle $\vartheta$ measured from the $x$-axis.
We select the principal axis associated with the larger eigenvalue of the in-plane stress tensor and define its orientation by
\begin{equation}
    \vartheta
    := \frac{1}{2}\operatorname{Arg}
    \left[
    (\sigma_{xx}-\sigma_{yy})+2i\sigma_{xy}
    \right] ,
\end{equation}
where $i$ is the imaginary unit satisfying $i^2=-1$, and $\operatorname{Arg}z$ denotes the principal argument of a nonzero complex number $z$, with values in $(-\pi,\pi]$, so that $\vartheta\in(-\pi/2,\pi/2]$.
The principal-axis orientation is understood modulo $\pi$ and is undefined when $\sigma_{xx}=\sigma_{yy}$ and $\sigma_{xy}=0$.
Figure~\ref{fig:comp_angle} shows $\vartheta-\pi/4$ as a function of the volume fraction and restitution coefficient.
The principal-axis angle approaches $\pi/4$ as the volume fraction increases or the restitution coefficient approaches unity.
The theory captures these trends and agrees reasonably well with the simulation results over much of the range examined, although quantitative deviations remain at some parameter values.

%%%%%%%%%%%%%%%%%%%%%%%%%%%%%%
\subsection{Comparison between the present Grad-type theory and the anisotropic Gaussian theory}\label{sec:comp_SA}
%%%%%%%%%%%%%%%%%%%%%%%%%%%%%%
In the present study, we have employed a Grad-type approximation for the velocity distribution function.
As shown in the previous subsection, the resulting theory reproduces the simulation results with reasonable accuracy over a wide range of parameters.
Previous studies, on the other hand, have developed moment theories based on an anisotropic Gaussian or anisotropic Gaussian reference state and demonstrated their usefulness for describing the non-Newtonian rheology of sheared granular systems~\cite{Lutsko05,Saha14,Saha16,Saha17,Alam19,Saha20}.

We now compare the present Grad-type theory with the anisotropic Gaussian theory of Saha and Alam~\cite{Saha16} for three-dimensional shear flows.
Unlike the dilute Grad approximation considered in Ref.~\cite{Saha14}, the present formulation includes the Enskog correction and retains all orders in the dimensionless shear-rate parameter.
We compare the two theories for the temperature, viscosity, and normal stress differences.

%%%%%%%%%%%%%%%%%%%%%%%%%%%%%%
\begin{figure}[hbtp]
    \centering
    \includegraphics[width=\linewidth]{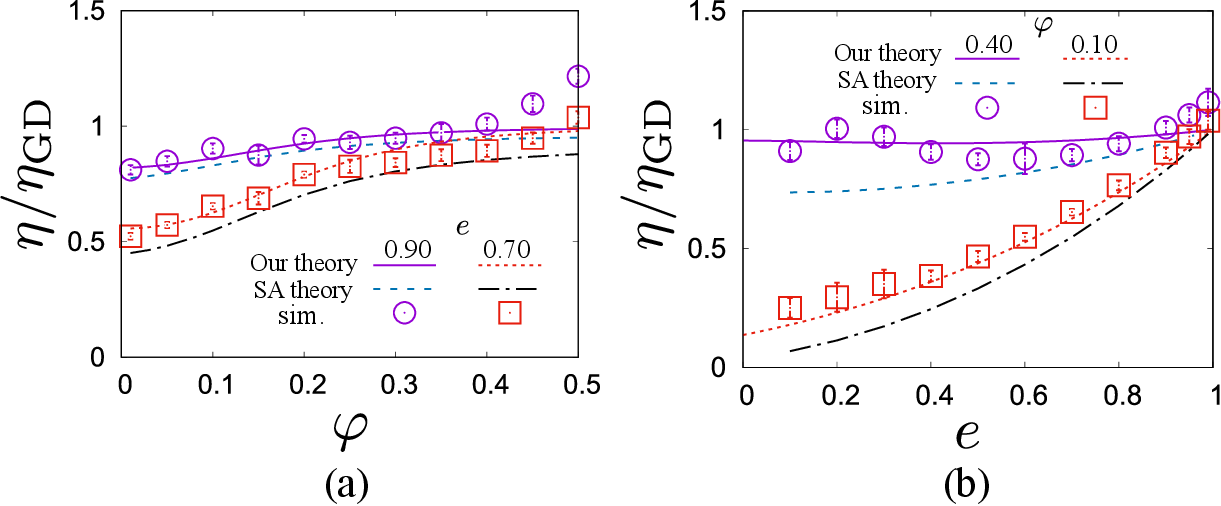}
    \caption{Plots of $\eta^*$ against (a) the volume fraction $\varphi$ for $e=0.70$ and $0.90$ and (b) the restitution coefficient $e$ for $\varphi=0.10$ and $0.40$.
    The solid and dotted lines represent our theoretical results and the dashed and dot-dashed lines correspond to the theory by Saha and Alam~\cite{Saha16}.
    The symbols indicate the simulation results.}
    \label{fig:comp_SA_eta}
\end{figure}
%%%%%%%%%%%%%%%%%%%%%%%%%%%%%%

\begin{figure}[hbtp]
    \centering
    \includegraphics[width=\linewidth]{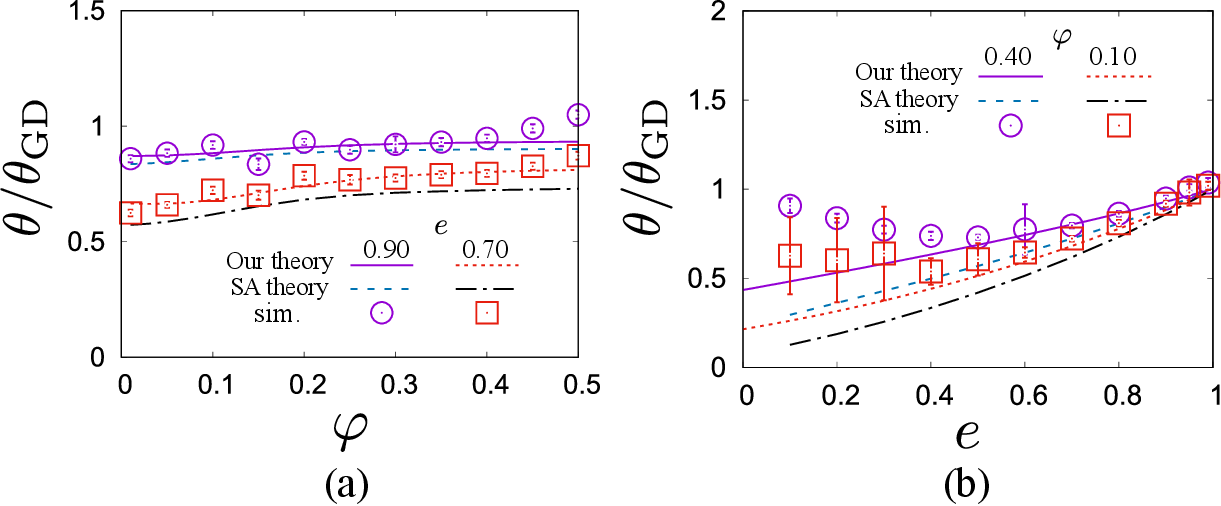}
    \caption{Plots of $\theta$ against (a) the volume fraction $\varphi$ for $e=0.70$ and $0.90$ and (b) the restitution coefficient $e$ for $\varphi=0.10$ and $0.40$.
    The solid and dotted lines represent our theoretical results and the dashed and dot-dashed lines correspond to the theory by Saha and Alam~\cite{Saha16}.
    The symbols indicate the simulation results.}
    \label{fig:comp_SA_theta}
\end{figure}

To quantify the comparisons of $\eta^*$ and $\theta$ shown in Figs.~\ref{fig:comp_SA_eta}(b) and \ref{fig:comp_SA_theta}(b), Table~\ref{table:comparison} lists the relative errors at $\varphi=0.10$ and $0.40$ for selected restitution coefficients.
For each quantity $X$, the relative error is defined as
$|X_{\mathrm{theory}}-X_{\mathrm{sim}}|/|X_{\mathrm{sim}}|$, where $X_{\mathrm{sim}}$ denotes the central value of the simulation results.
These data quantify the improvement relative to the anisotropic Gaussian theory while also showing the remaining limitations of the present predictions.

%%%%%%%%%%%%%%%%%%%%%%%%%%%%%%
\begin{table}[htbp]
    \centering
    \caption{Relative errors of the temperature $\theta$, viscosity $\eta^*$, and the normal stress differences $N_1$ and $N_2$ predicted by the present theory and the theory of Saha and Alam~\cite{Saha16}.
    The relative error of a quantity $X$ is defined as
    $|X_{\mathrm{theory}}-X_{\mathrm{sim}}|/|X_{\mathrm{sim}}|$.}
    \begin{tabular}{c|c|cc|cc|cc|cc}
        \hline
        \multirow{2}{*}{$\varphi$}
        & \multirow{2}{*}{$e$}
        & \multicolumn{2}{c|}{$\theta$}
        & \multicolumn{2}{c|}{$\eta^*$}
        & \multicolumn{2}{c|}{$N_1$}
        & \multicolumn{2}{c}{$N_2$}\\
        \cline{3-10}
        & & Our theory & Saha \& Alam
          & Our theory & Saha \& Alam
          & Our theory & Saha \& Alam
          & Our theory & Saha \& Alam\\
        \hline
        \multirow{10}{*}{$0.10$}
        & $0.10$ & $5.819\times10^{-1}$ & $7.969\times10^{-1}$ & $2.820\times10^{-1}$ & $7.273\times10^{-1}$ & $3.576\times10^{-1}$ & $3.479\times10^{-1}$ & $6.490\times10^{-3}$ & $1.101\times10^{0}$\\
        & $0.20$ & $4.758\times10^{-1}$ & $6.885\times10^{-1}$ & $2.171\times10^{-1}$ & $6.148\times10^{-1}$ & $3.051\times10^{-1}$ & $2.843\times10^{-1}$ & $3.122\times10^{-1}$ & $6.503\times10^{-1}$\\
        & $0.30$ & $4.127\times10^{-1}$ & $5.983\times10^{-1}$ & $1.714\times10^{-1}$ & $5.081\times10^{-1}$ & $3.167\times10^{-1}$ & $2.917\times10^{-1}$ & $3.724\times10^{-1}$ & $6.552\times10^{-1}$\\
        & $0.40$ & $1.803\times10^{-1}$ & $3.793\times10^{-1}$ & $6.469\times10^{-2}$ & $3.622\times10^{-1}$ & $2.390\times10^{-1}$ & $2.100\times10^{-1}$ & $6.022\times10^{-1}$ & $1.371\times10^{-1}$\\
        & $0.50$ & $1.531\times10^{-1}$ & $3.081\times10^{-1}$ & $6.077\times10^{-2}$ & $2.887\times10^{-1}$ & $2.105\times10^{-1}$ & $1.817\times10^{-1}$ & $6.550\times10^{-1}$ & $7.152\times10^{-2}$\\
        & $0.60$ & $7.895\times10^{-2}$ & $2.023\times10^{-1}$ & $4.119\times10^{-2}$ & $2.124\times10^{-1}$ & $1.818\times10^{-1}$ & $1.553\times10^{-1}$ & $7.184\times10^{-1}$ & $3.492\times10^{-2}$\\
        & $0.70$ & $5.672\times10^{-2}$ & $1.443\times10^{-1}$ & $4.282\times10^{-2}$ & $1.614\times10^{-1}$ & $1.507\times10^{-1}$ & $1.283\times10^{-1}$ & $7.657\times10^{-1}$ & $8.522\times10^{-2}$\\
        & $0.80$ & $3.888\times10^{-2}$ & $9.445\times10^{-2}$ & $3.823\times10^{-2}$ & $1.121\times10^{-1}$ & $1.366\times10^{-1}$ & $1.202\times10^{-1}$ & $8.068\times10^{-1}$ & $8.912\times10^{-2}$\\
        & $0.90$ & $3.778\times10^{-2}$ & $6.404\times10^{-2}$ & $4.671\times10^{-2}$ & $8.098\times10^{-2}$ & $1.124\times10^{-1}$ & $1.034\times10^{-1}$ & $8.641\times10^{-1}$ & $1.285\times10^{-1}$\\
        & $0.95$ & $4.235\times10^{-2}$ & $5.509\times10^{-2}$ & $4.383\times10^{-2}$ & $6.049\times10^{-2}$ & $6.907\times10^{-2}$ & $6.418\times10^{-2}$ & $9.073\times10^{-1}$ & $2.401\times10^{-1}$\\
        \hline
        \multirow{10}{*}{$0.40$}
        & $0.10$ & $4.679\times10^{-1}$ & $6.742\times10^{-1}$ & $4.572\times10^{-2}$ & $1.913\times10^{-1}$ & $7.014\times10^{-1}$ & $8.232\times10^{-1}$ & $8.273\times10^{-1}$ & $1.293\times10^{0}$\\
        & $0.20$ & $3.654\times10^{-1}$ & $5.678\times10^{-1}$ & $5.596\times10^{-2}$ & $2.630\times10^{-1}$ & $6.544\times10^{-1}$ & $7.660\times10^{-1}$ & $7.140\times10^{-1}$ & $1.075\times10^{0}$\\
        & $0.30$ & $2.472\times10^{-1}$ & $4.445\times10^{-1}$ & $2.638\times10^{-2}$ & $2.258\times10^{-1}$ & $6.295\times10^{-1}$ & $7.216\times10^{-1}$ & $5.316\times10^{-1}$ & $7.945\times10^{-1}$\\
        & $0.40$ & $1.414\times10^{-1}$ & $3.250\times10^{-1}$ & $4.070\times10^{-2}$ & $1.521\times10^{-1}$ & $6.228\times10^{-1}$ & $6.925\times10^{-1}$ & $4.035\times10^{-1}$ & $5.952\times10^{-1}$\\
        & $0.50$ & $5.837\times10^{-2}$ & $2.201\times10^{-1}$ & $7.788\times10^{-2}$ & $9.667\times10^{-2}$ & $5.700\times10^{-1}$ & $6.266\times10^{-1}$ & $3.392\times10^{-1}$ & $4.795\times10^{-1}$\\
        & $0.60$ & $4.024\times10^{-2}$ & $1.686\times10^{-1}$ & $7.938\times10^{-2}$ & $6.667\times10^{-2}$ & $4.920\times10^{-1}$ & $5.369\times10^{-1}$ & $2.982\times10^{-1}$ & $3.961\times10^{-1}$\\
        & $0.70$ & $7.749\times10^{-3}$ & $9.153\times10^{-2}$ & $6.887\times10^{-2}$ & $4.416\times10^{-2}$ & $5.099\times10^{-1}$ & $5.365\times10^{-1}$ & $1.745\times10^{-1}$ & $2.309\times10^{-1}$\\
        & $0.80$ & $1.621\times10^{-3}$ & $6.353\times10^{-2}$ & $2.583\times10^{-2}$ & $4.943\times10^{-2}$ & $5.039\times10^{-1}$ & $5.182\times10^{-1}$ & $1.594\times10^{-1}$ & $1.848\times10^{-1}$\\
        & $0.90$ & $1.890\times10^{-2}$ & $5.073\times10^{-2}$ & $2.823\times10^{-2}$ & $6.531\times10^{-2}$ & $2.920\times10^{-1}$ & $2.998\times10^{-1}$ & $4.285\times10^{-2}$ & $3.922\times10^{-2}$\\
        & $0.95$ & $4.481\times10^{-2}$ & $6.033\times10^{-2}$ & $6.771\times10^{-2}$ & $8.585\times10^{-2}$ & $2.637\times10^{-1}$ & $2.672\times10^{-1}$ & $7.464\times10^{-2}$ & $8.954\times10^{-2}$\\
        \hline
    \end{tabular}
    \label{table:comparison}
\end{table}
%%%%%%%%%%%%%%%%%%%%%%%%%%%%%%

Figures~\ref{fig:comp_SA_eta}(a) and \ref{fig:comp_SA_eta}(b) compare the predictions of the two theories for the viscosity $\eta^*$ as functions of the volume fraction and the restitution coefficient, respectively.
The present Grad-type theory generally gives smaller deviations from the simulated viscosity than the anisotropic Gaussian theory. 
The exceptions among the points listed in Table~\ref{table:comparison} can be found only at $\varphi=0.40$ with $e=0.60$ and $0.70$.
In particular, the discrepancy between the simulation and the present theory remains relatively small over the range of volume fractions shown in Fig.~\ref{fig:comp_SA_eta}(a).
The difference between the two theories becomes more pronounced in the strongly dissipative regime, as shown in Fig.~\ref{fig:comp_SA_eta}(b), where the anisotropic Gaussian prediction deviates substantially from the simulation data while the present theory still gives reasonable agreement.

%%%%%%%%%%%%%%%%%%%%%%%%%%%%%%
%%%%%%%%%%%%%%%%%%%%%%%%%%%%%%

Figures \ref{fig:comp_SA_theta}(a) and \ref{fig:comp_SA_theta}(b) compare $\theta$ predicted by the present theory and the anisotropic Gaussian theory of Ref.~\cite{Saha16} with the simulation results. At all points listed in Table \ref{table:comparison}, the present theory gives a smaller relative temperature error. 
This improvement does not imply uniformly small errors: substantial deviations remain under strongly inelastic conditions. The comparison concerns the three-dimensional system and parameter values examined here and does not establish a general ordering of the two closures for other observables or flow conditions. 
Because errors in the numerator and denominator of a stress ratio can partially cancel, we assess the temperature and viscosity separately.

%%%%%%%%%%%%%%%%%%%%%%%%%%%%%%
\begin{figure}[hbtp]
    \centering
    \includegraphics[width=\linewidth]{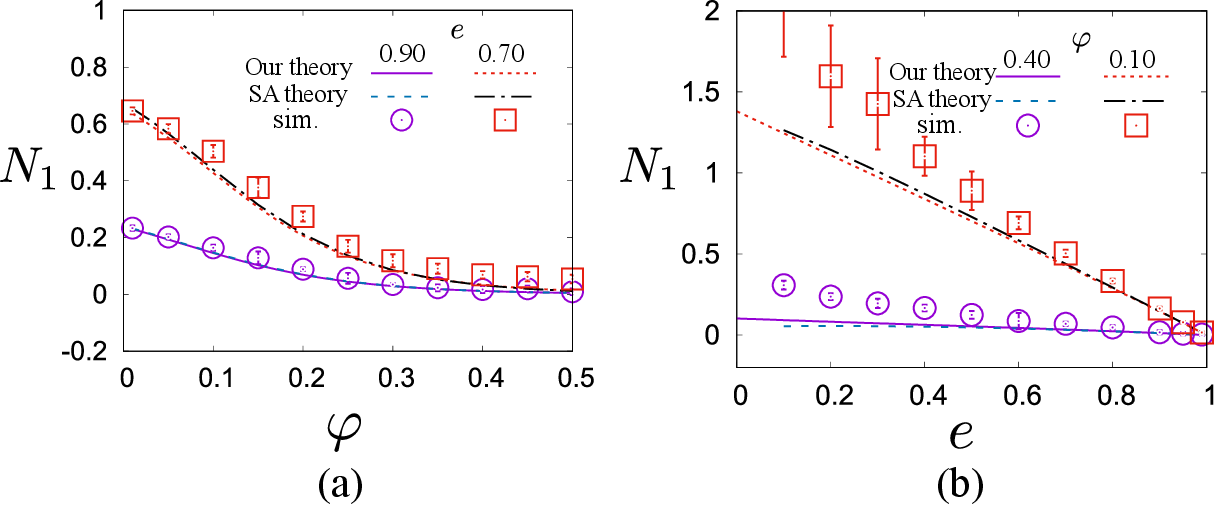}
    \caption{Plots of $N_1$ against (a) the volume fraction $\varphi$ for $e=0.70$ and $0.90$ and (b) the restitution coefficient $e$ for $\varphi=0.10$ and $0.40$.
    The solid and dotted lines represent our theoretical results, and the dashed and dot-dashed lines correspond to the theory by Saha and Alam~\cite{Saha16}.
    The symbols indicate the simulation results.}
    \label{fig:comp_SA_N1}
\end{figure}
%%%%%%%%%%%%%%%%%%%%%%%%%%%%%%

Figure~\ref{fig:comp_SA_N1} shows the dependence of the first normal stress difference $N_1$ on the volume fraction for $e=0.90$ and $0.70$ in panel (a), and on the restitution coefficient for fixed $\varphi=0.10$ and $0.40$ in panel (b).
Both theories capture the overall parameter dependence of $N_1$, but appreciable quantitative deviations remain, particularly at $\varphi=0.40$.
As shown in Table~\ref{table:comparison}, the anisotropic Gaussian theory gives smaller relative errors
at all ten tabulated points for $\varphi=0.10$, whereas the present theory gives smaller errors at all ten points for $\varphi=0.40$.
In the latter case, however, the improvement is modest compared with the remaining deviations from the simulation results.

%%%%%%%%%%%%%%%%%%%%%%%%%%%%%%
\begin{figure}[hbtp]
    \centering
    \includegraphics[width=\linewidth]{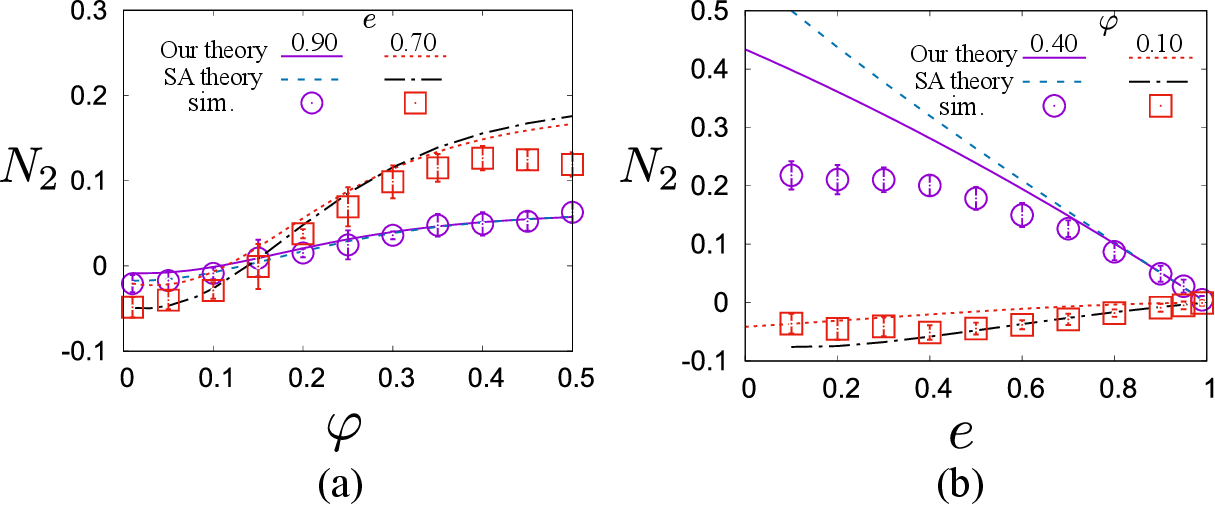}
    \caption{Plots of $N_2$ against (a) the volume fraction $\varphi$ for $e=0.70$ and $0.90$ and (b) the restitution coefficient $e$ for $\varphi=0.10$ and $0.40$.
    The solid and dotted lines represent our theoretical results, and the dashed and dot-dashed lines correspond to the theory by Saha and Alam~\cite{Saha16}.
    The symbols indicate the simulation results.}
    \label{fig:comp_SA_N2}
\end{figure}
%%%%%%%%%%%%%%%%%%%%%%%%%%%%%%

Similarly, Fig.~\ref{fig:comp_SA_N2} shows $N_2$ as a function of $\varphi$ for fixed $e=0.70$ and $0.90$ in panel (a), and as a function of $e$ for fixed $\varphi=0.10$ and $0.40$ in panel (b).
For $N_2$, the relative accuracy depends on both $\varphi$ and $e$. 
The present theory gives smaller errors at most tabulated points for $\varphi=0.40$, whereas the anisotropic Gaussian theory performs better for $\varphi=0.10$ over $0.40\le e\le0.95$. At $\varphi=0.10$ and $e\le0.30$, however,
the present theory gives smaller relative errors.
The remaining errors vary considerably:
they are approximately $0.65\%$, $31.2\%$, and $37.2\%$ for $e=0.10$, $0.20$, and $0.30$, respectively.

These comparisons indicate that neither closure is uniformly superior for all rheological quantities.
The present theory gives smaller relative temperature errors at all $20$ points listed in Table~\ref{table:comparison} and smaller viscosity errors at $18$ points.
The viscosity exceptions occur at $\varphi=0.40$ with $e=0.60$ and $0.70$.
For the normal stress differences, the relative performance depends on the volume fraction and restitution coefficient.
These relative improvements do not imply uniformly small errors: substantial deviations remain for the temperature at small $e$ and for the normal stress differences at some parameter values.

%%%%%%%%%%%%%%%%%%%%%%%%%%%%%%
\section{Relaxation process}\label{Sec:relaxation}
%%%%%%%%%%%%%%%%%%%%%%%%%%%%%%
In this section, we examine whether the present theory reproduces relaxation following a sudden change in the volume fraction.
This protocol is motivated by the variation of the bulk packing fraction in granular flows down inclined planes.
Silbert et al.~\cite{Silbert01} reported approximately uniform bulk packing fractions between $\varphi\approx0.55$ and $0.59$ for inclination angles from $20^\circ$ to $25^\circ$.
At $26^\circ$, they reported $\varphi\approx0.53$, although the region with an approximately uniform packing fraction becomes narrower.
Changing the inclination angle generally changes the shear rate and stress as well as the volume fraction.
We therefore consider an idealized volume-fraction quench, rather than a direct representation of a change in inclination, and examine a wider range of volume fractions than that established in Ref.~\cite{Silbert01}.

The system is initially prepared in a steady sheared state at $\varphi_{\mathrm{ini}}$ for $t<0$.
At $t=0$, the volume fraction is suddenly reduced to $\varphi_{\mathrm{tar}}$ by expanding the simulation box.
The particle number, diameter, and mass are kept fixed.
The box dimensions and particle coordinates are rescaled isotropically according to
\begin{equation}
    L_{\alpha}^{+}=sL_{\alpha}^{-},
    \qquad
    \bm r_i^{+}=s\bm r_i^{-},
    \qquad
    s:=\left(
    \frac{\varphi_{\mathrm{ini}}}{\varphi_{\mathrm{tar}}}
    \right)^{1/3},
\end{equation}
where the superscripts $-$ and $+$ denote quantities immediately before and after the quench, respectively.
The peculiar velocities are preserved, while the physical velocities are assigned consistently with the unchanged shear rate:
\begin{equation}
    \bm V_i^{+}=\bm V_i^{-},
    \qquad
    \bm v_i^{+}
    =\bm V_i^{-}+\dot{\gamma}y_i^{+}\bm e_x.
\end{equation}
The restitution coefficient is also unchanged.
Note that we analyze only the cases for $\varphi_\mathrm{ini}>\varphi_\mathrm{tar}$.
Thus, a sudden change of the volume fraction does not cause any overlap of particles at the instant of the box size change.

Because the peculiar velocities and particle masses are preserved, the kinetic temperature is continuous across the quench.
The kinetic stress per unit volume changes in proportion to the number density:
\begin{equation}
    T(0^+)=T(0^-),
    \qquad
    \sigma_{\alpha\beta}^{k}(0^+)
    =\frac{n_{\mathrm{tar}}}{n_{\mathrm{ini}}}
    \sigma_{\alpha\beta}^{k}(0^-),
\end{equation}
where $n_{\mathrm{tar}}/n_{\mathrm{ini}}
=\varphi_{\mathrm{tar}}/\varphi_{\mathrm{ini}}$.
Consequently, the dimensionless kinetic variables satisfy
\begin{equation}
    \theta(0^+)=\theta(0^-),
    \qquad
    \Delta\theta(0^+)=\Delta\theta(0^-),
    \qquad
    \delta\theta(0^+)=\delta\theta(0^-),
    \qquad
    \Pi_{\alpha\beta}^{*}(0^+)
    =\Pi_{\alpha\beta}^{*}(0^-).
    \label{eq:relation_0-_0+}
\end{equation}
We initialize the theoretical evolution using the kinetic moments measured in the simulation immediately before the quench.
According to Eq.~\eqref{eq:relation_0-_0+}, these values are preserved across the quench and are used as the initial conditions for Eqs.~\eqref{eq:dynamic_eqs} at $\varphi=\varphi_{\mathrm{tar}}$.
Both the collisional moments and the collisional stress are evaluated with the contact value $g_0(\varphi_\mathrm{tar})$.

The instantaneous Enskog closure does not, however, retain the pair structure produced by the expansion. 
Since all pair separations are multiplied by $s$ while the particle diameter remains unchanged, the region $d<r<sd$ contains no particle pairs immediately after the quench. For the protocol considered here, $s=(0.35/0.20)^{1/3}\simeq1.205$. 
The subsequent recovery of near-contact pairs is not described by the closure. 
This structural transient is distinct from the continuity of the kinetic variables in Eq.~\eqref{eq:relation_0-_0+} and can affect the earliest collisional-stress measurements.

\begin{figure}[htbp]
    \centering
    \includegraphics[width=\linewidth]{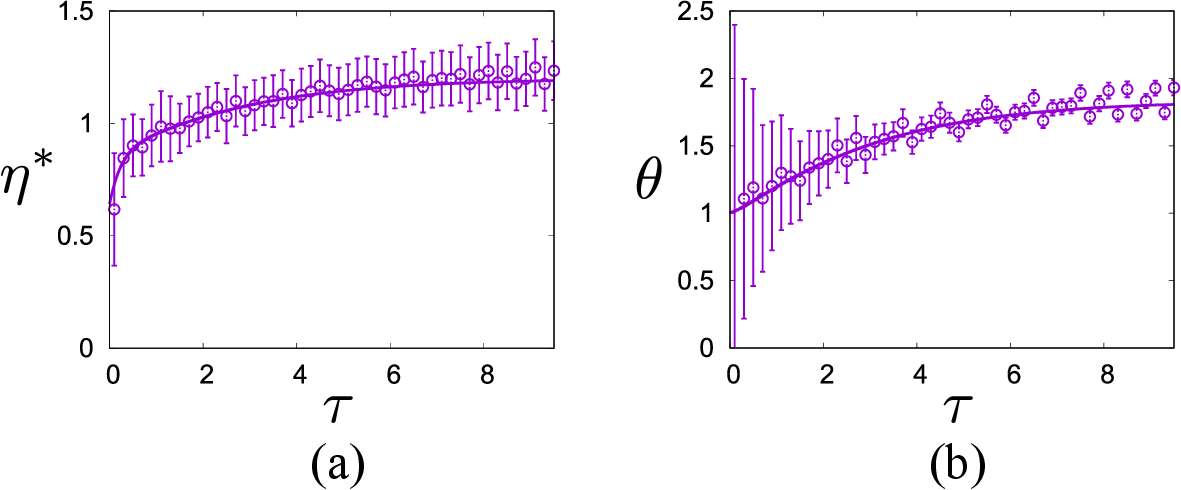}
    \caption{
    Time evolution of (a) the dimensionless viscosity $\eta^*$
    and (b) the dimensionless temperature $\theta$
    following an instantaneous expansion of the simulation box,
    which reduces the volume fraction from
    $\varphi_{\mathrm{ini}}=0.35$ to
    $\varphi_{\mathrm{tar}}=0.20$ at $e=0.90$.
    Lines show the theoretical predictions, and symbols show
    the simulation results.
    }
    \label{fig:evol}
\end{figure}

Figure~\ref{fig:evol} shows the time evolution of (a) the dimensionless viscosity $\eta^*$ and (b) the dimensionless temperature $\theta$ following a quench from $\varphi_{\mathrm{ini}}=0.35$ to $\varphi_{\mathrm{tar}}=0.20$ at $e=0.90$.
The theoretical predictions obtained by solving Eq.~\eqref{eq:dynamic_eqs} agree well with the simulation results for both quantities.
This comparison supports the applicability of the present theory to the relaxation process examined here.

%%%%%%%%%%%%%%%%%%%%%%%%%%%%%%
%%%%%%%%%%%%%%%%%%%%%%%%%%%%%%

%%%%%%%%%%%%%%%%%%%%%%%%%%%%%%
\section{Discussion}\label{Sec:Discussion}
%%%%%%%%%%%%%%%%%%%%%%%%%%%%%%
Let us discuss our results. 
In the first subsection, we argue for a long-range correlation function for dilute situations.
In the second subsection, we discuss a long-range correlation function for dense situations.
In the third subsection, we discuss the effect of mutual friction between grains.

%%%%%%%%%%%%%%%%%%%%%%%%%%%%%%
\subsection{Long-range correlation for relatively dilute cases}
%%%%%%%%%%%%%%%%%%%%%%%%%%%%%%
Otsuki and Hayakawa \cite{Otsuki09a} demonstrated the existence of a long-range correlation for sheared granular flow.
Such a long-range correlation is generic for sheared liquids.
The mechanism of the long-range correlation function has been argued using the fluctuating hydrodynamics~\cite{Kirkpatrick82,Ronis82,Lutsko85,Wada03,Wada04} and a stochastic lattice gas model~\cite{Spohn83}.
Recently, the existence of a long-range correlation function has been verified by extensive molecular dynamics simulations~\cite {Nakano21,Nakano22}, although they also indicate that the correlation effects are not relevant to hydrodynamics.

The present comparisons assess the accuracy of the Enskog closure for mean rheological quantities, but do not establish that spatial correlation effects are absent.
Moreover, our study is consistent with the applicability of the fluctuating hydrodynamics for granular flows as in Ref.~\cite{Otsuki09a}, in which the transport coefficients can be determined by the Enskog kinetic theory without consideration of the long-range correlation.

%%%%%%%%%%%%%%%%%%%%%%%%%%%%%%
\subsection{Long-range correlation for denser flows}
%%%%%%%%%%%%%%%%%%%%%%%%%%%%%%
Next, let us discuss the phenomenological correlation length introduced by Jenkins and Berzi for $\varphi>0.50$ \cite{Jenkins10, Berzi18,Berzi20}.
Although a proper kinetic theory for dense granular flow for $\varphi>0.50$ is out of scope for this paper, we should consider correlation effects in the dense region.
To include correlation effects, we can also look to Refs.~\cite{Suzuki15, Suzuki19}. 

%%%%%%%%%%%%%%%%%%%%%%%%%%%%%%
\subsection{Effects of mutual friction between grains}
%%%%%%%%%%%%%%%%%%%%%%%%%%%%%%
In this paper, the system consists of frictionless particles.
It would be possible to extend the theory to systems with friction by considering the tangential restitution coefficient~\cite{Garzo}. 
However, this model might be inadequate to simulate real systems. 
This is because the contact time is infinitesimal as long as a hard-core model is adopted.
The finite duration of contact is important, for example, for discontinuous shear-thickening~\cite{Otsuki11}.

Models of kinetic theory incorporating finite-time contact have been partially incorporated by Refs.~\cite{Sugimoto20, Ishikawa21, Takada22a, Takada22b}, which treated granular flow and inertial suspensions. 
In these references, the repulsive potential between particles is assumed to be described by harmonic potentials.

%%%%%%%%%%%%%%%%%%%%%%%%%%%%%%
\section{Summary}\label{Sec:Summary}
%%%%%%%%%%%%%%%%%%%%%%%%%%%%%%
We have developed a kinetic theory for dry granular particles under uniform simple shear by combining the Enskog approximation with Grad's moment method. 
The collisional moments and stresses are expressed in terms of generalized hypergeometric functions, retaining all orders in the dimensionless shear-rate parameter within the adopted Grad closure.
Appendix~\ref{sec:truncation} shows that, at $\varphi=0.30$ and across the range of restitution coefficients examined, the steady-state predictions from finite-order expansions converge to those from the generalized hypergeometric expressions as the truncation order increases.
The formulation provides closed evolution equations for the kinetic stress and an expression for the collisional stress in terms of the kinetic moments, without introducing fitting parameters. 
Retaining the shear-induced anisotropy allows the theory to describe normal stress differences, which vanish for simple shear at Navier--Stokes order about an isotropic reference state.

Comparisons with event-driven simulations for $\varphi\le 0.5$ show that the theory reproduces the viscosity reasonably well over much of the range examined. 
Relative to the anisotropic Gaussian theory, it gives smaller temperature errors at all points listed in Table \ref{table:comparison} and smaller viscosity errors at most of these points. 
The viscosity exceptions occur at $\varphi=0.40$ with $e=0.60$ and $0.70$. 
Substantial temperature deviations remain under strongly inelastic conditions. 
For the normal stress differences, the two theories show comparable overall accuracy, with their relative performance depending on the volume fraction and restitution coefficient.
These results identify the strengths and limitations of the present closure for each observable rather than establishing its uniform superiority.

We have also compared the predicted relaxation of the temperature and viscosity with simulations following an instantaneous expansion from $\varphi=0.35$ to $0.20$ at $e=0.90$. 
The agreement supports the use of the closure for the relaxation protocol examined here. 
However, the closure immediately uses the contact value at the target density and does not describe the initial recovery of near-contact pairs after the expansion. 
The comparison therefore does not establish the accuracy of the theory for arbitrarily early times or for general transient protocols.

The formulation provides a basis for studying steady rheology and relaxation within the Grad--Enskog approximation. 
Future work may examine a wider range of transient protocols and extensions that include velocity-dependent restitution, interparticle friction, and finite contact duration.

%%%%%%%%%%%%%%%%%%%%%%%%%%%%%%
\section*{Acknowledgments}
%%%%%%%%%%%%%%%%%%%%%%%%%%%%%%
We thank Andr\'{e}s Santos, Vicente Garz\'{o}, Hiroyoshi Nakano, and Stefan Luding for fruitful discussions.
This research was partially supported by the Grant-in-Aid of MEXT for Scientific Research (Grants Nos.~JP24K06974, JP24K07193, JP25K01063, and JP26K06960).
HH thanks the Kyoto University Foundation for its support.

During the revision of this manuscript, the authors used ChatGPT (OpenAI, GPT-5) to assist with checking algebraic expressions and revising scientific explanations.
The authors verified the resulting expressions and reviewed and edited all adopted text. 
The authors take full responsibility for the final manuscript.

%%%%%%%%%%%%%%%%%%%%%%%%%%%%%%
\section*{Author Contributions}
%%%%%%%%%%%%%%%%%%%%%%%%%%%%%%
S. Iizuka performed the initial theoretical calculations and numerical simulations and prepared the first draft of the manuscript. 
S. Takada checked the initial calculations and simulation results, performed additional theoretical calculations and numerical simulations, and led the revision and preparation of the final manuscript. 
H. Hayakawa supervised the study and critically reviewed and edited the manuscript.

%%%%%%%%%%%%%%%%%%%%%%%%%%%%%%
\section*{Conflict of Interest}
%%%%%%%%%%%%%%%%%%%%%%%%%%%%%%
The authors declare no conflicts of interest.

%%%%%%%%%%%%%%%%%%%%%%%%%%%%%%
\section*{Data Availability}
%%%%%%%%%%%%%%%%%%%%%%%%%%%%%%
The numerical data underlying the figures and tables, together with the analysis and plotting scripts, are openly available in the GitHub repository associated with this article~\cite{IizukaTakadaHayakawaGithub}.

%%%%%%%%%%%%%%%%%%%%%%%%%%%%%%
\appendix

%%%%%%%%%%%%%%%%%%%%%%%%%%%%%%
\section{Finite-size effects in the simulations}\label{sec:finite_size}
\begin{figure}[hbtp]
    \centering
    \includegraphics[width=0.5\linewidth]{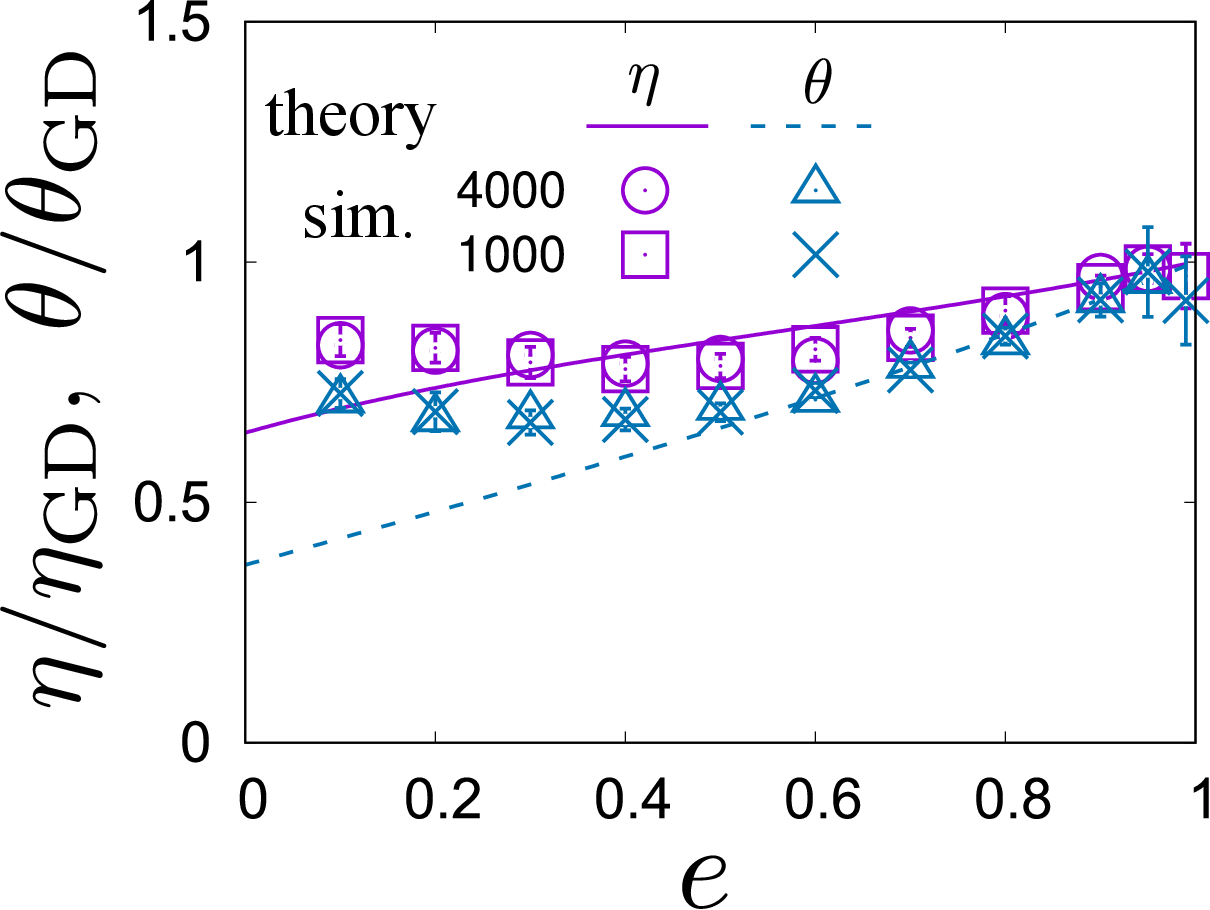}
    \caption{Normalized viscosity $\eta/\eta_{\mathrm{GD}}$ and temperature $\theta/\theta_{\mathrm{GD}}$ as functions of the restitution coefficient $e$ at $\varphi=0.30$.
    Open circles and squares represent the viscosity obtained from simulations with $N=4000$ and $1000$, respectively.
    Open triangles and crosses represent the corresponding temperature results.
    The solid and blue dashed curves show the theoretical predictions for the viscosity and temperature, respectively.
    }
    \label{fig:finite_size}
\end{figure}

%%%%%%%%%%%%%%%%%%%%%%%%%%%%%%
To assess finite-size effects in the simulations, we compare the viscosity and temperature obtained for $N=1000$ and $4000$ at a fixed volume fraction of $\varphi=0.30$.
Figure~\ref{fig:finite_size} shows the restitution-coefficient dependence of the normalized viscosity $\eta/\eta_{\mathrm{GD}}$ and temperature $\theta/\theta_{\mathrm{GD}}$.
For both quantities, the results for the two particle numbers nearly coincide over the range of restitution coefficients examined.
Increasing the particle number by a factor of four produces no appreciable systematic change in either quantity.
These results indicate that finite-size effects on the viscosity and temperature are small under the conditions considered here.

%%%%%%%%%%%%%%%%%%%%%%%%%%%%%%
%%%%%%%%%%%%%%%%%%%%%%%%%%%%%%

%%%%%%%%%%%%%%%%%%%%%%%%%%%%%%
\section{Derivation of the expression of $\Lambda_{\alpha\beta}$}\label{sec:Derivation_Lambda}
%%%%%%%%%%%%%%%%%%%%%%%%%%%%%%

In this subsection, we derive the expression of $\Lambda_{\alpha\beta}$ under the Enskog approximation~\eqref{eq:decoupling} with Grad's expansion~\eqref{eq:Grad} for the distribution function.
$\Lambda_{\alpha\beta}$ can be rewritten as~\cite{Hayakawa17, Takada20}
\begin{equation}
    \Lambda_{\alpha\beta}
    = \overline{\Lambda}_{\alpha\beta}
    -\dot\gamma \left(\delta_{\alpha x}\sigma_{y\beta}^c + \delta_{\beta x}\sigma_{y\alpha}^c\right),
\end{equation}
where $\overline{\Lambda}_{\alpha\beta}$ is given by
\begin{align}
    \overline{\Lambda}_{\alpha\beta}
    &= \frac{1+e}{4}md^2 g_0
    \int d\bm{V}_1 \int d\bm{V}_2 \int d\hat{\bm{k}}
    \Theta(\bm{V}_{12}\cdot \hat{\bm{k}})
    (\bm{V}_{12}\cdot \hat{\bm{k}})^2\nonumber\\
    %%%
    &\hspace{7em}\times
    \left[V_{12,\alpha}\hat{k}_\beta
    + V_{12,\beta}\hat{k}_\alpha
    -(1+e)(\bm{V}_{12}\cdot \hat{\bm{k}})
    \hat{k}_\alpha \hat{k}_\beta\right]
    f\left(\bm{V}_1+\frac{1}{2}\dot\gamma d \hat{k}_y \bm{e}_x\right)
    f\left(\bm{V}_2-\frac{1}{2}\dot\gamma d\hat{k}_y \bm{e}_x\right),
    \label{eq:def_barLambda}
\end{align}
and the collisional contribution of the stress $\sigma_{\alpha\beta}^c$ is given by Eq.~\eqref{eq:Pc}.

Let us introduce $I^{(\ell)}(\hat{\bm{k}})$ and $I_\alpha^{(\ell)}(\hat{\bm{k}})$ ($\alpha=x, y, z$) as~\cite{Takada20}
\begin{subequations}\label{eq:def_I_Ia}
\begin{align}
    I^{(\ell)}(\hat{\bm{k}})
    &:= \int d\bm{V}_1 \int d\bm{V}_2 
    \Theta\left(\hat{\bm{k}}\cdot \bm{V}_{12}\right)
    \left(\hat{\bm{k}}\cdot \bm{V}_{12}\right)^\ell
    f\left(\bm{V}_1+\frac{1}{2}\dot\gamma d \hat{k}_y \bm{e}_x\right)
    f\left(\bm{V}_2-\frac{1}{2}\dot\gamma d \hat{k}_y \bm{e}_x\right)\nonumber\\
    %%%
    &= \int d\bm{V}_1 \int d\bm{V}_2 
    \Theta\left(\hat{\bm{k}}\cdot \bm{V}_{12}-\dot\gamma d \hat{k}_x\hat{k}_y\right)
    \left(\hat{\bm{k}}\cdot \bm{V}_{12}-\dot\gamma d \hat{k}_x\hat{k}_y\right)^\ell
    f\left(\bm{V}_1\right) f\left(\bm{V}_2\right),\\
    %%%
    %%%
    %%%
    I_\alpha^{(\ell)}(\hat{\bm{k}})
    &:= \int d\bm{V}_1 \int d\bm{V}_2 
    \Theta\left(\hat{\bm{k}}\cdot \bm{V}_{12}\right)
    \left(\hat{\bm{k}}\cdot \bm{V}_{12}\right)^\ell
    V_{12,\alpha}
    f\left(\bm{V}_1+\frac{1}{2}\dot\gamma d\hat{k}_y \bm{e}_x\right)
    f\left(\bm{V}_2-\frac{1}{2}\dot\gamma d \hat{k}_y \bm{e}_x\right)\nonumber\\
    %%%
    &= \int d\bm{V}_1 \int d\bm{V}_2 
    \Theta\left(\hat{\bm{k}}\cdot \bm{V}_{12}-\dot\gamma d \hat{k}_x\hat{k}_y\right)
    \left(\hat{\bm{k}}\cdot \bm{V}_{12}-\dot\gamma d \hat{k}_x\hat{k}_y\right)^\ell
    \left(V_{12,\alpha}-\dot\gamma d \hat{k}_y\delta_{\alpha x}\right)
    f\left(\bm{V}_1\right) f\left(\bm{V}_2\right),
\end{align}
\end{subequations}
where we have used the changes of the variables $\bm{V}_1+(1/2)\dot\gamma d \hat{k}_y \bm{e}_x \to \bm{V}_1$ and $\bm{V}_2-(1/2)\dot\gamma d \hat{k}_y \bm{e}_x \to \bm{V}_2$ in the integrand of the second lines in Eqs.~\eqref{eq:def_I_Ia}.
Using $I^{(\ell)}(\hat{\bm{k}})$ and $I_\alpha^{(\ell)}(\hat{\bm{k}})$, $\sigma_{\alpha\beta}^c$ in Eq.~\eqref{eq:Pc} and $\overline{\Lambda}_{\alpha\beta}$ in Eq.~\eqref{eq:def_barLambda} are, respectively, expressed as
\begin{align}
  \sigma_{\alpha\beta}^c
    &= -\frac{1+e}{4}md^3 g_0(\varphi) 
    \int d\hat{\bm{k}} \hat{k}_\alpha \hat{k}_\beta I^{(2)}(\hat{\bm{k}}),\\
    %%%
    \overline{\Lambda}_{\alpha\beta}
    &= \frac{1+e}{4}md^2 g_0(\varphi) 
    \int d\hat{\bm{k}} 
    \left[\hat{k}_\alpha I_\beta^{(2)}(\hat{\bm{k}})
    + \hat{k}_\beta I_\alpha^{(2)}(\hat{\bm{k}})
    - (1+e)\hat{k}_\alpha \hat{k}_\beta 
    I^{(3)}(\hat{\bm{k}})\right].
\end{align}
Introducing $L_{\alpha\beta}$ and $M_{\alpha\beta}$ defined as
\begin{subequations}\label{L_ab&M_ab}
\begin{align}
    L_{\alpha\beta}
    &:= \int d\hat{\bm{k}} 
    \left[\hat{k}_\alpha I_\beta^{(2)}(\hat{\bm{k}})
    + \hat{k}_\beta I_\alpha^{(2)}(\hat{\bm{k}})
    - 2\hat{k}_\alpha \hat{k}_\beta 
    I^{(3)}(\hat{\bm{k}})
    + \dot\gamma d \hat{k}_y\left(\delta_{\alpha x}\hat{k}_\beta
    + \delta_{\beta x}\hat{k}_\alpha\right)I^{(2)}(\hat{\bm{k}})\right],\\
    %%%
    M_{\alpha\beta}
    &:= \int d\hat{\bm{k}} \hat{k}_\alpha \hat{k}_\beta I^{(3)}(\hat{\bm{k}}) ,
\end{align}
\end{subequations}
the collisional moment $\Lambda_{\alpha\beta}$ can be rewritten as
\begin{equation}\label{Lambda_L_M}
    \Lambda_{\alpha\beta}
    = \frac{1+e}{4}md^2 g_0
    \left[L_{\alpha\beta} + (1-e)M_{\alpha\beta}\right].
\end{equation}
Thus, the first step of obtaining $\Lambda_{\alpha\beta}$ is to evaluate $I^{(\ell)}(\hat{\bm{k}})$ and $I_\alpha^{(\ell)}(\hat{\bm{k}})$ introduced in Eq.~\eqref{eq:def_I_Ia}, and the second step is to obtain $L_{\alpha\beta}$ and $M_{\alpha\beta}$ introduced in Eq.~\eqref{L_ab&M_ab} before the evaluation of the integral in Eq.~\eqref{Lambda_L_M}. 

With the aid of Grad's approximation~\eqref{eq:Grad}, the product of the one-body velocity distribution functions is written as
\begin{align}
    f(\bm{V}_1) f(\bm{V}_2)
    &\approx
    n^2\left(\frac{m}{2\pi T}\right)^3
    \exp \left(-2G^2 - \frac{g^2}{2}\right)
    \nonumber\\
    %%%
    &\hspace{1em}\times
    \left[
    1 
    + \Pi_{\alpha\beta}^*
    \left(2G_\alpha G_\beta 
    + \frac{1}{2}g_\alpha g_\beta\right)
    + \Pi_{\alpha\beta}^* \Pi_{\gamma\delta}^*
    \left(G_\alpha + \frac{g_\alpha}{2}\right)
    \left(G_\beta + \frac{g_\beta}{2}\right)
    \left(G_\gamma - \frac{g_\gamma}{2}\right)
    \left(G_\delta - \frac{g_\delta}{2}\right)
    \right],
    \label{eq:product_f}
\end{align}
where we have introduced $G_\alpha:=(V_{1,\alpha}+V_{2,\alpha})/2\vt$ and $g_\alpha:=(V_{1,\alpha}-V_{2,\alpha})/\vt$.
Let us introduce the dimensionless variables $\widetilde{I}^{(\ell)}\left(\hat{\bm{k}}\right):= I^{(\ell)}\left(\hat{\bm{k}}\right)/(n^2 \vt^\ell)$ and $\widetilde{I}_\alpha^{(\ell)}\left(\hat{\bm{k}}\right) := I_\alpha^{(\ell)}\left(\hat{\bm{k}}\right)/(n^2 \vt^{\ell+1})$ (see Eqs.~\eqref{eq:def_I_Ia}), which can be written as
\begin{subequations}\label{eq:nondim_I_Ia}
\begin{align}
    \widetilde{I}^{(\ell)}\left(\hat{\bm{k}}\right)
    &= \frac{1}{(2\pi)^{3/2}}
    \int d\bm{g} \Theta\left(\hat{\bm{k}}\cdot\bm{g}-\frac{1}{\sqrt{2\theta}}\hat{k}_x\hat{k}_y\right)
    \left(\hat{\bm{k}}\cdot\bm{g}-\frac{1}{\sqrt{2\theta}}\hat{k}_x\hat{k}_y\right)^\ell
    \exp\left(-\frac{1}{2}g^2\right)
    \mathcal{P}_1\left(\{\bm{g}\}; \hat{\bm{k}}\right),\\
    %%%
    %%%
    %%%
    \widetilde{I}_{\alpha}^{(\ell)}\left(\hat{\bm{k}}\right)
    &= \frac{1}{(2\pi)^{3/2}}
    \int d\bm{g} \Theta\left(\hat{\bm{k}}\cdot\bm{g}-\frac{1}{\sqrt{2\theta}}\hat{k}_x\hat{k}_y\right)
    \left(\hat{\bm{k}}\cdot\bm{g}-\frac{1}{\sqrt{2\theta}}\hat{k}_x\hat{k}_y\right)^\ell
    \left(g_\alpha-\frac{1}{\sqrt{2\theta}}\hat{k}_y\delta_{\alpha x}\right)
    \exp\left(-\frac{1}{2}g^2\right)
    \mathcal{P}_1\left(\{\bm{g}\}; \hat{\bm{k}}\right),
\end{align}
\end{subequations}
where we have introduced $\mathcal{P}\left(\{\bm{g}\}; \hat{\bm{k}}\right)$ defined as
\begin{align}
    \mathcal{P}_1\left(\{\bm{g}\}; \hat{\bm{k}}\right) 
    &:= 1 + \frac{1}{8} \Pi_{\beta\gamma}^*\Pi_{\beta\gamma}^* 
    + \frac{1}{2}\Pi_{\beta\gamma}^* g_\beta g_\gamma
    - \frac{1}{4}\Pi_{\beta\delta}^*\Pi_{\gamma\delta}^* g_\beta g_\gamma 
    + \frac{1}{16} \Pi_{\beta\gamma}^*\Pi_{\delta\mu}^* g_\beta g_\gamma g_\delta g_\mu.
\end{align}
We have used Eq.~\eqref{eq:product_f} for the derivation.

To evaluate the integrals over $\bm{g}$ for $\widetilde{I}^{(\ell)}\left(\hat{\bm{k}}\right)$ and $\widetilde{I}_{\alpha}^{(\ell)}\left(\hat{\bm{k}}\right)$ in Eqs.~\eqref{eq:nondim_I_Ia}, introducing the variable transformation specific to the three-dimensional system, $g_\alpha=U_{\alpha \beta}\overline{g}_\beta$ with \cite{Takada20} 
\begin{align}
    U_{\alpha \beta}=
    \begin{pmatrix}
    U_{xx} & U_{xy} & U_{xz}\\
    U_{yx} & U_{yy} & U_{yz}\\
    U_{zx} & U_{zy} & U_{zz}
    \end{pmatrix}
    =
    \begin{pmatrix}
    \displaystyle \frac{\hat{k}_y}{\sqrt{\hat{k}_x^2+\hat{k}_y^2}} & \displaystyle \frac{\hat{k}_x\hat{k}_z}{\sqrt{\hat{k}_x^2+\hat{k}_y^2}} & \hat{k}_x\\
      &\\
    \displaystyle -\frac{\hat{k}_x}{\sqrt{\hat{k}_x^2+\hat{k}_y^2}} & \displaystyle \frac{\hat{k}_y\hat{k}_z}{\sqrt{\hat{k}_x^2+\hat{k}_y^2}} & \hat{k}_y\\
      &\\
    0 & -\sqrt{\hat{k}_x^2+\hat{k}_y^2} & \hat{k}_z
    \end{pmatrix} ,
    \label{eq:def_U}
\end{align}
$\widetilde{I}^{(\ell)}\left(\hat{\bm{k}}\right)$ and $\widetilde{I}_\alpha^{(\ell)}\left(\hat{\bm{k}}\right)$ in Eq.~\eqref{eq:nondim_I_Ia} can be rewritten as
\begin{subequations}\label{eq:nondim_I_Ia_ver2}
\begin{align}
    \widetilde{I}^{(\ell)}\left(\hat{\bm{k}}\right)
    &= \frac{1}{(2\pi)^{3/2}}
    \int d\bm{g} \Theta\left(\overline{g}_3-\frac{1}{\sqrt{2\theta}}\hat{k}_x\hat{k}_y\right)
    \left(\overline{g}_3-\frac{1}{\sqrt{2\theta}}\hat{k}_x\hat{k}_y\right)^\ell
    \exp\left(-\frac{1}{2}\overline{g}^2\right)
    \mathcal{P}_2\left(\{\overline{g}_m\}; \hat{\bm{k}}\right),\\
    %%%
    \widetilde{I}_{\alpha}^{(\ell)}\left(\hat{\bm{k}}\right)
    &= \frac{1}{(2\pi)^{3/2}}
    \int d\bm{g} \Theta\left(\overline{g}_3-\frac{1}{\sqrt{2\theta}}\hat{k}_x\hat{k}_y\right)
    \left(\overline{g}_3-\frac{1}{\sqrt{2\theta}}\hat{k}_x\hat{k}_y\right)^\ell
    \left(U_{\alpha i}\overline{g}_i-\frac{1}{\sqrt{2\theta}}\hat{k}_y\delta_{\alpha x}\right)
    \exp\left(-\frac{1}{2}\overline{g}^2\right)
    \mathcal{P}_2\left(\{\overline{g}_m\}; \hat{\bm{k}}\right),
\end{align}
\end{subequations}
where we have introduced $\mathcal{P}_2\left(\{\overline{g}_m\}; \hat{\bm{k}}\right)$ as
\begin{equation}
    \mathcal{P}_2\left(\{\overline{g}_m\}; \hat{\bm{k}}\right) 
    := 1 
    + \frac{1}{8} \Pi_{\beta\gamma}^*\Pi_{\beta\gamma}^* 
    + \frac{1}{2}\Pi_{\beta\gamma}^* U_{\beta i} U_{\gamma j} \overline{g}_i \overline{g}_j
    - \frac{1}{4}\Pi_{\beta\delta}^*\Pi_{\gamma\delta}^* U_{\beta i} U_{\gamma j} \overline{g}_i \overline{g}_j 
    + \frac{1}{16} \Pi_{\beta\gamma}^*\Pi_{\delta\mu}^* U_{\beta i} U_{\gamma j} U_{\delta k} U_{\mu l} \overline{g}_i \overline{g}_j \overline{g}_k \overline{g}_l.
    \label{eq:P2}
\end{equation}
Now, let us decompose $\widetilde{I}_{\alpha}^{(\ell)}\left(\hat{\bm{k}}\right)$ as
\begin{equation}
    \widetilde{I}_{\alpha}^{(\ell)}\left(\hat{\bm{k}}\right)
    =
    U_{\alpha 1}\widetilde{\overline{I}}_1^{(\ell)}\left(\hat{\bm{k}}\right)
    +U_{\alpha 2}\widetilde{\overline{I}}_2^{(\ell)}\left(\hat{\bm{k}}\right)
    +\hat{k}_\alpha\widetilde{I}^{(\ell+1)}\left(\hat{\bm{k}}\right)
    +\frac{1}{\sqrt{2\theta}}\hat{k}_y \left(\hat{k}_\alpha \hat{k}_x
    -\delta_{\alpha x}\right)
    \widetilde{I}^{(\ell)}\left(\hat{\bm{k}}\right),
    \label{eq:relation_tI_tbI}
\end{equation}
where
\begin{equation}
    \widetilde{\overline{I}}_i^{(\ell)}\left(\hat{\bm{k}}\right)
    %%%
    := \frac{1}{(2\pi)^{3/2}}
    \int d\bm{g} \Theta\left(\overline{g}_3-\frac{1}{\sqrt{2\theta}}\hat{k}_x\hat{k}_y\right)
    \left(\overline{g}_3-\frac{1}{\sqrt{2\theta}}\hat{k}_x\hat{k}_y\right)^\ell
    \overline{g}_i
    \exp\left(-\frac{1}{2}\overline{g}^2\right)
    \mathcal{P}_2\left(\{\overline{g}_m\}; \hat{\bm{k}}\right),\quad
    (i=1, 2).
    \label{eq:def_tilde_bar_I}
\end{equation}
After the integration over $\overline{g}_1$ and $\overline{g}_2$, $\widetilde{I}^{(\ell)}\left(\hat{\bm{k}}\right)$ and $\widetilde{\overline{I}}_i^{(\ell)}\left(\hat{\bm{k}}\right)$ are rewritten as
\begin{subequations}\label{eq:nondim_I_Ia_ver3}
\begin{align}
    \widetilde{I}^{(\ell)}\left(\hat{\bm{k}}\right)
    &= 
    \frac{1}{\sqrt{2\pi}}
    \int_{-\infty}^\infty d\overline{g}_3
    \Theta\left(\overline{g}_3-\frac{1}{\sqrt{2\theta}}\hat{k}_x\hat{k}_y\right)
    \left(\overline{g}_3-\frac{1}{\sqrt{2\theta}}\hat{k}_x\hat{k}_y\right)^\ell
    \exp\left(-\frac{\overline{g}_3^2}{2}\right)
    \mathcal{P}_3(\overline{g}_3;\hat{\bm{k}}),\\
    %%%
    \widetilde{\overline{I}}_i^{(\ell)}\left(\hat{\bm{k}}\right)
    &= 
    \frac{1}{\sqrt{2\pi}}
    \int_{-\infty}^\infty d\overline{g}_3
    \Theta\left(\overline{g}_3-\frac{1}{\sqrt{2\theta}}\hat{k}_x\hat{k}_y\right)
    \left(\overline{g}_3-\frac{1}{\sqrt{2\theta}}\hat{k}_x\hat{k}_y\right)^\ell
    \exp\left(-\frac{\overline{g}_3^2}{2}\right)
    \mathcal{P}_{3,i}(\overline{g}_3;\hat{\bm{k}}),
\end{align}
\end{subequations}
where we have introduced $\mathcal{P}_3(\overline{g}_3;\hat{\bm{k}})$ and $\mathcal{P}_{3,i}(\overline{g}_3;\hat{\bm{k}})$ defined as
\begin{subequations}\label{eq:P3}
\begin{align}
    \mathcal{P}_3(\overline{g}_3;\hat{\bm{k}})
    &:= \frac{1}{2\pi}
    \int_{-\infty}^\infty d\overline{g}_1 
    \int_{-\infty}^\infty d\overline{g}_2 
    \exp\left(-\frac{\overline{g}_1^2+\overline{g}_2^2}{2}\right)
    \mathcal{P}_2\left(\{\overline{\bm{g}}\}; \hat{\bm{k}}\right),\\
    %%%
    \mathcal{P}_{3,i}(\overline{g}_3;\hat{\bm{k}})
    &:= \frac{1}{2\pi}
    \int_{-\infty}^\infty d\overline{g}_1 
    \int_{-\infty}^\infty d\overline{g}_2 
    \exp\left(-\frac{\overline{g}_1^2+\overline{g}_2^2}{2}\right)
    \overline{g}_i
    \mathcal{P}_2\left(\{\overline{\bm{g}}\}; \hat{\bm{k}}\right).
\end{align}
\end{subequations}
Substituting Eq.~\eqref{eq:P2} into Eqs.~\eqref{eq:P3}, we obtain
\begin{subequations}
\begin{align}
    \mathcal{P}_3(\overline{g}_3;\hat{\bm{k}})
    &=
    \mathcal{P}_3^{(0)}(\overline{g}_3;\hat{\bm{k}})
    +\mathcal{P}_3^{(1)}(\overline{g}_3;\hat{\bm{k}})
    +\mathcal{P}_3^{(2)}(\overline{g}_3;\hat{\bm{k}})
    +\mathcal{P}_3^{(3)}(\overline{g}_3;\hat{\bm{k}}),\\
    %%%
    %%%
    \mathcal{P}_{3,i}(\overline{g}_3;\hat{\bm{k}})
    &=
    \mathcal{P}_{3,i}^{(1)}(\overline{g}_3;\hat{\bm{k}})
    +\mathcal{P}_{3,i}^{(2)}(\overline{g}_3;\hat{\bm{k}})
    +\mathcal{P}_{3,i}^{(3)}(\overline{g}_3;\hat{\bm{k}}),\quad (i=1, 2),
\end{align}
\end{subequations}
where we have introduced
\begin{subequations}
\begin{align}
    \mathcal{P}_3^{(0)}(\overline{g}_3;\hat{\bm{k}})
    &:=
    \frac{1}{2\pi}
    \int_{-\infty}^\infty d\overline{g}_1 
    \int_{-\infty}^\infty d\overline{g}_2 
    \exp\left(-\frac{\overline{g}_1^2+\overline{g}_2^2}{2}\right)
    \left(1+\frac{1}{8}\Pi_{\beta\gamma}^*\Pi_{\beta\gamma}^*
    \right),\\
    %%%
    %%%
    %%%
    \mathcal{P}_3^{(1)}(\overline{g}_3;\hat{\bm{k}})
    &:=
    \frac{1}{4\pi}
    \int_{-\infty}^\infty d\overline{g}_1 
    \int_{-\infty}^\infty d\overline{g}_2 
    \exp\left(-\frac{\overline{g}_1^2+\overline{g}_2^2}{2}\right)
    \Pi_{\beta\gamma}^* U_{\beta i}U_{\gamma j}\overline{g}_i\overline{g}_j,\\
    %%%
    \mathcal{P}_{3,1}^{(1)}(\overline{g}_3;\hat{\bm{k}})
    &:=
    \frac{1}{4\pi}
    \int_{-\infty}^\infty d\overline{g}_1 
    \int_{-\infty}^\infty d\overline{g}_2 
    \exp\left(-\frac{\overline{g}_1^2+\overline{g}_2^2}{2}\right)
    \overline{g}_1
    \Pi_{\beta\gamma}^* U_{\beta i}U_{\gamma j}\overline{g}_i\overline{g}_j,\\
    %%%
    \mathcal{P}_{3,2}^{(1)}(\overline{g}_3;\hat{\bm{k}})
    &:=
    \frac{1}{4\pi}
    \int_{-\infty}^\infty d\overline{g}_1 
    \int_{-\infty}^\infty d\overline{g}_2 
    \exp\left(-\frac{\overline{g}_1^2+\overline{g}_2^2}{2}\right)
    \overline{g}_2
    \Pi_{\beta\gamma}^* U_{\beta i}U_{\gamma j}\overline{g}_i\overline{g}_j,\\
    %%%
    %%%
    %%%
    \mathcal{P}_3^{(2)}(\overline{g}_3;\hat{\bm{k}})
    &:=
    -\frac{1}{8\pi}
    \int_{-\infty}^\infty d\overline{g}_1 
    \int_{-\infty}^\infty d\overline{g}_2 
    \exp\left(-\frac{\overline{g}_1^2+\overline{g}_2^2}{2}\right)
    \Pi_{\beta\delta}^*\Pi_{\gamma\delta}^* U_{\beta i}U_{\gamma j}\overline{g}_i\overline{g}_j,\\
    %%%
    \mathcal{P}_{3,1}^{(2)}(\overline{g}_3;\hat{\bm{k}})
    &:=
    -\frac{1}{8\pi}
    \int_{-\infty}^\infty d\overline{g}_1 
    \int_{-\infty}^\infty d\overline{g}_2 
    \exp\left(-\frac{\overline{g}_1^2+\overline{g}_2^2}{2}\right)
    \overline{g}_1
    \Pi_{\beta\delta}^*\Pi_{\gamma\delta}^* U_{\beta i}U_{\gamma j}\overline{g}_i\overline{g}_j,\\
    %%%
    \mathcal{P}_{3,2}^{(2)}(\overline{g}_3;\hat{\bm{k}})
    &:=
    -\frac{1}{8\pi}
    \int_{-\infty}^\infty d\overline{g}_1 
    \int_{-\infty}^\infty d\overline{g}_2 
    \exp\left(-\frac{\overline{g}_1^2+\overline{g}_2^2}{2}\right)
    \overline{g}_2
    \Pi_{\beta\delta}^*\Pi_{\gamma\delta}^* U_{\beta i}U_{\gamma j}\overline{g}_i\overline{g}_j,\\
    %%%
    %%%
    %%%
    \mathcal{P}_3^{(3)}(\overline{g}_3;\hat{\bm{k}})
    &:=
    \frac{1}{32\pi}
    \int_{-\infty}^\infty d\overline{g}_1 
    \int_{-\infty}^\infty d\overline{g}_2 
    \exp\left(-\frac{\overline{g}_1^2+\overline{g}_2^2}{2}\right)
    \Pi_{\beta\gamma}^* \Pi_{\delta\mu}^* 
    U_{\beta i}U_{\gamma j}U_{\delta k}U_{\mu l}
    \overline{g}_i\overline{g}_j\overline{g}_k\overline{g}_l,\\
    %%%
    \mathcal{P}_{3,1}^{(3)}(\overline{g}_3;\hat{\bm{k}})
    &:=
    \frac{1}{32\pi}
    \int_{-\infty}^\infty d\overline{g}_1 
    \int_{-\infty}^\infty d\overline{g}_2 
    \exp\left(-\frac{\overline{g}_1^2+\overline{g}_2^2}{2}\right)
    \overline{g}_1
    \Pi_{\beta\gamma}^* \Pi_{\delta\mu}^* 
    U_{\beta i}U_{\gamma j}U_{\delta k}U_{\mu l}
    \overline{g}_i\overline{g}_j\overline{g}_k\overline{g}_l,\\
    %%%
    \mathcal{P}_{3,2}^{(3)}(\overline{g}_3;\hat{\bm{k}})
    &:=
    \frac{1}{32\pi}
    \int_{-\infty}^\infty d\overline{g}_1 
    \int_{-\infty}^\infty d\overline{g}_2 
    \exp\left(-\frac{\overline{g}_1^2+\overline{g}_2^2}{2}\right)
    \overline{g}_2
    \Pi_{\beta\gamma}^* \Pi_{\delta\mu}^* 
    U_{\beta i}U_{\gamma j}U_{\delta k}U_{\mu l}
    \overline{g}_i\overline{g}_j\overline{g}_k\overline{g}_l.
\end{align}
\end{subequations}

The procedure to evaluate $\mathcal{P}_{3,1}(\overline{g}_3;\hat{\bm{k}})$, $\mathcal{P}_{3,2}(\overline{g}_3;\hat{\bm{k}})$, and $\mathcal{P}_3(\overline{g}_3;\hat{\bm{k}})$ follows that in Ref.~\cite{Takada20}.
Now, let us focus on $\mathcal{P}_{3,1}(\overline{g}_3;\hat{\bm{k}})$.
We can evaluate $\mathcal{P}_{3,1}^{(1)}(\overline{g}_3;\hat{\bm{k}})$ and $\mathcal{P}_{3,1}^{(2)}(\overline{g}_3;\hat{\bm{k}})$ as
\begin{align}
    \mathcal{P}_{3,1}^{(1)}(\overline{g}_3;\hat{\bm{k}})
    &=
    \frac{1}{4\pi}
    \int_{-\infty}^\infty d\overline{g}_1 
    \int_{-\infty}^\infty d\overline{g}_2 
    \exp\left(-\frac{\overline{g}_1^2+\overline{g}_2^2}{2}\right)
    \Pi_{\beta\gamma}^* 2U_{\beta 1}U_{\gamma 3}\overline{g}_1^2\overline{g}_3\nonumber\\
    %%%
    &= \overline{g}_3
    \Pi_{\gamma\delta}^* U_{\gamma 1}U_{\delta 3}\nonumber\\
    %%%
    &=
    \frac{\overline{g}_3}{\sqrt{\hat{k}_x^2 + \hat{k}_y^2}}
    \left[\hat{k}_x \hat{k}_y \left(\Pi_{xx}^*-\Pi_{yy}^*\right)
    - \left(\hat{k}_x^2 - \hat{k}_y^2\right)\Pi_{xy}^*\right],
    \label{eq:P_31_1}\\
    %%%
    %%%
    %%%
    \mathcal{P}_{3,1}^{(2)}(\overline{g}_3;\hat{\bm{k}})
    &=
    -\frac{1}{8\pi}
    \int_{-\infty}^\infty d\overline{g}_1 
    \int_{-\infty}^\infty d\overline{g}_2 
    \exp\left(-\frac{\overline{g}_1^2+\overline{g}_2^2}{2}\right)
    \Pi_{\beta\delta}^*\Pi_{\gamma\delta}^* 
    2U_{\beta 1}U_{\gamma 3}\overline{g}_1^2 \overline{g}_3\nonumber\\
    %%%
    &= -\frac{1}{2}
    \overline{g}_3
    \Pi_{\gamma\mu}^*\Pi_{\delta\mu}^* U_{\gamma 1}U_{\delta 3}\nonumber\\
    %%%
    &=  -\frac{1}{2}
    \frac{\overline{g}_3}{\sqrt{\hat{k}_x^2 + \hat{k}_y^2}}
    \left[\hat{k}_x \hat{k}_y \left(\Pi_{xx}^*-\Pi_{yy}^*\right)
    - \left(\hat{k}_x^2 - \hat{k}_y^2\right)\Pi_{xy}^*\right]
    \left(\Pi_{xx}^* + \Pi_{yy}^*\right).
    \label{eq:P_31_2}
\end{align}
Here, $\mathcal{P}_{3,1}^{(3)}(\overline{g}_3;\hat{\bm{k}})$ can be decomposed into multiple terms and computed as follows:
\begin{equation}
    \mathcal{P}_{3,1}^{(3)}(\overline{g}_3;\hat{\bm{k}})
    =
     \mathcal{P}_{3,1}^{(3,1)}(\overline{g}_3;\hat{\bm{k}})
    +\mathcal{P}_{3,1}^{(3,2)}(\overline{g}_3;\hat{\bm{k}})
    +\mathcal{P}_{3,1}^{(3,3)}(\overline{g}_3;\hat{\bm{k}})
    +\mathcal{P}_{3,1}^{(3,4)}(\overline{g}_3;\hat{\bm{k}}),
\end{equation}
where we have introduced:
\begin{align}
    \mathcal{P}_{3,1}^{(3,1)}(\overline{g}_3;\hat{\bm{k}})
    &:=
    \frac{1}{32\pi}
    \int_{-\infty}^\infty d\overline{g}_1 
    \int_{-\infty}^\infty d\overline{g}_2 
    \exp\left(-\frac{\overline{g}_1^2+\overline{g}_2^2}{2}\right)
    \Pi_{\beta\gamma}^* \Pi_{\delta\mu}^* 
    4U_{\beta 1}U_{\gamma 1}U_{\delta 1}U_{\mu 3}
    \overline{g}_1^4 \overline{g}_3\nonumber\\
    %%%
    &=
    \frac{3}{4}
    \overline{g}_3 \Pi_{\beta\gamma}^* \Pi_{\delta\mu}^* 
    U_{\beta 1}U_{\gamma 1}U_{\delta 1}U_{\mu 3}
    \nonumber\\
    %%%
    &=
    \frac{3}{4}
    \frac{\overline{g}_3}{\left(\hat{k}_x^2 + \hat{k}_y^2\right)^{3/2}}
    \left[\hat{k}_x \hat{k}_y \left(\Pi_{xx}^*-\Pi_{yy}^*\right)
    - \left(\hat{k}_x^2 - \hat{k}_y^2\right)\Pi_{xy}^*\right]
    \left[-\hat{k}_\beta \hat{k}_\gamma \Pi_{\beta\gamma}^*
    + \left(1-2\hat{k}_z^2\right)\left(\Pi_{xx}^* + \Pi_{yy}^*\right)\right],
    \label{eq:P_31_31}\\
    %%%
    %%%
    %%%
    \mathcal{P}_{3,1}^{(3,2)}(\overline{g}_3;\hat{\bm{k}})
    &:=
    \frac{1}{32\pi}
    \int_{-\infty}^\infty d\overline{g}_1 
    \int_{-\infty}^\infty d\overline{g}_2 
    \exp\left(-\frac{\overline{g}_1^2+\overline{g}_2^2}{2}\right)
    \Pi_{\beta\gamma}^* \Pi_{\delta\mu}^* 
    4U_{\beta 1}U_{\gamma 3}U_{\delta 3}U_{\mu 3}
    \overline{g}_1^2 \overline{g}_3^3
    \nonumber\\
    %%%
    &=
    \frac{1}{4}
    \overline{g}_3^3 \Pi_{\beta\gamma}^* \Pi_{\delta\mu}^* 
    U_{\beta 1}U_{\gamma 3}U_{\delta 3}U_{\mu 3}
    \nonumber\\
    %%%
    &=
    \frac{1}{4}
    \frac{\overline{g}_3^3}{\sqrt{\hat{k}_x^2 + \hat{k}_y^2}}
    \left[\hat{k}_x \hat{k}_y \left(\Pi_{xx}^*-\Pi_{yy}^*\right)
    - \left(\hat{k}_x^2 - \hat{k}_y^2\right)\Pi_{xy}^*\right]
    \hat{k}_\beta \hat{k}_\gamma \Pi_{\beta\gamma}^*,
    \label{eq:P_31_32}\\
    %%%
    %%%
    %%%
    \mathcal{P}_{3,1}^{(3,3)}(\overline{g}_3;\hat{\bm{k}})
    &:=
    \frac{1}{32\pi}
    \int_{-\infty}^\infty d\overline{g}_1 
    \int_{-\infty}^\infty d\overline{g}_2 
    \exp\left(-\frac{\overline{g}_1^2+\overline{g}_2^2}{2}\right)
    \Pi_{\beta\gamma}^* \Pi_{\delta\mu}^* 
    8U_{\beta 1}U_{\gamma 2}U_{\delta 2}U_{\mu 3}
    \overline{g}_1^2 \overline{g}_2^2 \overline{g}_3
    \nonumber\\
    %%%
    &=
    \frac{1}{2}
    \overline{g}_3 \Pi_{\beta\gamma}^* \Pi_{\delta\mu}^* 
    U_{\beta 1}U_{\gamma 2}U_{\delta 2}U_{\mu 3}
    \nonumber\\
    %%%
    &=
    \frac{1}{2}
    \frac{\overline{g}_3 \hat{k}_z^2}{\left(\hat{k}_x^2 + \hat{k}_y^2\right)^{3/2}}
    \left[\hat{k}_x \hat{k}_y \left(\Pi_{xx}^*-\Pi_{yy}^*\right)
    - \left(\hat{k}_x^2 - \hat{k}_y^2\right)\Pi_{xy}^*\right]
    \left(\hat{k}_\beta \hat{k}_\gamma \Pi_{\beta\gamma}^*
    + \Pi_{xx}^* + \Pi_{yy}^*\right),
    \label{eq:P_31_33}\\
    %%%
    %%%
    %%%
    \mathcal{P}_{3,1}^{(3,4)}(\overline{g}_3;\hat{\bm{k}})
    &:=
    \frac{1}{32\pi}
    \int_{-\infty}^\infty d\overline{g}_1 
    \int_{-\infty}^\infty d\overline{g}_2 
    \exp\left(-\frac{\overline{g}_1^2+\overline{g}_2^2}{2}\right)
    \Pi_{\beta\gamma}^* \Pi_{\delta\mu}^*
    4U_{\beta 1}U_{\gamma 3}U_{\delta 2}U_{\mu 2}
    \overline{g}_1^2 \overline{g}_2^2 \overline{g}_3
    \nonumber\\
    %%%
    &=
    \frac{1}{4}
    \overline{g}_3 \Pi_{\beta\gamma}^* \Pi_{\delta\mu}^* 
    U_{\beta 1}U_{\gamma 3}U_{\delta 2}U_{\mu 2}
    \nonumber\\
    %%%
    &=
    \frac{1}{4}
    \frac{\overline{g}_3}{\left(\hat{k}_x^2 + \hat{k}_y^2\right)^{3/2}}
    \left[\hat{k}_x \hat{k}_y \left(\Pi_{xx}^*-\Pi_{yy}^*\right)
    - \left(\hat{k}_x^2 - \hat{k}_y^2\right)\Pi_{xy}^*\right]
    \left[
    \hat{k}_z^2
    \hat{k}_\beta \hat{k}_\gamma \Pi_{\beta\gamma}^*
    - \left(1 - 2\hat{k}_z^2\right)
    \left(\Pi_{xx}^* + \Pi_{yy}^*\right)
    \right].
    \label{eq:P_31_34}
\end{align}
Summing up Eqs.~\eqref{eq:P_31_1}, \eqref{eq:P_31_2}, \eqref{eq:P_31_31}--\eqref{eq:P_31_34}, we obtain 
\begin{equation}
    \mathcal{P}_{3,1}(\overline{g}_3;\hat{\bm{k}})
    =
    \frac{\overline{g}_3}{\sqrt{\hat{k}_x^2 + \hat{k}_y^2}}
    \left[\hat{k}_x \hat{k}_y \left(\Pi_{xx}^*-\Pi_{yy}^*\right)
    - \left(\hat{k}_x^2 - \hat{k}_y^2\right)\Pi_{xy}^*\right]
    \left(1+\frac{\overline{g}_3^2 - 3}{4}
    \hat{k}_\beta \hat{k}_\gamma \Pi_{\beta\gamma}^*
    \right).
    \label{eq:P3_1_}
\end{equation}

Similarly, we can evaluate $\mathcal{P}_{3,2}(\overline{g}_3;\hat{\bm{k}})$ and $\mathcal{P}_3(\overline{g}_3;\hat{\bm{k}})$, respectively, as
\begin{align}
    \mathcal{P}_{3,2}(\overline{g}_3;\hat{\bm{k}})
    &= \frac{\overline{g}_3 \hat{k}_z}{\sqrt{\hat{k}_x^2 + \hat{k}_y^2}}
    \left(\hat{k}_\beta \hat{k}_\gamma \Pi_{\beta\gamma}^*
    + \Pi_{xx}^* + \Pi_{yy}^*\right)
    \left(
    1+\frac{\overline{g}_3^2 -3}{4}
    \hat{k}_\beta \hat{k}_\gamma \Pi_{\beta\gamma}^*
    \right),
    \label{eq:P3_2_}\\
    %%%
    %%%
    %%%
    \mathcal{P}_3(\overline{g}_3;\hat{\bm{k}})
    &=
    1
    + \frac{\overline{g}_3^2-1}{2}
    \hat{k}_\beta \hat{k}_\gamma \Pi_{\beta\gamma}^*
    +
    \frac{\overline{g}_3^4 - 6\overline{g}_3^2 + 3}{16}
    \left(\hat{k}_\beta \hat{k}_\gamma \Pi_{\beta\gamma}^*\right)^2.
    \label{eq:P3_}
\end{align}

Substituting Eqs.~\eqref{eq:P3_1_}, \eqref{eq:P3_2_}, and \eqref{eq:P3_} into Eq.~\eqref{eq:nondim_I_Ia_ver3} with \eqref{eq:relation_tI_tbI}, and integrating them over $\overline{g}_3$ from $-\infty$ to $\infty$, we obtain the expressions of $\widetilde{I}^{(2)}\left(\hat{\bm{k}}\right)$, $\widetilde{I}^{(3)}\left(\hat{\bm{k}}\right)$, $\widetilde{\overline{I}}_1^{(2)}\left(\hat{\bm{k}}\right)$, and $\widetilde{\overline{I}}_2^{(2)}\left(\hat{\bm{k}}\right)$, respectively, as
\begin{align}
    \widetilde{I}^{(2)}\left(\hat{\bm{k}}\right)
    &= -\frac{1}{2\sqrt{\pi\theta}}
    \hat{k}_x\hat{k}_y 
    \exp \left(-\frac{1}{4\theta}\hat{k}_x^2\hat{k}_y^2\right)
    +\left(\frac{1}{2}+\frac{1}{4\theta}\hat{k}_x^2\hat{k}_y^2\right)
    \erfc \left(\frac{1}{\sqrt{4\theta}}\hat{k}_x\hat{k}_y\right)
    +\frac{1}{2}
    \erfc \left(\frac{1}{\sqrt{4\theta}}\hat{k}_x\hat{k}_y\right)
    \hat{k}_\beta \hat{k}_\gamma \Pi_{\beta\gamma}^*
    \nonumber\\
    %%%
    &\hspace{1em}
    + \frac{1}{16\sqrt{\pi\theta}}
    \hat{k}_x\hat{k}_y
    \exp \left(-\frac{1}{4\theta}\hat{k}_x^2\hat{k}_y^2\right)
    \left(\hat{k}_\beta \hat{k}_\gamma \Pi_{\beta\gamma}^*\right)^2,
    \\
    %%%
    %%%
    %%%
    \widetilde{I}^{(3)}\left(\hat{\bm{k}}\right)
    &= \sqrt{\frac{2}{\pi}}\left(1+ \frac{1}{4\theta}\hat{k}_x^2\hat{k}_y^2\right)
    \exp \left(-\frac{1}{4\theta}\hat{k}_x^2\hat{k}_y^2\right)
    -\frac{1}{2\sqrt{2\theta}}\hat{k}_x\hat{k}_y 
    \left(3 + \frac{1}{2\theta}\hat{k}_x^2\hat{k}_y^2\right) 
    \erfc \left(\frac{1}{\sqrt{4\theta}}\hat{k}_x\hat{k}_y\right)
    \nonumber\\
    %%%
    &\hspace{1em}
    + 3
    \left[
    \frac{1}{\sqrt{2\pi}}
    \exp \left(-\frac{1}{4\theta}\hat{k}_x^2\hat{k}_y^2\right)
    - \frac{1}{2\sqrt{2\theta}}
    \hat{k}_x\hat{k}_y\erfc \left(\frac{1}{\sqrt{4\theta}}\hat{k}_x\hat{k}_y\right)
    \right]
    \hat{k}_\beta \hat{k}_\gamma \Pi_{\beta\gamma}^*
    + \frac{3}{8\sqrt{2\pi}}
    \exp \left(-\frac{1}{4\theta}\hat{k}_x^2\hat{k}_y^2\right)
    \left(\hat{k}_\beta \hat{k}_\gamma \Pi_{\beta\gamma}^*\right)^2,\\
    %%%
    %%%
    %%%
    \widetilde{\overline{I}}_{1}^{(2)}\left(\hat{\bm{k}}\right)
    &=
    \frac{1}{\sqrt{\hat{k}_x^2 + \hat{k}_y^2}}
    \left[\hat{k}_x \hat{k}_y \left(\Pi_{xx}^*-\Pi_{yy}^*\right)
    - \left(\hat{k}_x^2 - \hat{k}_y^2\right)\Pi_{xy}^*\right]\nonumber\\
    %%%
    &\hspace{1em} \times
    \left[
    \sqrt{\frac{2}{\pi}} 
    \left(
    1 
    + \frac{1}{4}
    \hat{k}_\beta \hat{k}_\gamma \Pi_{\beta\gamma}^*
    \right)
    \exp \left(-\frac{1}{4\theta}\hat{k}_x^2\hat{k}_y^2\right)
    - \frac{1}{\sqrt{2\theta}}
    \hat{k}_x\hat{k}_y \erfc \left(\frac{1}{\sqrt{4\theta}}\hat{k}_x\hat{k}_y\right)
    \right],
    \label{eq:tilde_bar_I1}\\
    %%%
    %%%
    %%%
    \widetilde{\overline{I}}_{2}^{(2)}\left(\hat{\bm{k}}\right)
    &=
    \frac{1}{\sqrt{\hat{k}_x^2 + \hat{k}_y^2}}
    \hat{k}_z
    \left(\hat{k}_\beta \hat{k}_\gamma \Pi_{\beta\gamma}^*
    + \Pi_{xx}^* + \Pi_{yy}^*\right)
    \nonumber\\
    %%%
    &\hspace{1em} \times
    \left[
    \sqrt{\frac{2}{\pi}} 
    \left(
    1 + \frac{1}{4}
    \hat{k}_\beta \hat{k}_\gamma \Pi_{\beta\gamma}^*
    \right)
    \exp \left(-\frac{1}{4\theta}\hat{k}_x^2\hat{k}_y^2\right)
    - \frac{1}{\sqrt{2\theta}}
    \hat{k}_x\hat{k}_y \erfc \left(\frac{1}{\sqrt{4\theta}}\hat{k}_x\hat{k}_y\right)
    \right],
    \label{eq:tilde_bar_I2}
\end{align}
where $\mathrm{erfc}(z)$ is the complementary error function~\cite{Abramowitz} defined by
\begin{align}
    \mathrm{erfc} (z)
    &:= \frac{2}{\sqrt{\pi}}\int_z^\infty 
    \mathrm{e}^{-t^2} \mathrm{d}t 
    = 1-\mathrm{erf}(z).
    \label{eq:error_func}
\end{align}

Now, we turn to evaluate $\Lambda_{\alpha\beta}$.
Let us introduce the dimensionless variables $\widetilde{\Lambda}_{\alpha\beta}$ and $\widetilde{\overline{\Lambda}}_{\alpha\beta}$, respectively, as
\begin{equation}
    \widetilde{\Lambda}_{\alpha\beta}\\
    := \frac{1}
    {\displaystyle \frac{1+e}{4}md^2n^2g_0\vt^3}
    \Lambda_{\alpha\beta},\quad
    %%%
    \widetilde{\overline{\Lambda}}_{\alpha\beta}
    := \frac{1}
    {\displaystyle \frac{1+e}{4}md^2n^2g_0\vt^3}
    \overline{\Lambda}_{\alpha\beta}.
\end{equation}
From the definition of $\widetilde{I}^{(\ell)}\left(\hat{\bm{k}}\right)$ and $\widetilde{I}_\alpha^{(2)}\left(\hat{\bm{k}}\right)$ in Eqs.~\eqref{eq:nondim_I_Ia}, $\widetilde{\Lambda}_{\alpha\beta}$ and $\widetilde{\overline{\Lambda}}_{\alpha\beta}$ are, respectively, written as
\begin{subequations}
\begin{align}
    \widetilde{\Lambda}_{\alpha\beta}
    &= \widetilde{L}_{\alpha\beta}
    + (1-e)\widetilde{M}_{\alpha\beta},\\
    %%%
    \widetilde{\overline{\Lambda}}_{\alpha\beta}
    &= \int d\hat{\bm{k}}
    \left[\widetilde{I}_\alpha^{(2)}
    \left(\hat{\bm{k}}\right)\hat{k}_\beta
    +\widetilde{I}_\beta^{(2)}\left(\hat{\bm{k}}\right)\hat{k}_\alpha
    - (1+e) \widetilde{I}^{(3)}\left(\hat{\bm{k}}\right)
    \hat{k}_\alpha\hat{k}_\beta\right],
\end{align}
\end{subequations}
where we have introduced
\begin{subequations}\label{eq:def_L_M}
\begin{align}
    \widetilde{L}_{\alpha\beta}
    &:= \int d\hat{\bm{k}}
    \left[\hat{k}_\alpha \widetilde{J}_\beta\left(\hat{\bm{k}}\right)
    + \hat{k}_\beta \widetilde{J}_\alpha\left(\hat{\bm{k}}\right)
    + \sqrt{\frac{2}{\theta}}\hat{k}_x\hat{k}_y\hat{k}_\alpha \hat{k}_\beta
    \widetilde{I}^{(2)}\left(\hat{\bm{k}}\right)\right],
    \label{eq:def_L}\\
    %%%
    %%%
    %%%
    \widetilde{M}_{\alpha\beta}
    &:= \int d\hat{\bm{k}}
    \hat{k}_\alpha \hat{k}_\beta
    \widetilde{I}^{(3)}\left(\hat{\bm{k}}\right)
    \label{eq:def_M}
\end{align}
\end{subequations}
with
\begin{align}
    \widetilde{J}_\alpha\left(\hat{\bm{k}}\right)
    &:= 
    U_{\alpha 1}\widetilde{\overline{I}}_1^{(2)}\left(\hat{\bm{k}}\right)
    +U_{\alpha 2}\widetilde{\overline{I}}_2^{(2)}\left(\hat{\bm{k}}\right).
\end{align}
Here, from Eqs.~\eqref{eq:tilde_bar_I1} and \eqref{eq:tilde_bar_I2}, $J_\alpha\left(\hat{\bm{k}}\right)$ ($\alpha=x,y,z$) is given by
\begin{subequations}
\begin{align}
    \widetilde{J}_x\left(\hat{\bm{k}}\right)
    &= -\hat{k}_x \left(\hat{k}_\beta \hat{k}_\gamma \Pi_{\beta\gamma}^* - \Pi_{xx}^*-\dfrac{\hat{k}_y}{\hat{k}_x}\Pi_{xy}^*\right)
    \left[
    \sqrt{\frac{2}{\pi}} 
    \left(
    1 + \frac{1}{4}\hat{k}_\beta \hat{k}_\gamma \Pi_{\beta\gamma}^*
    \right)
    \exp \left(-\frac{1}{4\theta}\hat{k}_x^2\hat{k}_y^2\right)
    - \frac{1}{\sqrt{2\theta}}
    \hat{k}_x\hat{k}_y \erfc \left(\frac{1}{\sqrt{4\theta}}\hat{k}_x\hat{k}_y\right)
    \right],\\
    %%%
    %%%
    %%%
    \widetilde{J}_y\left(\hat{\bm{k}}\right)
    &= -\hat{k}_y \left(\hat{k}_\beta \hat{k}_\gamma \Pi_{\beta\gamma}^* - \Pi_{yy}^*-\dfrac{\hat{k}_x}{\hat{k}_y}\Pi_{xy}^*\right)
    \left[
    \sqrt{\frac{2}{\pi}} 
    \left(
    1 + \frac{1}{4}
    \hat{k}_\beta \hat{k}_\gamma \Pi_{\beta\gamma}^*
    \right)
    \exp \left(-\frac{1}{4\theta}\hat{k}_x^2\hat{k}_y^2\right)
    - \frac{1}{\sqrt{2\theta}}
    \hat{k}_x\hat{k}_y \erfc \left(\frac{1}{\sqrt{4\theta}}\hat{k}_x\hat{k}_y\right)
    \right],\\
    %%%
    %%%
    %%%
    \widetilde{J}_z\left(\hat{\bm{k}}\right)
    &= -\hat{k}_z \left(\hat{k}_\beta \hat{k}_\gamma \Pi_{\beta\gamma}^*-\Pi_{zz}^*\right)
    \left[
    \sqrt{\frac{2}{\pi}} 
    \left(
    1 + \frac{1}{4}
    \hat{k}_\beta \hat{k}_\gamma \Pi_{\beta\gamma}^*
    \right)
    \exp \left(-\frac{1}{4\theta}\hat{k}_x^2\hat{k}_y^2\right)
    - \frac{1}{\sqrt{2\theta}}
    \hat{k}_x\hat{k}_y \erfc \left(\frac{1}{\sqrt{4\theta}}\hat{k}_x\hat{k}_y\right)
    \right].
\end{align}
\end{subequations}

In the following, we evaluate $\widetilde{L}_{\alpha\beta}$ and $\widetilde{M}_{\alpha\beta}$ in Eq.~\eqref{eq:def_L_M}.
First, let us consider $\widetilde{L}_{\alpha\alpha}$.
It should be noted that $\widetilde{L}_{\alpha\alpha}=\sqrt{2/\theta}\int d\hat{\bm{k}}\hat{k}_x\hat{k}_y\widetilde{I}^{(2)}$ holds because of the relation $\hat{k}_\alpha \widetilde{J}_\alpha=0$.
Therefore, the trace $\widetilde{L}_{\alpha\alpha}(=\widetilde{L}_{xx}+\widetilde{L}_{yy}+\widetilde{L}_{zz})$ becomes
\begin{align}
    \widetilde{L}_{\alpha\alpha}
    &=
    \int d\hat{\bm{k}}\left[
    -\frac{1}{\sqrt{2\pi}\theta}
    \hat{k}_x^2\hat{k}_y^2 \exp \left(-\frac{1}{4\theta}\hat{k}_x^2\hat{k}_y^2\right)
    +\frac{1}{\sqrt{2\theta}}
    \left(1+\frac{1}{2\theta}\hat{k}_x^2\hat{k}_y^2\right) 
    \hat{k}_x\hat{k}_y
    \erfc \left(\frac{1}{\sqrt{4\theta}}\hat{k}_x\hat{k}_y\right)\right]\nonumber\\
    %%%
    &\hspace{1em}
    + \int d\hat{\bm{k}}
    \frac{1}{\sqrt{2\theta}}
    \hat{k}_x\hat{k}_y\erfc \left(\frac{1}{\sqrt{4\theta}}\hat{k}_x\hat{k}_y\right)
    \hat{k}_\beta \hat{k}_\gamma \Pi_{\beta\gamma}^*
    %%%
    + \int d\hat{\bm{k}}
    \frac{1}{8\sqrt{2\pi}\theta}
    \hat{k}_x^2\hat{k}_y^2
    \exp \left(-\frac{1}{4\theta}\hat{k}_x^2\hat{k}_y^2\right)
    \left(\hat{k}_\beta \hat{k}_\gamma \Pi_{\beta\gamma}^*\right)^2\nonumber\\
    %%%
    %%%
    %%%
    &= -\frac{1}{\sqrt{2\pi}\theta}
    \mathcal{A}_{2,2,0}(\theta)
    +\frac{1}{\sqrt{2\theta}}
    \mathcal{B}_{1,1,0}(\theta)
    +\frac{1}{(2\theta)^{3/2}}
    \mathcal{B}_{3,3,0}(\theta)
    + \frac{1}{\sqrt{2\theta}}
    \mathcal{Q}_{1,1,0}\left(\theta, \overleftrightarrow{\Pi^*}\right)
    + \frac{1}{8\sqrt{2\pi}\theta}
    \mathcal{R}_{2,2,0}\left(\theta, \overleftrightarrow{\Pi^*}\right),
    \label{eq:Laa}
\end{align}
where we have introduced $\mathcal{A}_{s,t,u}(\theta)$, $\mathcal{B}_{s,t,u}(\theta)$ for nonnegative integers $s$, $t$, $u$ as
\begin{subequations}\label{eq:def_A_B_Q_R}
\begin{align}
    \mathcal{A}_{s,t,u}(\theta)
    &:= \int d\hat{\bm{k}}
    \hat{k}_x^s
    \hat{k}_y^t
    \hat{k}_z^u
    \exp\left(-\frac{1}{4\theta}\hat{k}_x^2\hat{k}_y^2\right),
    \label{eq:def_A}\\
    %%%
    \mathcal{B}_{s,t,u}(\theta)
    &:= \int d\hat{\bm{k}}
    \hat{k}_x^s
    \hat{k}_y^t
    \hat{k}_z^u
    \erfc\left(\frac{1}{\sqrt{4\theta}}\hat{k}_x\hat{k}_y\right),
    \label{eq:def_B}\\
    %%%
    \mathcal{Q}_{s,t,u}\left(\theta, \overleftrightarrow{\Pi^*}\right)
    &:= \int d\hat{\bm{k}}
    \hat{k}_x^s
    \hat{k}_y^t
    \hat{k}_z^u
    \erfc\left(\frac{1}{\sqrt{4\theta}}\hat{k}_x\hat{k}_y\right)
    \hat{k}_\beta \hat{k}_\gamma \Pi_{\beta\gamma}^*,\\
    %%%
    \mathcal{R}_{s,t,u}\left(\theta, \overleftrightarrow{\Pi^*}\right)
    &:= \int d\hat{\bm{k}}
    \hat{k}_x^s
    \hat{k}_y^t
    \hat{k}_z^u
    \exp\left(-\frac{1}{4\theta}\hat{k}_x^2\hat{k}_y^2\right)
    \left(\hat{k}_\beta \hat{k}_\gamma \Pi_{\beta\gamma}^*\right)^2.
\end{align}
\end{subequations}
Similarly, $\widetilde{M}_{\alpha\alpha}$ is rewritten as
\begin{align}
    \widetilde{M}_{\alpha\alpha}
    &= \int d\hat{\bm{k}}
    \left[\sqrt{\frac{2}{\pi}}\left(1+ \frac{1}{4\theta}\hat{k}_x^2\hat{k}_y^2\right)
    \exp \left(-\frac{1}{4\theta}\hat{k}_x^2\hat{k}_y^2\right)
    -\frac{1}{2\sqrt{2\theta}}\hat{k}_x\hat{k}_y 
    \left(3+ \frac{1}{2\theta}\hat{k}_x^2\hat{k}_y^2\right) 
    \erfc \left(\frac{1}{\sqrt{4\theta}}\hat{k}_x\hat{k}_y\right)\right]
    \nonumber\\
    %%%
    &\hspace{1em}
    + 3\int d\hat{\bm{k}}
    \left[
    \frac{1}{\sqrt{2\pi}}
    \exp \left(-\frac{1}{4\theta}\hat{k}_x^2\hat{k}_y^2\right)
    - \frac{1}{2\sqrt{2\theta}}
    \hat{k}_x\hat{k}_y\erfc \left(\frac{1}{\sqrt{4\theta}}\hat{k}_x\hat{k}_y\right)
    \right]
    \hat{k}_\beta \hat{k}_\gamma \Pi_{\beta\gamma}^*
    \nonumber\\
    %%%
    &\hspace{1em}
    + \int d\hat{\bm{k}}
    \frac{3}{8\sqrt{2\pi}}
    \exp \left(-\frac{1}{4\theta}\hat{k}_x^2\hat{k}_y^2\right)
    \left(\hat{k}_\beta \hat{k}_\gamma \Pi_{\beta\gamma}^*\right)^2\nonumber\\
    %%%
    %%%
    %%%
    &= \sqrt{\frac{2}{\pi}}\mathcal{A}_{0,0,0}(\theta)+ \sqrt{\frac{2}{\pi}}\frac{1}{4\theta}\mathcal{A}_{2,2,0}(\theta)
    -\frac{3}{2\sqrt{2\theta}}\mathcal{B}_{1,1,0}(\theta)
    -\frac{1}{2(2\theta)^{3/2}}\mathcal{B}_{3,3,0}(\theta)
    \nonumber\\
    %%%
    &\hspace{1em}
    + \frac{3}{\sqrt{2\pi}}\mathcal{P}_{0,0,0}\left(\theta, \overleftrightarrow{\Pi^*}\right)
    - \frac{3}{2\sqrt{2\theta}}
    \mathcal{Q}_{1,1,0}\left(\theta, \overleftrightarrow{\Pi^*}\right)
    + \frac{3}{8\sqrt{2\pi}}\mathcal{R}_{0,0,0}\left(\theta, \overleftrightarrow{\Pi^*}\right),
    \label{eq:Maa}
\end{align}
with
\begin{equation}
    \mathcal{P}_{s,t,u}\left(\theta, \overleftrightarrow{\Pi}\right)
    := \int d\hat{\bm{k}}
    \hat{k}_x^s
    \hat{k}_y^t
    \hat{k}_z^u
    \exp\left(-\frac{1}{4\theta}\hat{k}_x^2\hat{k}_y^2\right)
    \hat{k}_\beta \hat{k}_\gamma \Pi_{\beta\gamma}^*.
\end{equation}

To obtain more explicit forms of $\widetilde{L}_{\alpha\alpha}$ and $\widetilde{M}_{\alpha\alpha}$, we further introduce the following quantity:
\begin{align}
    \mathcal{K}_{s,t,u}
    &:=
    \int d\hat{\bm{k}}
    \hat{k}_x^s
    \hat{k}_y^t
    \hat{k}_z^u\nonumber\\
    %%%
    &= \frac{[1+(-1)^s][1+(-1)^t][1+(-1)^u]}{4}
    B\left(\frac{s+1}{2}, \frac{t+1}{2}, \frac{u+1}{2}\right),
    \label{eq:def_K_stu}
\end{align}
with the multivariate Beta function
\begin{equation}
    B(a, b, c)
    := \frac{\Gamma(a)\Gamma(b)\Gamma(c)}{\Gamma(a+b+c)}.
\end{equation}
It is readily seen that $\mathcal{K}_{s,t,u}=0$ whenever at least one of $s$, $t$, and $u$ is odd.
Furthermore, the symmetry relation
\begin{equation}
    \mathcal{K}_{s,t,u}
    = \mathcal{K}_{t,s,u}
    = \mathcal{K}_{u,t,s},
\end{equation}
holds.

Using $\mathcal{K}_{s,t,u}$, $\mathcal{A}_{s,t,u}(\theta)$ and $\mathcal{B}_{s,t,u}(\theta)$ are, respectively, rewritten as
\begin{subequations}
\begin{align}
    \mathcal{A}_{s,t,u}(\theta)
    &= \sum_{n=0}^\infty \frac{1}{n!}
    \left(-\frac{1}{4\theta}\right)^n
    \int d\hat{\bm{k}}
    \hat{k}_x^{2n+s}
    \hat{k}_y^{2n+t}
    \hat{k}_z^u\nonumber\\
    %%%
    &= \mathcal{K}_{s,t,u}\mathcal{F}_{s,t,u}(\theta),\\
    %%%
    %%%
    \mathcal{B}_{s,t,u}(\theta)
    &= \int d\hat{\bm{k}}
    \hat{k}_x^s
    \hat{k}_y^t
    \hat{k}_z^u
    - \frac{1}{\sqrt{\pi\theta}}
    \sum_{n=0}^\infty \frac{1}{(2n+1)n!}
    \left(-\frac{1}{4\theta}\right)^n\int d\hat{\bm{k}}
    \hat{k}_x^{2n+s+1}
    \hat{k}_y^{2n+t+1}
    \hat{k}_z^u\nonumber\\
    %%%
    &= \mathcal{K}_{s,t,u}
    - \frac{1}{\sqrt{\pi\theta}}\mathcal{K}_{s+1,t+1,u}
    \mathcal{G}_{s,t,u}(\theta),
\end{align}
\end{subequations}
where we have introduced
\begin{subequations}\label{eq:def_FG}
\begin{align}
    \mathcal{F}_{s,t,u}(\theta)
    &:= {}_2F_2\left(\frac{s+1}{2}, \frac{t+1}{2}; \frac{s+t+u+3}{4}, \frac{s+t+u+5}{4}; -\frac{1}{16\theta}\right),
    \label{eq:def_F}\\
    %%%
    \mathcal{G}_{s,t,u}(\theta)
    &:= 
    {}_3F_3\left(\frac{1}{2},\frac{s+2}{2}, \frac{t+2}{2}; \frac{3}{2}, \frac{s+t+u+5}{4}, \frac{s+t+u+7}{4};-\frac{1}{16\theta}\right).
    \label{eq:def_G}
\end{align}
\end{subequations}

\begin{figure}[hbtp]
    \centering
    \begin{tikzpicture}
        \tikzset{Terminal/.style={rounded rectangle,  draw,  text centered, text width=8cm}};
        \tikzset{Process/.style={rectangle, draw, text centered, text width=6cm}};
        \node[Terminal](a) at (4, 0){Output: $\widetilde{L}_{\alpha\beta}\left(\theta, \overleftrightarrow{\Pi^*}\right)$ and $\widetilde{M}_{\alpha\beta}\left(\theta, \overleftrightarrow{\Pi^*}\right)$};
        \node[Process](b) at (0, -2){Evaluate $\mathcal{P}_{s,t,u}\left(\theta, \overleftrightarrow{\Pi^*}\right)$, $\mathcal{Q}_{s,t,u}\left(\theta, \overleftrightarrow{\Pi^*}\right)$, and $\mathcal{R}_{s,t,u}\left(\theta, \overleftrightarrow{\Pi^*}\right)$.};
        \node[Process](c) at (8, -1.5){Evaluate $\mathcal{A}_{s,t,u}(\theta)$, $\mathcal{B}_{s,t,u}(\theta)$, $\mathcal{H}^{(1)}_{s,t,u}(\theta)$, $\mathcal{H}^{(2)}_{s,t,u}(\theta)$, and $\mathcal{H}^{(3)}_{s,t,u}(\theta)$};
        \node[Process](d) at (8, -3){Evaluate $\mathcal{F}_{s,t,u}(\theta)$ and $\mathcal{G}_{s,t,u}(\theta)$};
        \node[Terminal](e) at (4, -4){Input: $\theta$ and $\overleftrightarrow{\Pi^*}$};
        \draw[->, thick] (e)--(b);
        \draw[->, thick] (e)--(d);
        \draw[->, thick] (d)--(c);
        \draw[->, thick] (d)--(b);
        \draw[->, thick] (c)--(b);
        \draw[->, thick] (b)--(a);
        \draw[->, thick] (c)--(a);
        \draw[->, thick] (e)--(a);
    \end{tikzpicture}
    \caption{A flow chart of our calculation.}
    \label{fig:flow_chart}
\end{figure}
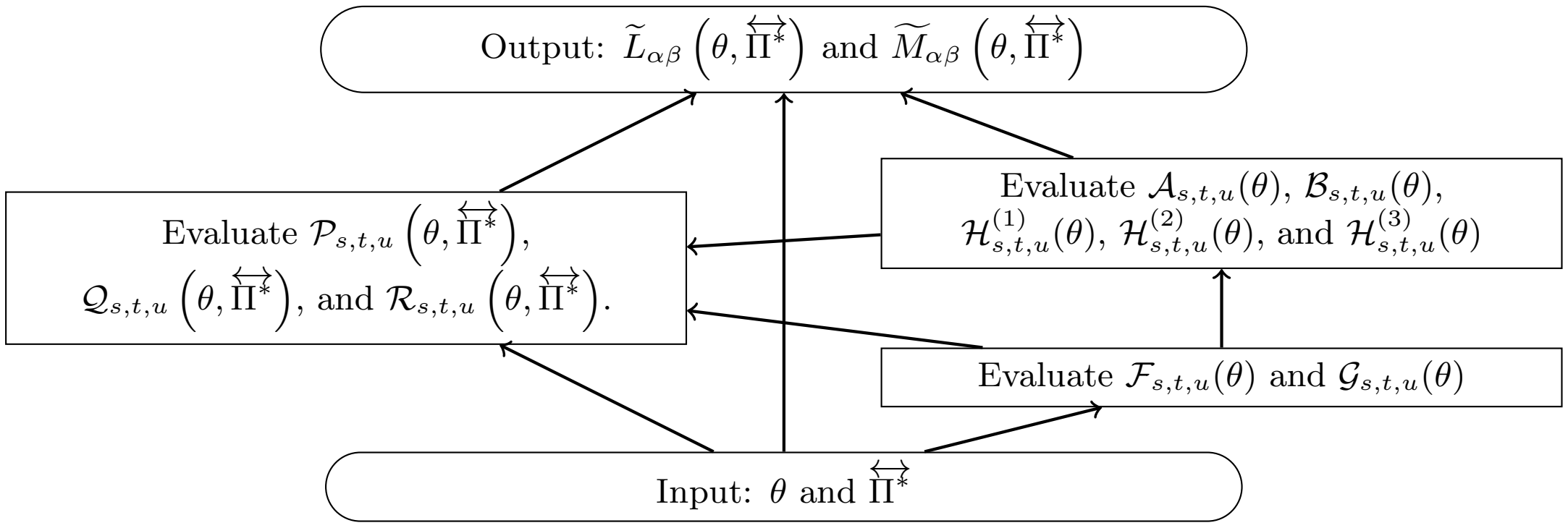

We can also rewrite $\mathcal{P}_{s,t,u}\left(\theta, \overleftrightarrow{\Pi^*}\right)$, $\mathcal{Q}_{s,t,u}\left(\theta, \overleftrightarrow{\Pi^*}\right)$, and $\mathcal{R}_{s,t,u}\left(\theta, \overleftrightarrow{\Pi^*}\right)$ as
\begin{subequations}
\begin{align}
    \mathcal{P}_{s,t,u}\left(\theta, \overleftrightarrow{\Pi^*}\right)
    &= \mathcal{K}_{s,t,u}
    \left[\mathcal{H}^{(1)}_{s,t,u}(\theta)\Pi_{xx}^*
    + \mathcal{H}^{(1)}_{t,s,u}(\theta)\Pi_{yy}^*\right]
    + 2\mathcal{K}_{s+1,t+1,u}\mathcal{F}_{s+1,t+1,u}(\theta)\Pi_{xy}^*,\\
    %%%
    %%%
    %%%
    \mathcal{Q}_{s,t,u}\left(\theta, \overleftrightarrow{\Pi^*}\right)
    &= \frac{\mathcal{K}_{s,t,u}}{3+s+t+u}
    \left[(s-u)\Pi_{xx}^*+ (t-u)\Pi_{yy}^*\right]
    - \frac{2}{\sqrt{\pi\theta}}\mathcal{K}_{s+2,t+2,u}
    \mathcal{G}_{s+1,t+1,u}(\theta)\Pi_{xy}^*\nonumber\\
    %%%
    &\hspace{1em}
    - \frac{1}{\sqrt{\pi\theta}} \mathcal{K}_{s+1,t+1,u}
    \left[\mathcal{H}^{(2)}_{s,t,u}(\theta)\Pi_{xx}^*
    + \mathcal{H}^{(2)}_{t,s,u}(\theta)\Pi_{yy}^*\right]
    + 2\mathcal{K}_{s+1,t+1,u}\Pi_{xy}^*,\\
    %%%
    %%%
    %%%
    \mathcal{R}_{s,t,u}\left(\theta, \overleftrightarrow{\Pi^*}\right)
    &= \mathcal{K}_{s,t,u}
    \left[\mathcal{H}^{(3)}_{s,t,u}(\theta)\left(\Pi_{xx}^*\right)^2
    + \mathcal{H}^{(3)}_{t,s,u}(\theta)\left(\Pi_{yy}^*\right)^2
    + 2\mathcal{H}^{(4)}_{s,t,u}(\theta)\Pi_{xx}^*\Pi_{yy}^*\right.\nonumber\\
    %%%
    &\hspace{4em}\left.
    + \frac{4(1+s)(1+t)}{(3+s+t+u)(5+s+t+u)}\mathcal{F}_{s+2,t+2,u}(\theta)\left(\Pi_{xy}^*\right)^2\right]\nonumber\\
    %%%
    &\hspace{1em}
    + 4\mathcal{K}_{s+1,t+1,u}
    \left[\mathcal{H}^{(1)}_{s+1,t+1,u}(\theta)\Pi_{xx}^*
    + \mathcal{H}^{(1)}_{t+1,s+1,u}(\theta)\Pi_{yy}^*\right]\Pi_{xy}^*,
\end{align}
\end{subequations}
where we have introduced
\begin{subequations}\label{eq:def_H}
\begin{align}
    \mathcal{H}^{(1)}_{s,t,u}(\theta)
    &:= \frac{1}{3+s+t+u}\left[(1+s)\mathcal{F}_{s+2,t,u}(\theta)
    -(1+u)\mathcal{F}_{s,t,u+2}(\theta)\right],\\
    %%%
    \mathcal{H}^{(2)}_{s,t,u}(\theta)
    &:= \frac{1}{5+s+t+u}
    \left[(2+s)\mathcal{G}_{s+2,t,u}(\theta)
    - (1+u)\mathcal{G}_{s,t,u+2}(\theta)\right],\\
    %%%
    \mathcal{H}^{(3)}_{s,t,u}(\theta)
    &:= \frac{1}{(3+s+t+u)(5+s+t+u)}\nonumber\\
    &\hspace{1em}\times
    \left[(1+s)(3+s)\mathcal{F}_{s+4,t,u}(\theta)
    -2(1+s)(1+u)\mathcal{F}_{s+2,t,u+2}(\theta)
    + (1+u)(3+u)\mathcal{F}_{s,t,u+4}(\theta)\right],\\
    %%%
    \mathcal{H}^{(4)}_{s,t,u}(\theta)
    &:= \frac{1}{(3+s+t+u)(5+s+t+u)}\nonumber\\
    &\hspace{1em}\times
    \left[(1+s)(1+t)\mathcal{F}_{s+2,t+2,u}(\theta)
    -(1+s)(1+u)\mathcal{F}_{s+2,t,u+2}(\theta)\right.\nonumber\\
    &\hspace{2em}\left.    
    -(1+t)(1+u)\mathcal{F}_{s,t+2,u+2}(\theta)
    +(1+u)(3+u)\mathcal{F}_{s,t,u+4}(\theta)\right].
\end{align}
\end{subequations}
Therefore, the final expressions of $\widetilde{L}_{\alpha\alpha}$ and $\widetilde{M}_{\alpha\alpha}$ are, respectively, given by
\begin{subequations}
\begin{align}
    \widetilde{L}_{\alpha\alpha}
    &= \frac{2\sqrt{2\pi}}{15\theta}
    \left\{
    -\mathcal{F}_{2,2,0}(\theta)
    -\mathcal{G}_{1,1,0}(\theta)
    -\frac{1}{14\theta}\mathcal{G}_{3,3,0}(\theta)
    -\mathcal{H}^{(2)}_{1,1,0}(\theta)
    \left(\Pi_{xx}^* + \Pi_{yy}^*\right)
    + 2\sqrt{\pi\theta}\Pi_{xy}^*\right.\nonumber\\
    %%%
    &\hspace{4.5em}\left.
    + \frac{1}{8}\mathcal{H}^{(3)}_{2,2,0}(\theta)
    \left[\left(\Pi_{xx}^*\right)^2+ \left(\Pi_{yy}^*\right)^2\right]
    + \frac{1}{4}\mathcal{H}^{(4)}_{2,2,0}(\theta)\Pi_{xx}^*\Pi_{yy}^*
    + \frac{1}{14}\mathcal{F}_{4,4,0}(\theta)\left(\Pi_{xy}^*\right)^2\right\},
    \label{eq:Laa2}\\
    %%%
    %%%
    \widetilde{M}_{\alpha\alpha}
    &= 4\sqrt{2\pi}\left\{\mathcal{F}_{0,0,0}(\theta)
    + \frac{1}{60\theta}\mathcal{F}_{2,2,0}(\theta)
    + \frac{1}{20\theta}
    \mathcal{G}_{1,1,0}(\theta)
    + \frac{1}{840\theta^2}
    \mathcal{G}_{3,3,0}(\theta)
    + \left[\frac{3}{2}\mathcal{H}^{(1)}_{0,0,0}(\theta)
    + \frac{1}{20\theta}\mathcal{H}^{(2)}_{1,1,0}(\theta)\right]
    \left(\Pi_{xx}^*+\Pi_{yy}^*\right)\right.
    \nonumber\\
    %%%
    &\hspace{4.5em}\left.
    - \frac{\sqrt{\pi}}{10\sqrt{\theta}}\Pi_{xy}^*
    + \frac{3}{16}\mathcal{H}^{(3)}_{0,0,0}(\theta)
    \left[\left(\Pi_{xx}^*\right)^2+\left(\Pi_{yy}^*\right)^2\right]
    + \frac{3}{8}\mathcal{H}^{(4)}_{0,0,0}(\theta)\Pi_{xx}^*\Pi_{yy}^*
    + \frac{1}{20}\mathcal{F}_{2,2,0}(\theta)\left(\Pi_{xy}^*\right)^2\right\}.
\end{align}
\end{subequations}
Here, we present the list of expressions of $\widetilde{\Lambda}_{\alpha\beta}^{(i)}$ ($i=0,1,\cdots$) and auxiliary functions in Table~\ref{table:list_of_functions}.
The procedure to evaluate $\widetilde{L}_{\alpha\beta}\left(\theta, \overleftrightarrow{\Pi^*}\right)$ and $\widetilde{M}_{\alpha\beta}\left(\theta, \overleftrightarrow{\Pi^*}\right)$ is shown as a flow chart in Fig.~\ref{fig:flow_chart}.

%%%%%%%%%%%%%%%%%%%%%%%%%%%%%%
\begin{table}[htbp]
    \centering
    \caption{List of the expressions of $\widetilde{\Lambda}_{\alpha\beta}^{(i)}$ ($i=0,1,\cdots$) and auxiliary functions.}
    \begin{tabular}{l}
        \hline
        $\widetilde{\Lambda}_{\alpha\alpha}
        = \widetilde{\Lambda}_{\alpha\alpha}^{(0)}(\theta)
        + \widetilde{\Lambda}_{\alpha\alpha}^{(1)}(\theta)\left(\Pi_{xx}^* + \Pi_{yy}^*\right)
        + \widetilde{\Lambda}_{\alpha\alpha}^{(2)}(\theta)\Pi_{xy}^*
        + \widetilde{\Lambda}_{\alpha\alpha}^{(3)}(\theta)
        \left[\left(\Pi_{xx}^*\right)^2+ \left(\Pi_{yy}^*\right)^2\right]
        + \widetilde{\Lambda}_{\alpha\alpha}^{(4)}(\theta)\Pi_{xx}^*\Pi_{yy}^*
        + \widetilde{\Lambda}_{\alpha\alpha}^{(5)}(\theta)\left(\Pi_{xy}^*\right)^2$\\
        %%%
        %%%
        $\widetilde{\Lambda}_{\alpha\alpha}^{(0)}(\theta)
        := -\dfrac{2\sqrt{2\pi}}{15\theta}
        \left[
        \mathcal{F}_{2,2,0}(\theta)
        +\mathcal{G}_{1,1,0}(\theta)
        +\dfrac{1}{14\theta}\mathcal{G}_{3,3,0}(\theta)\right]$\\
        $\hspace{4.5em}
        + (1-e)4\sqrt{2\pi}\left[\mathcal{F}_{0,0,0}(\theta)
        + \dfrac{1}{60\theta}\mathcal{F}_{2,2,0}(\theta)
        + \dfrac{1}{20\theta}
        \mathcal{G}_{1,1,0}(\theta)
        + \dfrac{1}{840\theta^2}
        \mathcal{G}_{3,3,0}(\theta)\right]$\\
        %%%
        %%%
        $\widetilde{\Lambda}_{\alpha\alpha}^{(1)}(\theta)
        := -\dfrac{2\sqrt{2\pi}}{15\theta}
        \mathcal{H}^{(2)}_{1,1,0}(\theta)
        + (1-e)4\sqrt{2\pi}
        \left[\dfrac{3}{2}\mathcal{H}^{(1)}_{0,0,0}(\theta)
        + \dfrac{1}{20\theta}\mathcal{H}^{(2)}_{1,1,0}(\theta)\right]$\\
        %%%
        %%%
        $\widetilde{\Lambda}_{\alpha\alpha}^{(2)}(\theta)
        := -\dfrac{2\sqrt{2}\pi}{15\sqrt{\theta}}(1-3e)$\\
        %%%
        %%%
        $\widetilde{\Lambda}_{\alpha\alpha}^{(3)}(\theta)
        := \dfrac{\sqrt{2\pi}}{60\theta}
        \mathcal{H}^{(3)}_{2,2,0}(\theta)
        + (1-e)\dfrac{3\sqrt{2\pi}}{4}\mathcal{H}^{(3)}_{0,0,0}(\theta)$\\
        %%%
        %%%
        $\widetilde{\Lambda}_{\alpha\alpha}^{(4)}(\theta)
        := \dfrac{\sqrt{2\pi}}{30\theta}
        \mathcal{H}^{(4)}_{2,2,0}(\theta)
        + (1-e)\dfrac{3\sqrt{2\pi}}{2}\mathcal{H}^{(4)}_{0,0,0}(\theta)$\\
        %%%
        %%%
        $\widetilde{\Lambda}_{\alpha\alpha}^{(5)}(\theta)
        := \dfrac{\sqrt{2\pi}}{105\theta}
        \mathcal{F}_{4,4,0}(\theta)
        + (1-e)\dfrac{\sqrt{2\pi}}{5}\mathcal{F}_{2,2,0}(\theta)$\\
        %%%
        \hline
        %%%
        $\widetilde{\Lambda}_{xy}
        = \widetilde{\Lambda}_{xy}^{(0)}(\theta)
        + \widetilde{\Lambda}_{xy}^{(1)}(\theta)\left(\Pi_{xx}^* + \Pi_{yy}^*\right)
        + \widetilde{\Lambda}_{xy}^{(2)}(\theta)\Pi_{xy}^*
        + \widetilde{\Lambda}_{xy}^{(3)}(\theta)
        \left(\Pi_{xx}^*+\Pi_{yy}^*\right)\Pi_{xy}^*$\\
        %%%
        %%%
        $\widetilde{\Lambda}_{xy}^{(0)}(\theta)
        := 
        -\dfrac{\sqrt{2}\pi}{15\sqrt{\theta}}
        \left[(1-3e) - \dfrac{1}{14\theta}(1+e)\right]$\\
        %%%
        %%%
        $\widetilde{\Lambda}_{xy}^{(1)}(\theta)
        := -\dfrac{2\sqrt{2}\pi}{105\sqrt{\theta}}
        (4-3e)$\\
        %%%
        %%%
        $\widetilde{\Lambda}_{xy}^{(2)}(\theta)
        := \sqrt{2\pi}
        \left[\dfrac{8}{3}\mathcal{F}_{2,0,0}(\theta)
        - \dfrac{16}{15}\mathcal{F}_{2,2,0}(\theta)
        + \dfrac{4}{35\theta}\mathcal{G}_{3,1,0}(\theta)
        - \dfrac{12}{105\theta}\mathcal{G}_{3,3,0}(\theta)\right]
        + (1-e)\sqrt{2\pi}
        \left[\dfrac{4}{5}\mathcal{F}_{2,2,0}(\theta)
        + \dfrac{2}{35\theta}\mathcal{G}_{3,3,0}(\theta)\right]$\\
        %%%
        %%%
        $\widetilde{\Lambda}_{xy}^{(3)}(\theta)
        := \sqrt{2\pi}\left[
        \dfrac{2}{15}
        \mathcal{F}_{2,2,0}(\theta)
        + \dfrac{1}{3}\mathcal{H}^{(1)}_{2,0,0}(\theta)
        + \dfrac{1}{3}\mathcal{H}^{(1)}_{0,2,0}(\theta)
        - \dfrac{8}{15}\mathcal{H}^{(1)}_{2,2,0}(\theta)
        + \dfrac{1}{105\theta}\mathcal{H}^{(1)}_{4,4,0}(\theta)\right]
        + (1-e)\dfrac{\sqrt{2\pi}}{5}\mathcal{H}^{(1)}_{2,2,0}(\theta)$\\
        %%%
        \hline
        %%%
        $\widetilde{\Lambda}_{xx}
        = \widetilde{\Lambda}_{xx}^{(0)}(\theta)
        + \widetilde{\Lambda}_{xx}^{(1)}(\theta)\Pi_{xx}^*
        + \widetilde{\Lambda}_{xx}^{(2)}(\theta)\Pi_{yy}^*
        + \widetilde{\Lambda}_{xx}^{(3)}(\theta)\Pi_{xy}^*
        + \widetilde{\Lambda}_{xx}^{(4)}(\theta)\left(\Pi_{xx}^*\right)^2
        + \widetilde{\Lambda}_{xx}^{(5)}(\theta)\left(\Pi_{yy}^*\right)^2
        + \widetilde{\Lambda}_{xx}^{(6)}(\theta)\Pi_{xx}^*\Pi_{yy}^*
        + \widetilde{\Lambda}_{xx}^{(7)}(\theta)\left(\Pi_{xy}^*\right)^2$\\
        %%%
        $\widetilde{\Lambda}_{yy}
        = \widetilde{\Lambda}_{yy}^{(0)}(\theta)
        + \widetilde{\Lambda}_{yy}^{(1)}(\theta)\Pi_{xx}^*
        + \widetilde{\Lambda}_{yy}^{(2)}(\theta)\Pi_{yy}^*
        + \widetilde{\Lambda}_{yy}^{(3)}(\theta)\Pi_{xy}^*
        + \widetilde{\Lambda}_{yy}^{(4)}(\theta)\left(\Pi_{xx}^*\right)^2
        + \widetilde{\Lambda}_{yy}^{(5)}(\theta)\left(\Pi_{yy}^*\right)^2
        + \widetilde{\Lambda}_{yy}^{(6)}(\theta)\Pi_{xx}^*\Pi_{yy}^*
        + \widetilde{\Lambda}_{yy}^{(7)}(\theta)\left(\Pi_{xy}^*\right)^2$\\
        %%%
        %%%
        $\widetilde{\Lambda}_{xx}^{(0)}(\theta)
        = \widetilde{\Lambda}_{yy}^{(0)}(\theta)
        = -\dfrac{2\sqrt{2\pi}}{35\theta}\left[\mathcal{F}_{4,2,0}(\theta)
        + \mathcal{G}_{3,1,0}(\theta)
        + \dfrac{5}{66\theta}\mathcal{G}_{5,3,0}(\theta)\right]$\\
        $\hspace{9em}
        + (1-e)\dfrac{4\sqrt{2\pi}}{3}
        \left[\mathcal{F}_{2,0,0}(\theta)
        + \dfrac{3}{140\theta}\mathcal{F}_{4,2,0}(\theta)
        + \dfrac{9}{140\theta}\mathcal{G}_{3,1,0}(\theta)
        + \dfrac{1}{616\theta^2}\mathcal{G}_{5,3,0}(\theta)\right]$\\
        %%%
        %%%
        $\widetilde{\Lambda}_{xx}^{(1)}(\theta)
        = \widetilde{\Lambda}_{yy}^{(2)}(\theta)
        = \sqrt{2\pi}
        \left[\dfrac{8}{3}\mathcal{F}_{2,0,0}(\theta)
        + \dfrac{4}{35\theta}\mathcal{G}_{3,1,0}(\theta)
        -\dfrac{8}{3}\mathcal{H}^{(1)}_{2,0,0}(\theta)
        -\dfrac{6}{35\theta}\mathcal{H}^{(2)}_{3,1,0}(\theta)\right]
        + (1-e)2\sqrt{2\pi}
        \left[\mathcal{H}^{(1)}_{2,0,0}(\theta)
        + \dfrac{3}{70\theta}\mathcal{H}^{(2)}_{3,1,0}(\theta)\right]$\\
        %%%
        %%%
        $\widetilde{\Lambda}_{xx}^{(2)}(\theta)
        = \widetilde{\Lambda}_{yy}^{(1)}(\theta)
        = -\sqrt{2\pi}
        \left[\dfrac{8}{3}\mathcal{H}^{(1)}_{0,2,0}(\theta)
        + \dfrac{6}{35\theta}\mathcal{H}^{(2)}_{1,3,0}(\theta)\right]
        + (1-e)2\sqrt{2\pi}
        \left[\mathcal{H}^{(1)}_{0,2,0}(\theta)
        + \dfrac{3}{70\theta}\mathcal{H}^{(2)}_{1,3,0}(\theta)\right]$\\
        %%%
        %%%
        $\widetilde{\Lambda}_{xx}^{(3)}(\theta)
        =\widetilde{\Lambda}_{yy}^{(3)}(\theta)
        = -\dfrac{2\sqrt{2}\pi}{105\sqrt{\theta}}(5-9e)$\\
        %%%
        %%%
        $\widetilde{\Lambda}_{xx}^{(4)}(\theta)
        =\widetilde{\Lambda}_{yy}^{(5)}(\theta)
        = \dfrac{2\sqrt{2\pi}}{3}
        \left[\mathcal{H}^{(1)}_{2,0,0}(\theta)
        - \mathcal{H}^{(3)}_{2,0,0}(\theta)
        + \dfrac{3}{280\theta}\mathcal{H}^{(3)}_{4,2,0}(\theta)\right]
        + (1-e)\dfrac{\sqrt{2\pi}}{4}
        \mathcal{H}^{(3)}_{2,0,0}(\theta)$\\
        %%%
        %%%
        $\widetilde{\Lambda}_{xx}^{(5)}(\theta)
        =\widetilde{\Lambda}_{yy}^{(4)}(\theta)
        = \dfrac{2\sqrt{2\pi}}{3}
        \left[- \mathcal{H}^{(3)}_{0,2,0}(\theta)
        + \dfrac{3}{280\theta}\mathcal{H}^{(3)}_{2,4,0}(\theta)\right]
        + (1-e)\dfrac{\sqrt{2\pi}}{4}
        \mathcal{H}^{(3)}_{0,2,0}(\theta)$\\
        %%%
        %%%
        $\widetilde{\Lambda}_{xx}^{(6)}(\theta)
        =\widetilde{\Lambda}_{yy}^{(6)}(\theta)
        = \dfrac{2\sqrt{2\pi}}{3}
        \left[\mathcal{H}^{(1)}_{0,2,0}(\theta)
        - 2\mathcal{H}^{(4)}_{2,0,0}(\theta)
        + \dfrac{3}{140\theta}\mathcal{H}^{(4)}_{4,2,0}(\theta)\right]
        + (1-e)\dfrac{\sqrt{2\pi}}{2}
        \mathcal{H}^{(4)}_{2,0,0}(\theta)$\\
        %%%
        %%%
        $\widetilde{\Lambda}_{xx}^{(7)}(\theta)
        =\widetilde{\Lambda}_{yy}^{(7)}(\theta)
        = \sqrt{2\pi}
        \left[\dfrac{4}{15}\mathcal{F}_{2,2,0}(\theta)
        - \dfrac{8}{35}\mathcal{F}_{4,2,0}(\theta)
        + \dfrac{1}{231\theta}\mathcal{F}_{6,4,0}(\theta)\right]
        + (1-e)\dfrac{3\sqrt{2\pi}}{35}\mathcal{F}_{4,2,0}(\theta)$\\
        %%%
        \hline\hline
        %%%
        $\mathcal{F}_{s,t,u}(\theta)
        = {}_2F_2\left(\dfrac{s+1}{2}, \dfrac{t+1}{2}; \dfrac{s+t+u+3}{4}, \dfrac{s+t+u+5}{4}; -\dfrac{1}{16\theta}\right)$\\
        %%%
        $\mathcal{G}_{s,t,u}(\theta)
        = {}_3F_3\left(\dfrac{1}{2},\dfrac{s+2}{2}, \dfrac{t+2}{2}; \dfrac{3}{2}, \dfrac{s+t+u+5}{4}, \dfrac{s+t+u+7}{4};-\dfrac{1}{16\theta}\right)$\\
        %%%
        \hline
        %%%
        $\mathcal{H}^{(1)}_{s,t,u}(\theta)
        = \dfrac{1}{3+s+t+u}\left[(1+s)\mathcal{F}_{s+2,t,u}(\theta)
        -(1+u)\mathcal{F}_{s,t,u+2}(\theta)\right]$\\
        %%%
        $\mathcal{H}^{(2)}_{s,t,u}(\theta)
        =\dfrac{1}{5+s+t+u}
        \left[(2+s)\mathcal{G}_{s+2,t,u}(\theta)
        - (1+u)\mathcal{G}_{s,t,u+2}(\theta)\right]$\\
        %%%
        $\mathcal{H}^{(3)}_{s,t,u}(\theta)
        = \dfrac{1}{(3+s+t+u)(5+s+t+u)}
        \left[(1+s)(3+s)\mathcal{F}_{s+4,t,u}(\theta)
        -2(1+s)(1+u)\mathcal{F}_{s+2,t,u+2}(\theta)
        + (1+u)(3+u)\mathcal{F}_{s,t,u+4}(\theta)\right]$\\
        %%%
        $\mathcal{H}^{(4)}_{s,t,u}(\theta)
        =\dfrac{1}{(3+s+t+u)(5+s+t+u)}$\\
        $\hspace{5em}\times\left[(1+s)(1+t)\mathcal{F}_{s+2,t+2,u}(\theta)
        -(1+s)(1+u)\mathcal{F}_{s+2,t,u+2}(\theta)
        -(1+t)(1+u)\mathcal{F}_{s,t+2,u+2}(\theta)
        +(1+u)(3+u)\mathcal{F}_{s,t,u+4}(\theta)\right]$\\
        %%%
        \hline
    \end{tabular}
    \label{table:list_of_functions}
\end{table}
%%%%%%%%%%%%%%%%%%%%%%%%%%%%%%

%%%%%%%%%%%%%%%%%%%%%%%%%%%%%%
%%%%%%%%%%%%%%%%%%%%%%%%%%%%%%

Similarly, we can evaluate each component of the collisional stress $\sigma_{\alpha\beta}^c$.
The expression of $\sigma_{xy}^c$ can be evaluated as follows:
\begin{align}
    \int d\hat{\bm{k}}\hat{k}_x\hat{k}_y 
    \widetilde{I}^{(2)}\left(\hat{\bm{k}}\right)
    &= -\frac{1}{2\sqrt{\pi\theta}}
    \mathcal{A}_{2,2,0}(\theta)
    +\frac{1}{2}\mathcal{B}_{1,1,0}(\theta)
    +\frac{1}{4\theta}\mathcal{B}_{3,3,0}(\theta)
    +\frac{1}{2}\mathcal{Q}_{1,1,0}\left(\theta, \overleftrightarrow{\Pi^*}\right)
    + \frac{1}{16\sqrt{\pi\theta}}
    \mathcal{R}_{2,2,0}\left(\theta, \overleftrightarrow{\Pi^*}\right)\nonumber\\
    %%%
    %%%
    &= -\frac{2\sqrt{\pi}}{15\sqrt{\theta}}
    \left[F_{2,2,0}(\theta) + \mathcal{G}_{1,1,0}(\theta)
    + \frac{1}{14\theta}
    \mathcal{G}_{3,3,0}(\theta)\right]
    - \frac{2\sqrt{\pi}}{15\sqrt{\theta}}
    \mathcal{H}^{(2)}_{1,1,0}(\theta)
    \left(\Pi_{xx}^* + \Pi_{yy}^*\right)
    + \frac{4\pi}{15}\Pi_{xy}^*\nonumber\\
    %%%
    &\hspace{1em}
    + \frac{\sqrt{\pi}}{60\sqrt{\theta}}\mathcal{H}^{(3)}_{2,2,0}(\theta)\left[\left(\Pi_{xx}^*\right)^2+\left(\Pi_{yy}^*\right)^2\right]
    + \frac{\sqrt{\pi}}{30\sqrt{\theta}}\mathcal{H}^{(4)}_{2,2,0}(\theta)\Pi_{xx}^*\Pi_{yy}^*
    + \frac{\sqrt{\pi}}{105\sqrt{\theta}}\mathcal{F}_{4,4,0}(\theta)\left(\Pi_{xy}^*\right)^2.
\end{align}
Table~\ref{table:list_of_functions2} lists the coefficient functions $\widetilde{\Pi}^{c(i)}(\theta)$ and $\widetilde{\Pi}_{\alpha\beta}^{c(i)}(\theta)$, together with the auxiliary functions.

%%%%%%%%%%%%%%%%%%%%%%%%%%%%%%
\begin{table}[htbp]
    \centering
    \caption{List of the expressions of $\widetilde{\Pi}_{xy}^{c(i)}$ ($i=0,1,\cdots$) and auxiliary functions.}
    \begin{tabular}{l}
        \hline
        $\widetilde{\Pi}^{c}
        = \widetilde{\Pi}^{c(0)}(\theta)
        + \widetilde{\Pi}^{c(1)}(\theta)\Pi_{xy}^*
        + \widetilde{\Pi}^{c(2)}(\theta)
        \left(\Pi_{xx}^* + \Pi_{yy}^*\right)\Pi_{xy}^*$\\
        %%%
        %%%
        $\widetilde{\Pi}^{c(0)}(\theta)
        = \dfrac{2\pi}{3}\left(1+\dfrac{1}{30\theta}\right)$\\
        %%%
        %%%
        $\widetilde{\Pi}^{c(1)}(\theta)
        = -\dfrac{4\sqrt{\pi}}{45\sqrt{\theta}}
        \mathcal{G}_{1,1,0}(\theta)$\\
        %%%
        %%%
        $\widetilde{\Pi}^{c(2)}(\theta)
        = \dfrac{\sqrt{\pi}}{45\sqrt{\theta}}
        \mathcal{H}_{2,2,0}^{(1)}(\theta)$\\
        %%%
        %%%
        %%%
        \hline
        $\widetilde{\Pi}_{xy}^{c}
        = \widetilde{\Pi}_{xy}^{c(0)}(\theta)
        + \widetilde{\Pi}_{xy}^{c(1)}(\theta)\left(\Pi_{xx}^* + \Pi_{yy}^*\right)
        + \widetilde{\Pi}_{xy}^{c(2)}(\theta)\Pi_{xy}^*
        + \widetilde{\Pi}_{xy}^{c(3)}(\theta)
        \left[\left(\Pi_{xx}^*\right)^2+ \left(\Pi_{yy}^*\right)^2\right]
        + \widetilde{\Pi}_{xy}^{c(4)}(\theta)\Pi_{xx}^*\Pi_{yy}^*
        + \widetilde{\Pi}_{xy}^{c(5)}(\theta)\left(\Pi_{xy}^*\right)^2$\\
        %%%
        %%%
        $\widetilde{\Pi}_{xy}^{c(0)}(\theta)
        = -\dfrac{2\sqrt{\pi}}{15\sqrt{\theta}}
        \left[F_{2,2,0}(\theta) + \mathcal{G}_{1,1,0}(\theta)
        + \dfrac{1}{14\theta}
        \mathcal{G}_{3,3,0}(\theta)\right]$\\
        %%%
        %%%
        $\widetilde{\Pi}_{xy}^{c(1)}(\theta)
        = - \dfrac{2\sqrt{\pi}}{15\sqrt{\theta}}
        \mathcal{H}^{(2)}_{1,1,0}(\theta)$\\
        %%%
        %%%
        $\widetilde{\Pi}_{xy}^{c(2)}(\theta)
        = \dfrac{4\pi}{15}$\\
        %%%
        %%%
        $\widetilde{\Pi}_{xy}^{c(3)}(\theta)
        = \dfrac{\sqrt{\pi}}{60\sqrt{\theta}}
        \mathcal{H}^{(3)}_{2,2,0}(\theta)$\\
        %%%
        %%%
        $\widetilde{\Pi}_{xy}^{c(4)}(\theta)
        = \dfrac{\sqrt{\pi}}{30\sqrt{\theta}}
        \mathcal{H}^{(4)}_{2,2,0}(\theta)$\\
        %%%
        %%%
        $\widetilde{\Pi}_{xy}^{c(5)}(\theta)
        = \dfrac{\sqrt{\pi}}{105\sqrt{\theta}}
        \mathcal{F}_{4,4,0}(\theta)$\\
        %%%
        %%%
        %%%
        \hline
        %%%
        $\widetilde{\Pi}_{xx}^{c}
        = \widetilde{\Pi}_{xx}^{c(0)}(\theta)
        + \widetilde{\Pi}_{xx}^{c(1)}(\theta)\Pi_{xx}^*
        + \widetilde{\Pi}_{xx}^{c(2)}(\theta)\Pi_{xy}^*
        + \widetilde{\Pi}_{xx}^{c(3)}(\theta)
        \Pi_{xx}^*\Pi_{xy}^*
        + \widetilde{\Pi}_{xx}^{c(4)}(\theta)
        \Pi_{yy}^*\Pi_{xy}^*$\\
        %%%
        %%%
        $\widetilde{\Pi}_{yy}^{c}
        = \widetilde{\Pi}_{yy}^{c(0)}(\theta)
        + \widetilde{\Pi}_{yy}^{c(1)}(\theta)\Pi_{yy}^*
        + \widetilde{\Pi}_{yy}^{c(2)}(\theta)\Pi_{xy}^*
        + \widetilde{\Pi}_{yy}^{c(3)}(\theta)
        \Pi_{xx}^*\Pi_{xy}^*
        + \widetilde{\Pi}_{yy}^{c(4)}(\theta)
        \Pi_{yy}^*\Pi_{xy}^*$\\
        %%%
        %%%
        $\widetilde{\Pi}_{xx}^{c(0)}(\theta)
        = \widetilde{\Pi}_{yy}^{c(0)}(\theta)
        = \dfrac{2\pi}{3}\left(1+\dfrac{3}{70\theta}\right)$\\
        %%%
        %%%
        $\widetilde{\Pi}_{xx}^{c(1)}(\theta)
        = \widetilde{\Pi}_{yy}^{c(1)}(\theta)
        = \dfrac{4\pi}{15}$\\
        %%%
        %%%
        $\widetilde{\Pi}_{xx}^{c(2)}(\theta)
        = \widetilde{\Pi}_{yy}^{c(2)}(\theta)
        = -\dfrac{4\sqrt{\pi}}{35\sqrt{\theta}}
        \mathcal{G}_{3,1,0}(\theta)$\\
        %%%
        %%%
        $\widetilde{\Pi}_{xx}^{c(3)}(\theta)
        = \widetilde{\Pi}_{yy}^{c(4)}(\theta)
        = \dfrac{\sqrt{\pi}}{35\sqrt{\theta}}
        \mathcal{H}^{(1)}_{4,2,0}(\theta)$\\
        %%%
        %%%
        $\widetilde{\Pi}_{xx}^{c(4)}(\theta)
        = \widetilde{\Pi}_{yy}^{c(3)}(\theta)
        = \dfrac{\sqrt{\pi}}{35\sqrt{\theta}}
        \mathcal{H}^{(1)}_{2,4,0}(\theta)$\\
        \hline
    \end{tabular}
    \label{table:list_of_functions2}
\end{table}
%%%%%%%%%%%%%%%%%%%%%%%%%%%%%%

%%%%%%%%%%%%%%%%%%%%%%%%%%%%%%
\section{Effect of the truncation}\label{sec:truncation}
%%%%%%%%%%%%%%%%%%%%%%%%%%%%%%
In Ref.~\cite{Takada20}, the collisional moments were expanded in powers of $\epsilon:=1/\sqrt{\theta}$, and the resulting expansions were truncated at sixth order.
In the present work, these quantities are expressed in terms of generalized hypergeometric functions \eqref{eq:Lambda_hypergeometric} and \eqref{eq:Pi_c_hypergeometric}, allowing their evaluation without finite-order truncation in $\epsilon$.
In this Appendix, we expand these expressions in powers of $\epsilon$ and truncate them at order $N_\mathrm{c}$.
We then examine how the resulting steady-state predictions converge to those obtained from the generalized hypergeometric representation and quantify the errors remaining at each truncation order.

For each $N_\mathrm{c}$, we solve the corresponding steady-state moment equations and evaluate the dimensionless temperature $\theta$, the dimensionless viscosity $\eta^*$, and the normal stress differences $N_1$ and $N_2$.
We define their relative errors as
\begin{equation}
    \mathcal E_X^{(N_{\mathrm c})}
    :=
    \frac{\left|X^{(N_{\mathrm c})}-X^{(\infty)}\right|}{\left|X^{(\infty)}\right|},
    \qquad
    X\in\{\theta,\eta^*,N_1,N_2\},
    \label{eq:truncation_error}
\end{equation}
where $X^{(N_\mathrm{c})}$ is obtained using the expressions truncated at order $N_\mathrm{c}$, and $X^{(\infty)}$ is obtained using the generalized hypergeometric representation.
For brevity, the relative error in $\eta^*$ is denoted by $\mathcal{E}_\eta^{(N_{\mathrm c})}$.

%%%%%%%%%%%%%%%%%%%%%%%%%%%%%%
\begin{figure}[htbp]
    \centering
    \includegraphics[width=0.95\linewidth]{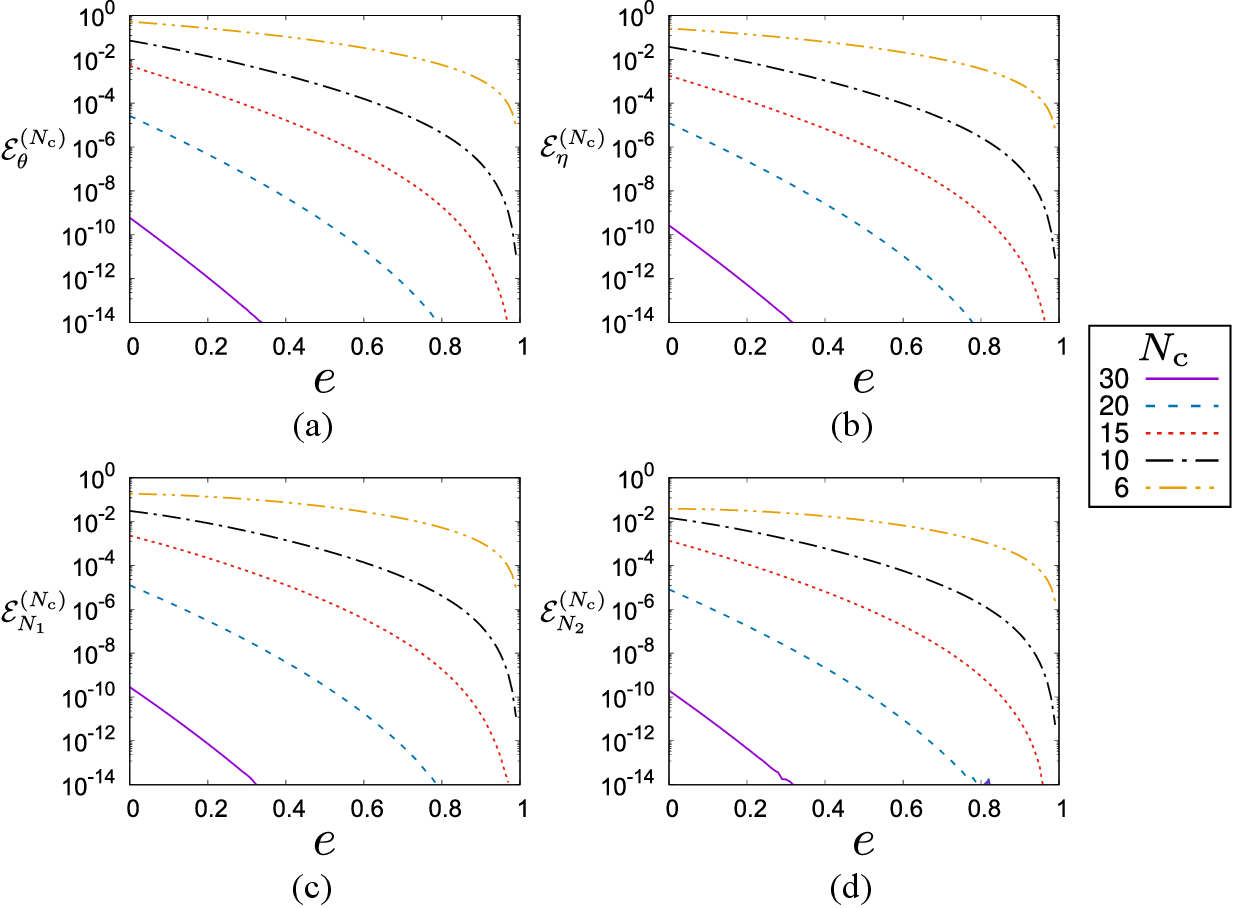}
    \caption{
    Relative errors defined in Eq.~\eqref{eq:truncation_error} for (a) the dimensionless temperature $\theta$, (b) the dimensionless viscosity $\eta^*$, (c) the normal stress difference $N_1$, and (d) the normal stress difference $N_2$, as functions of the restitution coefficient $e$ at $\varphi=0.30$.
    Results are shown for truncation orders $N_\mathrm{c}=6$, $10$, $15$, $20$, and $30$.
    The reference values are obtained using the generalized hypergeometric representation without finite-order truncation in $\epsilon$.
    }
    \label{fig:error}
\end{figure}
%%%%%%%%%%%%%%%%%%%%%%%%%%%%%%

Figure~\ref{fig:error} shows that increasing the truncation order rapidly reduces the relative errors in all four observables.
At a fixed truncation order, the errors generally increase as the restitution coefficient decreases.
This trend is consistent with the increase in the expansion parameter $\epsilon$ as $e$ decreases, as discussed in Appendix~\ref{sec:theta_e}.
Thus, although the sixth-order truncation gives small errors in the nearly elastic regime, appreciable errors remain in the strongly inelastic regime.

Higher truncation orders substantially improve the agreement with the generalized hypergeometric results over the entire range shown.
For $N_\mathrm{c}=20$, the relative errors are of order $10^{-5}$ or smaller for all four observables, while $N_\mathrm{c}=30$ reduces them to order $10^{-9}$ or smaller.
These results demonstrate that, at $\varphi=0.30$ and over the range of restitution coefficients examined, the steady-state predictions obtained from finite-order expansions converge to those obtained from the generalized hypergeometric expressions as the truncation order increases.
They also show the limitations of the sixth-order truncation under strongly inelastic conditions.

%%%%%%%%%%%%%%%%%%%%%%%%%%%%%%
\section{The relationship between $\theta$ and $e$}\label{sec:theta_e}
\begin{figure}[hbtp]
    \centering
    \includegraphics[width=\linewidth]{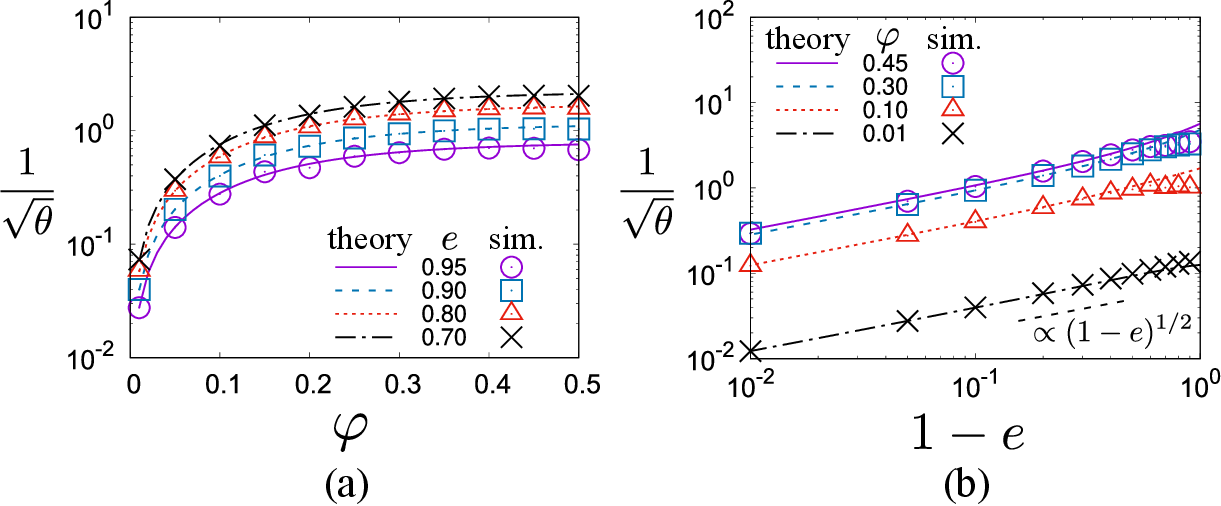}
    \caption{Plot of the expansion parameter $1/\sqrt{\theta}$ against (a) the volume fraction $\varphi$ for $e=0.70$ (inverted triangles and dash-dot line), $0.80$ (triangles and dotted line), $0.90$ (squares and dashed line), and $0.95$ (circles and solid line) and (b) the restitution coefficient $e$ for $\varphi=0.01$ (inverted triangles and dash-dot line), $0.10$ (triangles and dotted line), $0.30$ (squares and dashed line), and $0.45$ (circles and solid line).
    Here, the marks and lines represent the simulation and theoretical results, respectively.
    A guideline in panel (b) indicates $(1-e)^{1/2}$.}
    \label{fig:expansion}
\end{figure}

%%%%%%%%%%%%%%%%%%%%%%%%%%%%%%
There is a simple relation between $\theta$ and $e$ as follows.
Figure~\ref{fig:expansion} shows the dependence of the expansion parameter $\epsilon:=1/\sqrt{\theta}$ on the restitution coefficient and volume fraction.
In the dilute limit, retaining the leading collisional terms gives~\cite{Santos04}
\begin{equation}
    \theta
    =\frac{5\pi}{432}
    \frac{2+e}{(1-e)(1+e)^2(3-e)^2}
    \frac{1}{\varphi^2}.
\end{equation}
In the nearly elastic limit, this expression reduces to
\begin{equation}
    \theta
    \approx\frac{5\pi}{2304\varphi^2}\frac{1}{1-e},
    \qquad e\to1^-.
\end{equation}
Consequently,
\begin{equation}
    \epsilon
    \approx \frac{48\varphi}{\sqrt{5\pi}}\sqrt{1-e}.
\end{equation}
The prefactor above follows from the dilute approximation and is not assumed to apply at arbitrary volume fractions.
In the steady Bagnold regime, $T=md^2\dot{\gamma}^2\theta(\varphi,e)$.
Thus, the steady-state value of $\epsilon$ is independent of the imposed shear rate at fixed $\varphi$ and $e$.
Larger values of $\epsilon$ can require higher truncation orders, as assessed in Appendix~\ref{sec:truncation}.

%%%%%%%%%%%%%%%%%%%%%%%%%%%%%%
%%%%%%%%%%%%%%%%%%%%%%%%%%%%%%

%%%%%%%%%%%%%%%%%%%%%%%%%%%%%%

\end{document}